\documentclass[journal]{IEEEtran}
\usepackage{cite}
\usepackage{graphicx}
\usepackage{amsmath}
\usepackage{times}
\usepackage{multirow}
\usepackage{amssymb}
\usepackage{subfigure}
\usepackage[linesnumbered,ruled,vlined]{algorithm2e}
\usepackage{enumerate}
\usepackage{hyperref}
\usepackage{color}
\usepackage{array}
\usepackage{footnote}
\usepackage{bigstrut}
\usepackage{diagbox}
\usepackage{rotating}
\usepackage{makecell}
\usepackage{comment}

\begin{document}
	\title{Deep-JGAC: End-to-End Deep Joint Geometry and Attribute Compression for Dense Colored Point Clouds}
	\author{Yun Zhang,~\IEEEmembership{Senior Member,~IEEE}, Zixi Guo, Linwei Zhu, and C.-C. Jay Kuo,~\IEEEmembership{Life Fellow,~IEEE}

		\thanks{Yun Zhang and Zixi Guo are with the School of Electronics and Communication Engineering, Shenzhen Campus, Sun Yat-Sen University, Shenzhen 518107, China. (Email: {zhangyun2}@mail.sysu.edu.cn, guozx29@mail2.sysu.edu.cn).}
\thanks{L. Zhu is with the Shenzhen Institutes of Advanced Technology, Chinese Academy of Sciences, Shenzhen 518055, China (e-mail: lw.zhu@siat.ac.cn).}
\thanks{C.-C. Jay Kuo is with the Department of Electrical and Computer Engineering, University of Southern California, Los Angeles, CA 90089 USA (e-mail: jckuo@usc.edu).}
	}

\markboth{Submitted to IEEE Transactions on Circuits and Systems for Video Technology}%
{Guo \MakeLowercase{\textit{\textit{et al.}}}: Deep-JGAC: End-to-End Deep Joint Geometry and Attribute Compression for Dense Colored Point Clouds}

\maketitle
\begin{abstract}
		Colored point cloud becomes a fundamental representation in the realm of 3D vision. Effective Point Cloud Compression (PCC) is urgently needed due to huge amount of data. In this paper, we propose an end-to-end Deep Joint Geometry and Attribute point cloud Compression (Deep-JGAC) framework for dense colored point clouds, which exploits the correlation between the geometry and attribute for high compression efficiency. Firstly, we propose a flexible Deep-JGAC framework, where the geometry and attribute sub-encoders are compatible to either learning or non-learning based geometry and attribute encoders. Secondly, we propose an attribute-assisted deep geometry encoder that enhances the geometry latent representation with the help of attribute, where the geometry decoding remains unchanged. Moreover, Attribute Information Fusion Module (AIFM) is proposed to fuse attribute information in geometry coding. Thirdly, to solve the mismatch between the point cloud geometry and attribute caused by the geometry compression distortion, we present an optimized re-colorization module to attach the attribute to the geometrically distorted point cloud for attribute coding. It enhances the colorization and lowers the computational complexity.
Extensive experimental results demonstrate that in terms of the geometry quality metric D1-PSNR, the proposed Deep-JGAC achieves an average of 82.96\%, 36.46\%, 41.72\%, and {31.16\%} bit-rate reductions as compared to the state-of-the-art G-PCC, V-PCC, GRASP, and PCGCv2, respectively. In terms of perceptual joint quality metric MS-GraphSIM, the proposed Deep-JGAC achieves an average of 48.72\%, 14.67\%, and 57.14\% bit-rate reductions compared to the G-PCC, V-PCC, and IT-DL-PCC, respectively. The encoding/decoding time costs are also reduced by 94.29\%/24.70\%, and 96.75\%/91.02\% on average as compared with the V-PCC and IT-DL-PCC.

\end{abstract}
	
	
\section{Introduction}

\IEEEPARstart{C}{olored} point cloud represents a 3D scene with a large set of high-dimensional discrete points, where each point includes 3D geometry coordinates and high dimensional attributes, such as color, reflectance, transparency and velocity, etc. The point cloud has been one of the mainstream 3D visual representations, which has a wide scope of 3D applications, including autonomous driving \cite{auto_driving}, heritage preservation \cite{archaeological_preservation}, anomaly detection \cite{Industrial_Anomaly_Detection}, manufacturing, Virtual Reality (VR) and so on. To enable realistic visual applications, representing a high-quality 3D object typically requires millions or even billions of colored points. More points and bit-depth are required for representing larger scale of 3D scene and finer fidelity. The large volume of colored point cloud becomes one of the more critical and challenging issues that shall be solved for transmission over network, storage and computing. Moreover, the massive points in a point cloud are unstructured, non-uniform distributed and {irregularly} shaped. Meanwhile, the total number of points varies and each point has geometry and attribute components, which have different physiological properties and intrinsic correlation. To handle these challenging issues in Point Cloud Compression (PCC), a number of PCC schemes have been developed, which can generally be classified as traditional non-learning-based and learning-based schemes. In addition, since the colored point clouds {consist} of geometry and attribute, Point Cloud Geometry Coding (PCGC) and Point Cloud Attribute Coding (PCAC) schemes were developed.

\subsection{Related Work}
\subsubsection{Traditional PCC}
To compress the point cloud geometry, Garcia \textit{et al.} \cite{garcia2018intra} proposed {an} octree based PCGC exploiting correlations between parent and child nodes. However, as the tree depth increased, the number of bits representing the octree increased dramatically. Kathariya \textit{et al.} \cite{kathariya2018scalable} used binary trees to represent point clouds, which was less efficient in representing lower tree depths. Trisoup-based geometry coding schemes were combined with octrees and represented the object's surface as a series of triangular meshes. Essentially, trisoup-based methods reduced data dimensionality as the octree partitioned the point cloud into local plane-fitted blocks. Ainala \textit{et al.} \cite{ainala2016improved} combined octrees with projection methods. A coarse point cloud was presented with octrees first, which was refined with residuals from 2D planar projections. These algorithms mainly focus on geometry compression by exploiting geometry {properties}  of point clouds. However, the attribute of point clouds have not been considered and compressed.
		
To compress the point cloud attribute, Zhang \textit{et al.} \cite{zhang2014point} proposed a Graph Fourier Transform (GFT) to structure point sets with graph and transform attribute for decorrelation and higher energy compaction. To reduce the computational complexity of graph transform, De Queiroz \textit{et al.} \cite{de2016compression} proposed a Region-Adaptive Hierarchical Transform (RAHT) based attribute coding, which transformed attribute in {octree} based hierarchical order and
predicted attributes of higher level octree in Level of Detail (LoD) structured point clouds. Moreover, predictive and lifting transforms were also investigated based on the LoD representation. The predictive transform predicted higher LoD attributes with lower one, and then residuals were encoded. The lifting transform performed additional lift operations and adaptive quantization based on the predictive transform \cite{chen2023introduction}. Song \textit{et al.} \cite{song2023block} improved the predictive transform with progressive clustering and region-aware signal models. These methods optimized the PCAC by assuming geometry is lossless. The geometry distortion causing the attribute mismatch and quality degradation is not well considered.
		
Moving Picture Expert Group (MPEG) has integrated the PCGC and PCAC algorithms into two joint codecs, which are Geometry-based PCC (G-PCC) \cite{mammou2019g} and Video-based PCC (V-PCC) \cite{mammou2017video}. In geometry algorithms, the octree's simplicity and efficiency have made it one of the optional methods for G-PCC, alongside the trisoup algorithm. For attribute algorithms, RAHT, prediction transform, and lifting transform have been developed as optional in G-PCC attribute coding. In general, G-PCC voxelizes the point cloud, arranging them regularly in 3D space, thereby transforming the 3D point cloud into an octree structure. The octree structure is suitable for compression due to the high correlation between parent and child nodes.
V-PCC projects points onto different planes to form patches, which are then assembled into images containing occupancy, geometry, and attribute maps. These images are finally compressed using 2D codecs like High Efficiency Video Coding (HEVC) and the latest Versatile Video Coding (VVC).
Zhang \textit{et al.} \cite{zhang2023perceptually} proposed a perceptually weighted Rate-Distortion Optimization (RDO) scheme for V-PCC, where an adaptive Lagrange multiplier was adjusted by maximizing the perceptual quality of point cloud \cite{wu2021subjective}. Xiong \textit{et al.} \cite{9735359} designed an Error Projection Model (EPM) to solve the inconsistency between the sum of squared error based distortion and geometry quality metric for RDO. Gao \textit{et al.} \cite{10345479} improved the prediction  performance of coding unit partition by utilizing correlations of occupancy, geometry and attribute maps.
G-PCC and V-PCC eliminate the spatio-temporal redundancy in the geometry and attribute based on hybrid coding structures. However, the spatio-temporal correlation removed by G-PCC through quantization of the octree and transform coefficients is limited. The adaptation of 2D codecs to VPCC is suboptimal because 2D codecs are designed for natural images. In addition, G-PCC and V-PCC evaluate the quality of compressed point clouds using two independent geometry and attribute metrics. The perceptual quality degradation of PCC was scarcely considered.

\subsubsection{Learning-based PCC}
			
As learning-based image/video coding has witnessed a great success \cite{12,134545124,9360626}, a number of learning based PCGC and PCAC \cite{10380494,12351243} were investigated.

In learning based PCGC,
Huang \textit{et al.} \cite{huang20193d} developed a hierarchical autoencoder for point cloud geometry to achieve detail reconstruction. Gao \textit{et al.} \cite{5} proposed a neural graph sampling module that extracted latent keypoints by dynamically aggregating neighboring weights to expand associated features. Song \textit{et al.} \cite{song2023efficient} proposed a hierarchical attention structure with linear complexity in terms of context scale and maintained a global receptive field. It also introduced an intermediate grouping strategy to support parallel decoding. Wang \textit{et al.} \cite{4} introduced sparse convolution \cite{11} to accelerate computation and reduce memory costs in PCGC. Additionally, a hierarchical reconstruction method was proposed to improve compression performance. He \textit{et al.} \cite{he2022density} proposed density embedding, local position embedding and ancestor embedding to encode local geometry and density. Yu \textit{et al.} \cite{yu2023sparse} proposed a long-range-residual module in geometry compression and introduced a multi-scale geometry compression module to avoid the accumulation of reconstruction distortion. Wang \textit{et al.} \cite{wang2022sparse} proposed a hierarchical encoding and reconstruction network that utilizes multi-scale representations to extensively exploit cross-scale correlations for better context modeling. Pang \textit{et al.} \cite{pang2022grasp} proposed a Geometric Residual Analysis and Synthesis for PCC (GRASP), where a coarse layer was encoded with sparse convolutional network and an enhancement layer of residual was encoded point-based network. Generally, learning based PCGC used sparse or graph convolutions to construct learning networks for point clouds. Their performance has surpassed traditional G-PCC and V-PCC schemes \cite{yu2023sparse}. However, these PCGC schemes did not consider the correlation between attribute and geometry.

In learning based PCAC, Nguyen \textit{et al.} \cite{nguyen2023lossless} proposed a lossless PCAC network, which encoded color features sequentially and effectively utilized feature-level and point-level dependencies within the point cloud. Wang \textit{et al.} \cite{wang2023lossless} proposed a lightweight PCAC network with cross-scale, cross-group, and cross-color correlations for lossless compression, enabling parallel encoding and decoding. Sheng \textit{et al.} \cite{7} proposed a second-order point convolution module to expand the receptive field and incorporated multi-scale loss to focus attention on coarse-grained points covering the entire point cloud more effectively. Fang \textit{et al.} \cite{fang20223dac} employed the RAHT to obtain the transformed coefficients and used previously encoded attribute to model the probabilities of the un-coded coefficients. Pinheiro \textit{et al.} \cite{17} proposed a PCAC by using normalized flow and strictly reversible convolution, which enchanted the quality upper-bound. Wang \textit{et al.} \cite{8} proposed a Sparse convolution based PCAC (Sparse-PCAC), where a per-voxel autoregressive entropy model was proposed to improve the probability distribution prediction. Guo \textit{et al.} \cite{10693649} proposed a Transformer and Sparse Convolution based Module (TSCM) for PCAC, which enhanced the local dependency modeling and global representations. Meanwhile, TSCM based channel context module was proposed to predict probability distribution more accurately. These methods assumed the geometry information {was} losslessly encoded. In practice, the severe geometry distortion may degrade the performance of PCAC.

In joint PCC, Guarda \textit{et al.} \cite{guarda2023deep} proposed a deep learning-based lossy point cloud compression network, called IT-DL-PCC, to encode geometry and attribute into one bit stream. The geometry and attribute were jointly down-sampled and presented as a latent representation by feature extraction for coding. At the decoder, up-sampling and enhancement networks were used to reconstruct the geometry and attribute jointly. However, the coding efficiency could be further improved.

\subsection{Contributions of This Work}	
In this paper, we propose an end-to-end Deep Joint Geometry and Attribute Compression (Deep-JGAC) for dense colored point clouds. The main contributions are
\begin{itemize}
    \item We propose a learning-based flexible Deep-JGAC for joint PCC framework, which includes learned geometry and attribute codecs, and re-colorization. The learned attribute codec can be replaced as it is compatible with other learning or non-learning attribute codes.
	\item We propose an attribute-assisted geometry coding that enhances the geometry latent representation with the help of attribute. Moreover, an Attribute Information Fusion Module (AIFM) is proposed to fuse attribute information to geometry features.
	\item To address the attribute mismatch caused by geometry distortion, we present an optimized re-colorization module to attach the attribute to the geometrically distorted point cloud for attribute coding. It enhances the re-colorization and lowers the computational complexity.

\end{itemize}

The remainder of this paper is organized as follows. Section \ref{section2} presents the proposed Deep-JGAC, including framework, attribute-assisted geometry coding, re-colorization module and loss function. Experimental results and analysis are presented in Section \ref{exp}. Finally, Section \ref{section4} draws the conclusions.

\section{The Proposed Deep-JGAC}
\label{section2}

\subsection{Framework of The Proposed Deep-JGAC}
There are two general types of frameworks for colored PCC. One is one stream coding framework that represents geometry and attribute of point clouds jointly and {compresses} them into one bit-stream, like IT-DL-PCC. The other is the two-stream framework, which splits geometry and attribute and encodes them into two streams, such as G-PCC \cite{mammou2019g} and V-PCC\cite{mammou2017video}.
		
\begin{figure}[t]
	\centering
	\subfigure[] {\label{fig2c}
		\includegraphics[width=0.85\columnwidth]{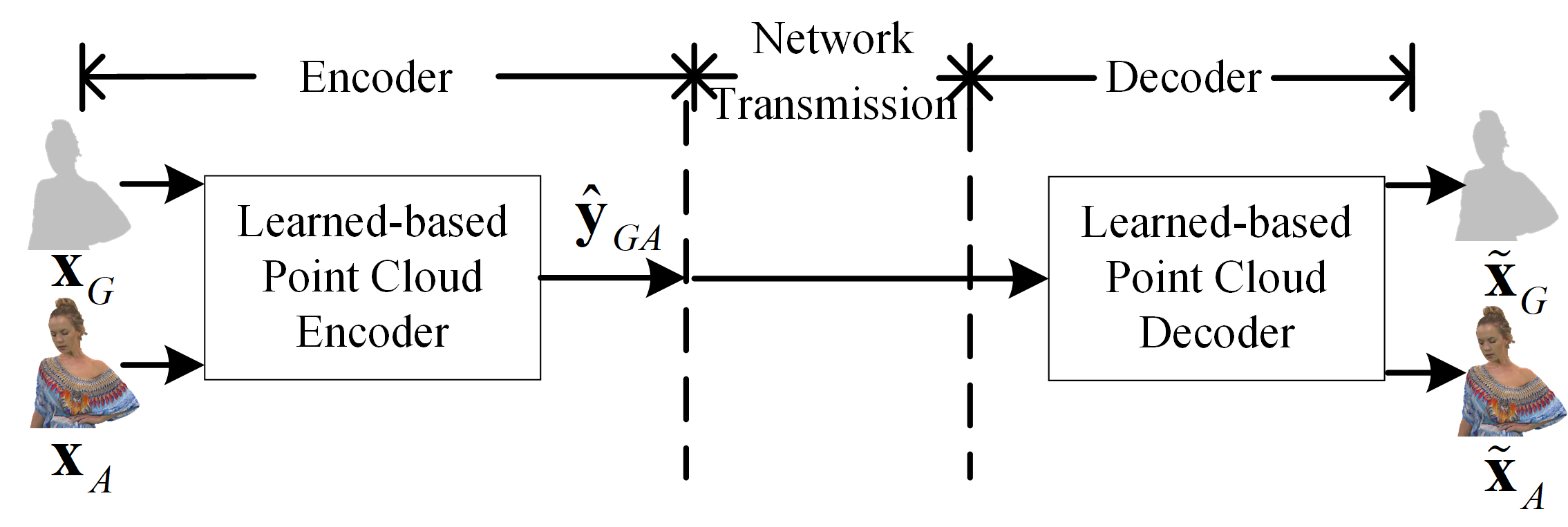}}
		\subfigure[] {\label{fig2a}
			\includegraphics[width=0.85\columnwidth]{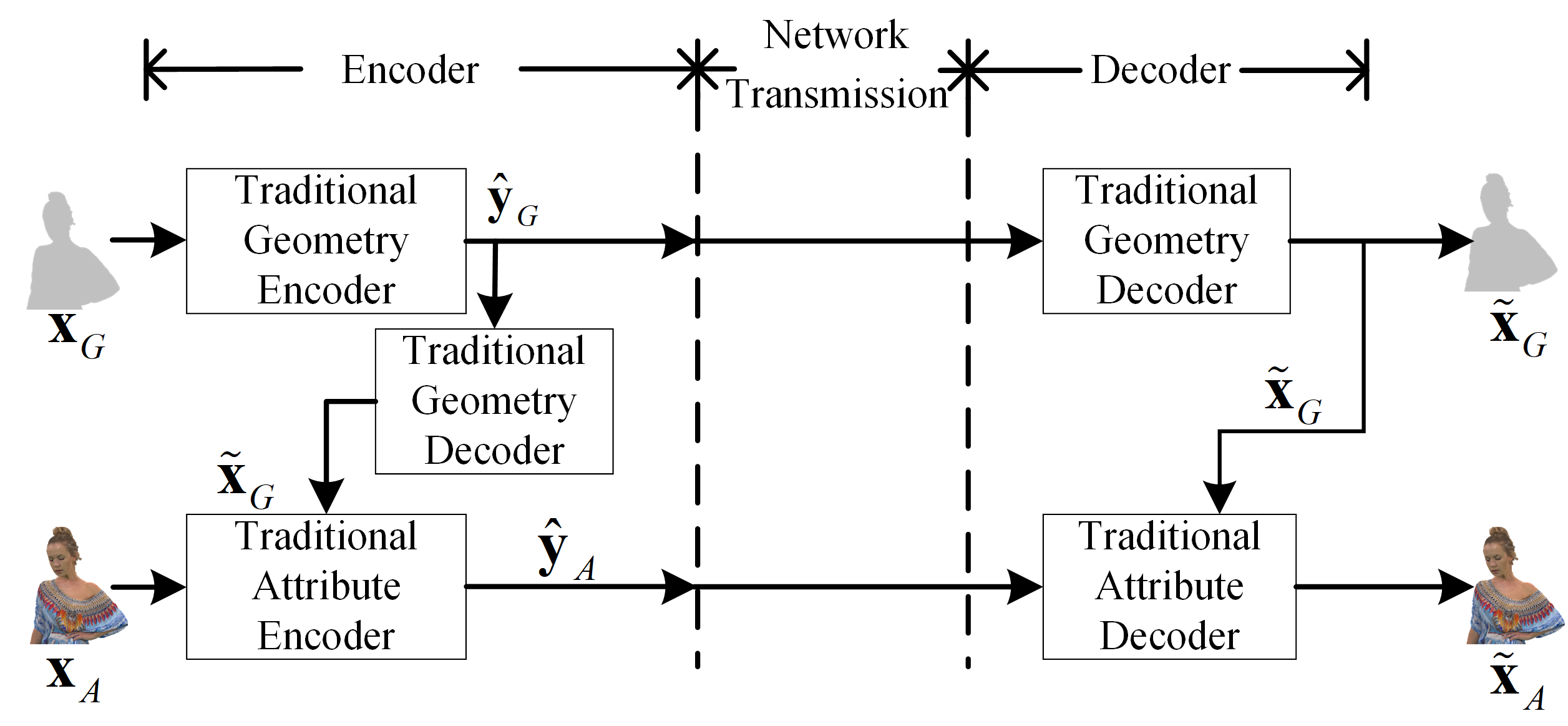}}
		\subfigure[] {\label{fig2b}
			\includegraphics[width=0.9\columnwidth]{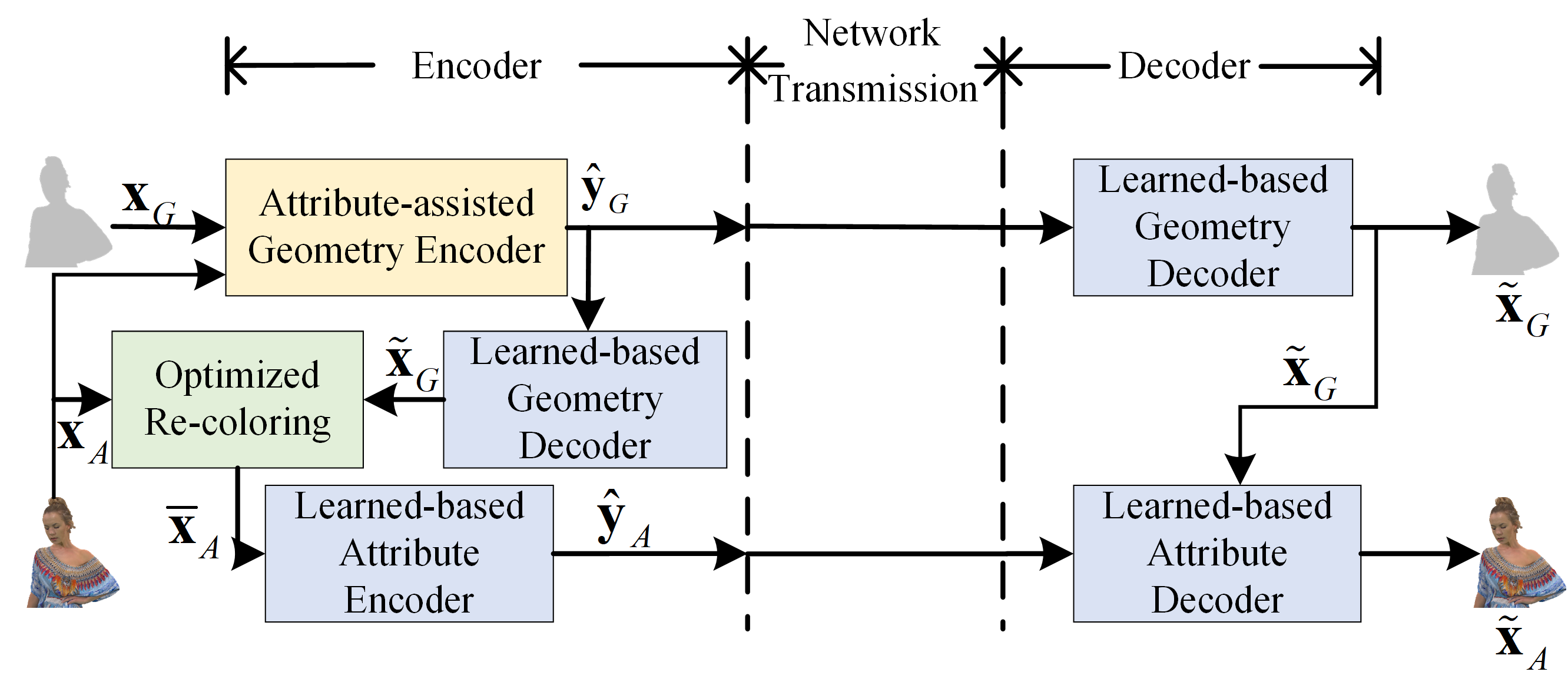}}
\caption{Frameworks of joint PCC. (a) Deep learning based one stream PCC framework  (b) General two-stream PCC framework. (c) Framework of the proposed Deep-JGAC .}
		\label{fig22}
\end{figure}
\subsubsection{Framework of IT-DL-PCC}
The IT-DL-PCC framework is shown in Fig. \ref{fig2c}. A point cloud consists of geometry and attribute, i.e., $\mathbf{x}_{GA}=\{\mathbf{x}_{G},\mathbf{x}_{A}\}$, is encoded and decoded as
\begin{equation}
\begin{cases}
    \begin{aligned}
		&\hat{\mathbf{y}}_{GA} = E(\mathbf{x}_{GA},	Q) \\
		&\tilde{\mathbf{x}}_{GA} = D(\hat{\mathbf{y}}_{GA})
	\end{aligned}
\end{cases},
\end{equation}
where $E(\cdot)$ and $D(\cdot)$ are point cloud encoder and decoder, $Q$ is a Quantization Parameter (QP). The geometry and attribute of the point cloud are treated as a whole and input to a learning-based point cloud encoder. Leveraging the powerful representation capabilities of deep learning, a latent representation $\hat{\mathbf{y}}_{GA}$ containing both geometry and attribute is obtained. At the decoder, $\tilde{\mathbf{x}}_{GA}$ is reconstructed from the latent representation $\hat{\mathbf{y}}_{GA}$ by the deep learning-based point cloud decoder. The framework is simple and straightforward. Also, the internal correlation and importance difference between geometry and attribute are implicitly exploited with deep convolution and sampling. However, the coding performance still could be improved as compared to the V-PCC.
\subsubsection{Framework of G-PCC and V-PCC}
Fig. \ref{fig2a} shows a general framework of G-PCC and V-PCC. Since point clouds consist of geometry $\mathbf{x}_{G}$ and attribute $\mathbf{x}_{A}$, they are separately represented and encoded. The coding process is modelled as
\begin{equation}
\begin{cases}
    \begin{aligned}
		&\hat{	\mathbf{y}}_{G} = E_{G}(	\mathbf{x}_G, Q_G) \\
	&\hat{	\mathbf{y}}_{A}=E_{A}({\mathbf{x}}_{A},{	\tilde{\mathbf{x}}_{G},Q_A})\\
		&\tilde{	\mathbf{x}}_{G} = D_{G}(\hat{\mathbf{y}}_{G}) \\
		&\tilde{	\mathbf{x}}_{A} =D_{A}(\hat{\mathbf{y}}_{A})
	\end{aligned}
\end{cases},
\end{equation}
where $E_G(\cdot)$ and $D_G(\cdot)$ are geometry encoder and decoder, $E_A(\cdot)$ and $D_A(\cdot)$ are attribute encoder and decoder, $Q_G$ and $Q_A$ are quantization for geometry and attribute. At the encoder, geometry compression is performed firstly. Then, attribute $\mathbf{x}_{A}$, is encoded with the support of the reconstructed geometry ${\tilde{\mathbf{x}}_{G}}$. Bit allocation is performed manually by adjusting $Q_G$ and $Q_A$. Two streams ${\hat{\mathbf{y}}_{G}}$ and ${\hat{\mathbf{y}}_{A}}$ are multiplexed and transmitted to remote client for sequential decoding. Finally, the point cloud is reconstructed with the decoded geometry and attribute. The attribute is available at encoder, however, it is not effectively exploited in geometry coding.

\begin{figure*}[!t]
\centering
  \includegraphics[width=6in]{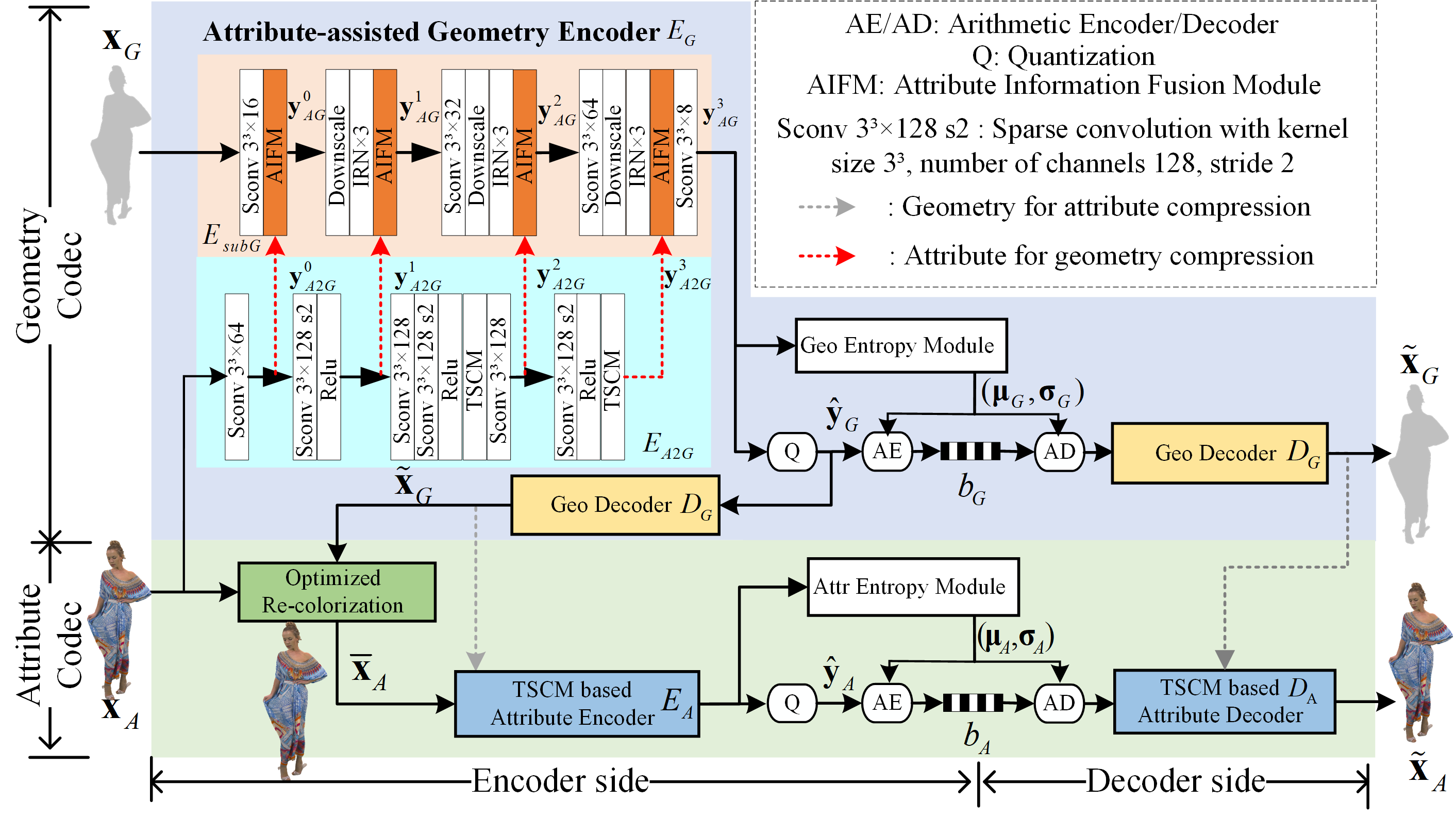}
  \caption{Network architecture of the proposed Deep-JGAC.}
  \label{fig3}
\end{figure*}			
\subsubsection{Framework of Deep-JGAC}
Based on the two-stream PCC framework for G-PCC and V-PCC, we propose the Deep-JGAC framework, as shown in Fig. \ref{fig2b}. Firstly, there is an explicit dependency between geometry and attribute which can be exploited for geometry coding. Secondly, learning based geometry and attribute encoders are developed. Thirdly, we proposed a re-colorization module to handle the mismatch caused by the geometry compression distortion. The Deep-JGAC framework is mathematically modeled as
\begin{equation}
\begin{cases}
	\begin{aligned}
		&\hat{	\mathbf{y}}_{G} = E_{G}(	\mathbf{x}_G,	\mathbf{x}_A,Q_G) \\
		&\tilde{	\mathbf{x}}_{G} = D_{G}(\hat{\mathbf{y}}_{G}) \\
		&{\overline{\mathbf{x}}_{A}} = C(	\mathbf{x}_{A},{	\tilde{\mathbf{x}}_{G}})\\
        &\hat{	\mathbf{y}}_{A} = E_{A}(\overline{\mathbf{x}}_{A},{	\tilde{\mathbf{x}}_{G},Q_A})\\
		&\tilde{	\mathbf{x}}_{A} =D_{A}( \hat{	\mathbf{y}}_{A} )
	\end{aligned}
\end{cases},
\end{equation}
where $E_G(\cdot)$ is an attribute-assisted geometry encoder, $C(\cdot)$ is re-colorization module. $Q_A$ and $Q_G$ denote quantization for geometry and attribute. $\hat{	\mathbf{y}}_{G}$ and $\hat{	\mathbf{y}}_{A}$ represent the quantified latent representations of the geometry and attribute. ${\overline{\mathbf{x}}_{A}}$ denotes the re-colored point cloud. $\tilde{\mathbf{x}}_{G}$ and $\tilde{\mathbf{x}}_{A}$ represent reconstructed geometry and attribute. Note that $E_G(\cdot)$ is input with both $\mathbf{x}_G$ and $\mathbf{x}_A$, while the geometry decoding remains unchanged with the output of $\tilde{	\mathbf{x}}_{G}$. Also, the learned attribute codec can be replaced as it is compatible with other learning or non-learning attribute codes.

\begin{figure}[t]
\centering
	\subfigure[] {\label{fig9aa}
	\includegraphics[scale=0.6]{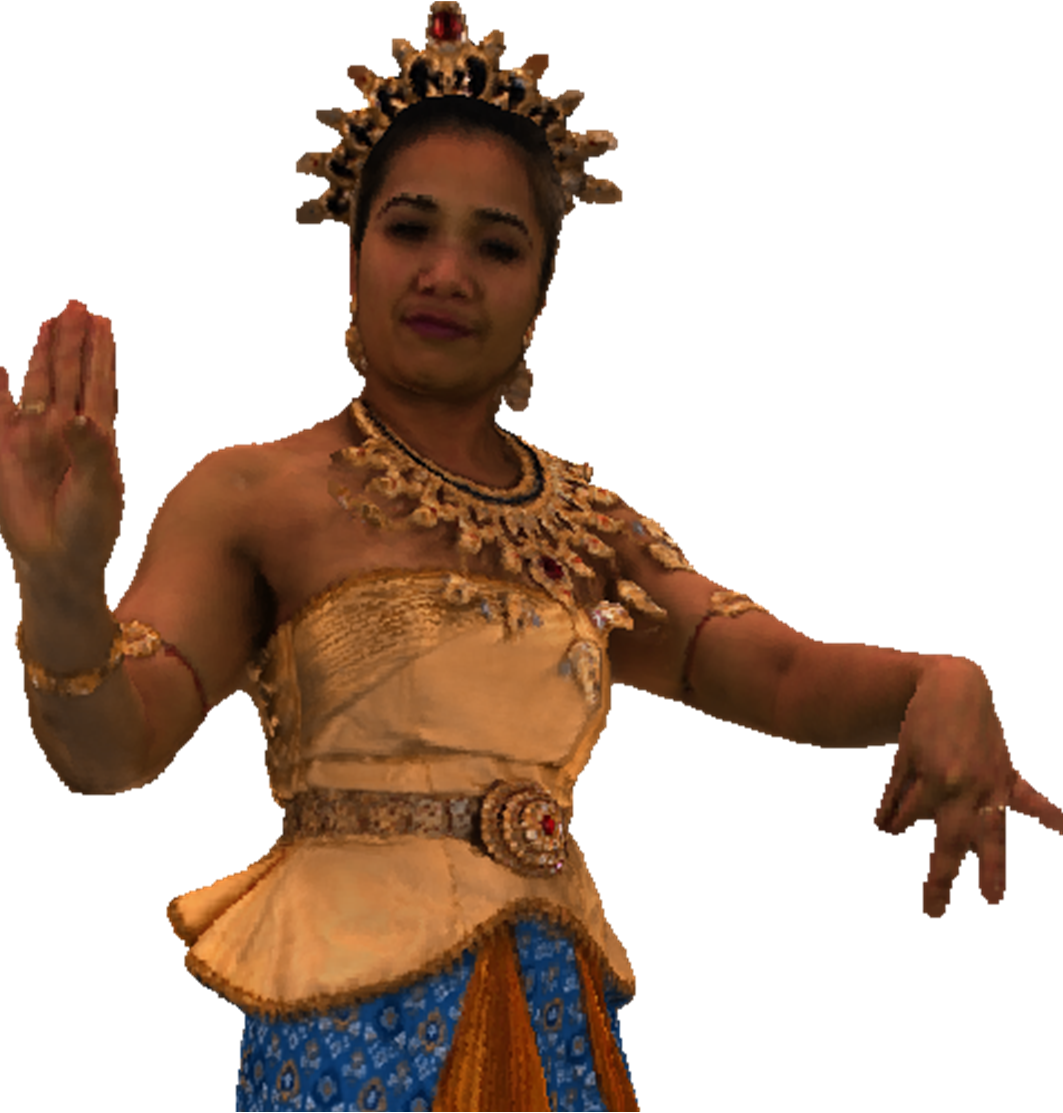}}
	\subfigure[] {\label{fig9b}
	\includegraphics[scale=0.6]{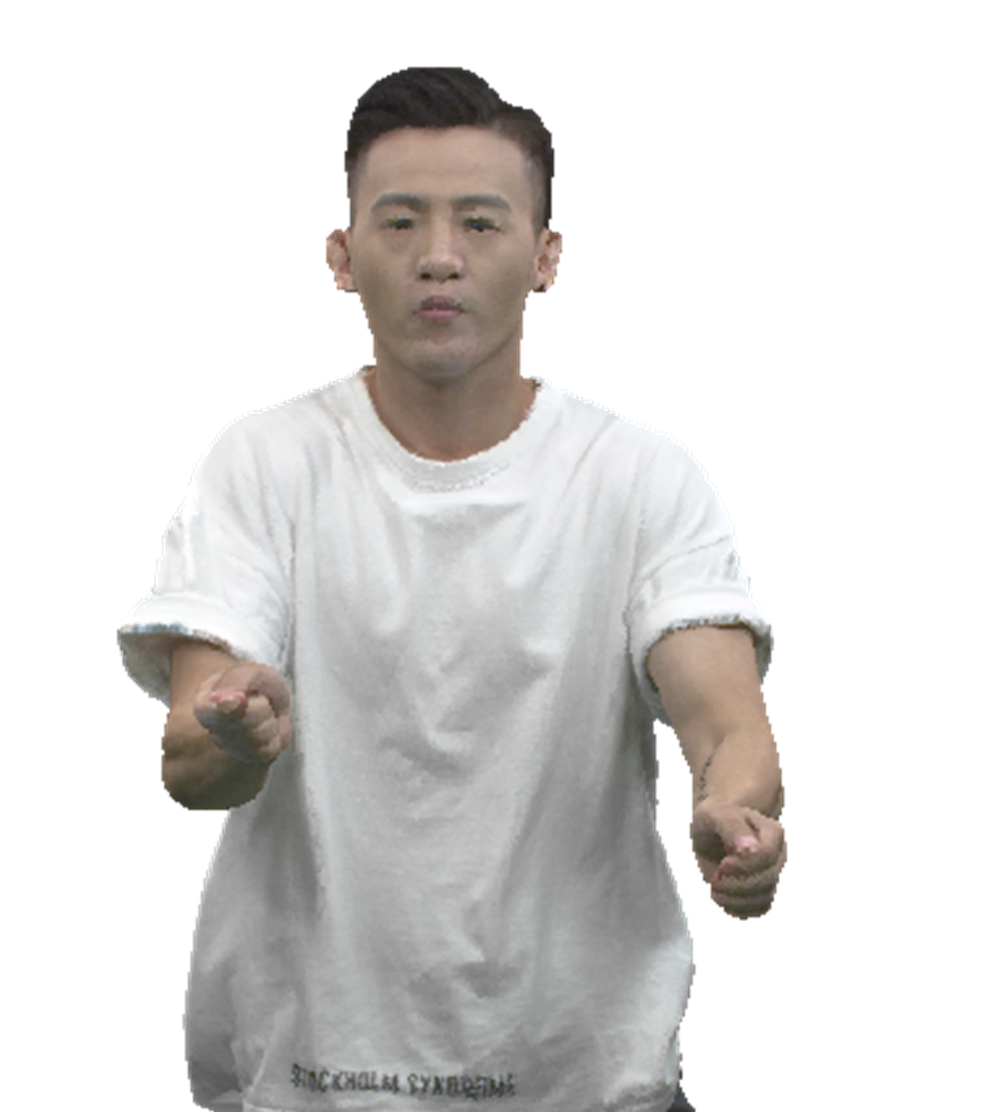}}		
	\subfigure[] {\label{fig9a}
	\includegraphics[scale=0.3]{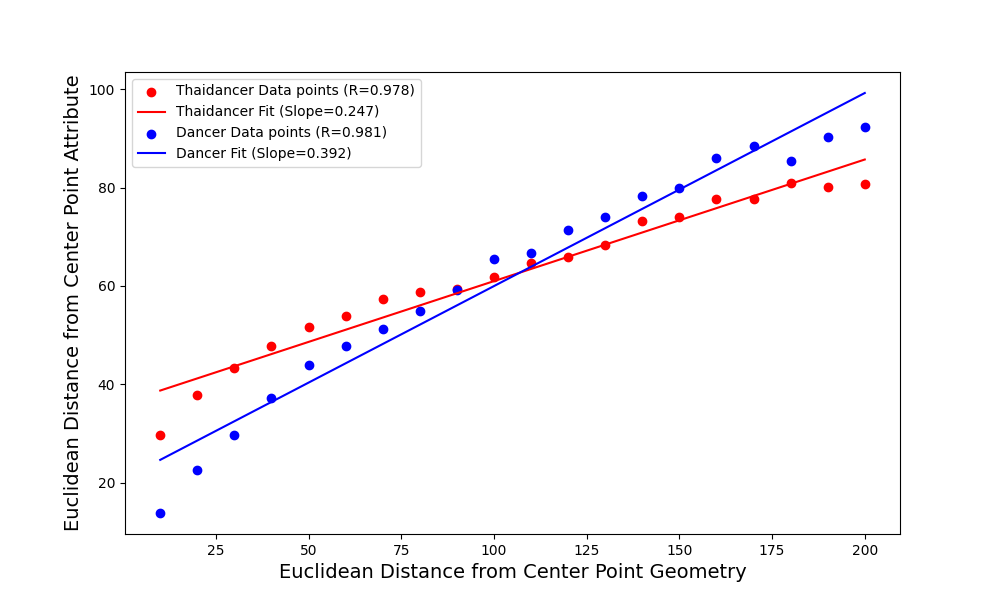}}
	\caption{Correlation analysis between attribute and geometry. (a) \textit{Thaidancer}.   (b) \textit{Dancer}.   (c) Scatter plot and fitting lines of geometry distance and attribute difference. $R^2$ represents the correlation coefficient.}
	\label{fig99}
\end{figure}
Fig. \ref{fig3} shows the structure of the Deep-JGAC. Firstly, the Deep-JGAC obtains $\hat{\mathbf{y}}_{G}$ through the attribute-assisted geometry encoder $E_G(\cdot)$, where attribute is exploited as prior information by AIFM.
$\hat{\mathbf{y}}_G$ is used to reconstruct $\tilde{\mathbf{x}}_{G}$ through the geometry decoder. Based on the reconstructed geometry $\tilde{\mathbf{x}}_{G}$ and the original point cloud attribute ${\mathbf{x}}_{A}$, the re-colored point cloud $\overline{\mathbf{x}}_{A}$ is obtained by the proposed re-colorization. $\overline{\mathbf{x}}_{A}$ is then passed through the attribute encoder to obtain $\mathbf{y}_A$. {Both the quantified geometry and attribute latent representations, $\mathbf{\hat{y}}_G$ and $\mathbf{\hat{y}}_A$, are transmitted to the decoder.
At the decoder, $\mathbf{\hat{y}}_G$ is used to reconstruct $\tilde{\mathbf{x}}_{G}$ through the geometry decoder. Finally, $\tilde{\mathbf{x}}_{G}$ serves as spatial coordinate support, and $\mathbf{\hat{y}}_A$} is used to reconstruct $\tilde{\mathbf{x}}_{A}$ through the attribute decoder.
		
Overall, our Deep-JGAC consists of geometry and attribute latent representations, which are different from other deep learning based PCC networks {only focusing on} one of the two components. Moreover, compared to the one-stream latent representation in IT-DL-PCC \cite{guarda2023deep}, our framework is more flexible and effective. Firstly, the geometry and attribute have different properties that shall be exploited with more efficient network and loss design. Secondly, the importance difference between geometry and attribute could be exploited with bit allocation. Thirdly, the deep-JGAC is compatible with non-learning based codecs, where the attribute codec can be replaced with V-PCC attribute codec for higher efficiency.

\subsection{ Attribute-assisted Geometry Coding}
\begin{figure}[t]
\centering
{
	\includegraphics[width=1\columnwidth]{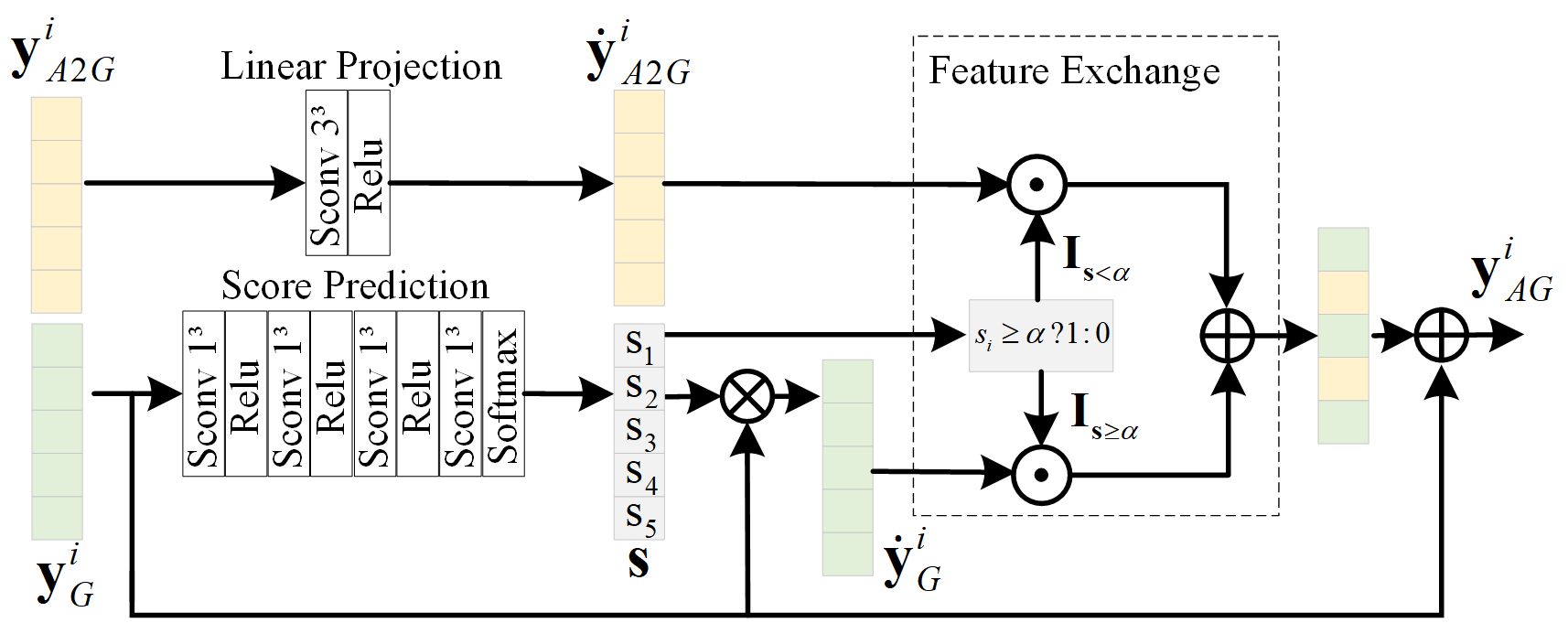}}
	\caption{ Structure of the proposed AIFM.}
\label{fig4}
\end{figure}
		
In point cloud attribute compression, the attribute is attached to the geometry as the geometry provides coordinate prior. Additionally, spatially neighboring points may have similar attribute, which is regarded as spatial correlation to be exploited in attribute compression. On the contrary, points with similar attribute may also lead to similar geometry, i.e., spatially neighboring. To validate this assumption, we randomly selected 100 anchor points in a point cloud and performed a correlation analysis between the attribute and geometry. Figs. \ref{fig99} (a) and (b) are two point clouds for statistical analysis. Fig. \ref{fig99} (c) shows the correlation for the selected 100 anchor points. The x-axis represents the radial distance from the selected anchor points, ranging from 10 to 200 with step 10. The $y$-axis represents the average Euclidean difference between anchor points and their 20 neighboring points within the radial distance. The dots are real collected data and lines are fitted from the dots. For the two tested point clouds, we can observe that the attribute difference increases linearly with the radial distance from the selected anchor points. The correlation coefficients of the fitting lines are 0.978 and 0.981 for the \emph{Thaidancer} and \emph{Dancer}, respectively, which indicate a high correlation between geometry and attribute. This correlation  actually could be exploited not only for attribute coding, but also for the geometry coding.
		
In addition, there are redundant features in extracted features from deep networks \cite{10283855}. This implies that replacing redundant features with more representative ones can enhance performance. In addition, due to the correlation between geometry and attribute, the attribute can help extract more effective geometry features. To this end, we propose the attribute-assisted geometry coding, which utilizes attribute information to extract more representative features for geometry compression. Fig. \ref{fig3} shows the framework of the proposed attribute-assisted geometry coding, which primarily consists of the geometry sub-encode $E_{subG}(\cdot)$ and the attribute prior extraction module $E_{A2G}(\cdot)$. The geometry sub-encoder $E_{subG}(\cdot)$ is adapted from geometry compression networks and the downscale Inception-Residual Network (IRN) \cite{4}. The $E_{A2G}(\cdot)$ is employed to extract multi-scale attribute prior features $\mathbf{y}^i_{A2G}$, $i\in\{0,1,2,3\}$. {TSCMs \cite{10693649} in $E_{A2G}(\cdot)$ are used to enhance the attribute prior feature extraction.} The prior attribute information $\mathbf{y}^i_{A2G}$ is integrated to the geometry encoder with an AIFM, which help to extract more effective multi-layer geometry features $\mathbf{y}^i_{AG}$ with the multi-modality feature interaction \cite{multimodal}. Consequently, a more compact and representative geometry latent representation $\hat{\mathbf{y}}_{G}$ is generated, which improves the geometry reconstruction and reduces coding bit rate.
	

 Fig. \ref{fig4} shows the structure of AIFM. Specifically, the attribute prior $\mathbf{y}^i_{A2G}$ is projected to $\dot{\mathbf{y}}^{i}_{A2G}$ which aligns with geometry features for subsequent feature exchange through a linear projection layer $LP(\cdot)$, which is $\dot{\mathbf{y}}^{i}_{A2G} = LP(\mathbf{y}^i_{A2G})$. $LP(\cdot)$ consists of one sparse convolution layer and one non-linear activation. Then, the initial geometry features $\mathbf{y}^i_{G}$ are processed with a score prediction $SP(\cdot)$ to predict importance weights $\mathbf{s}$, indicating which features in $\mathbf{y}^i_{G}$ are more informative. The weighted geometry feature $\dot{\mathbf{y}}^{i}_{G}$ is presented as
\begin{equation}
		\begin{cases}
			\begin{aligned}	
				&\dot{\mathbf{y}}^{i}_{G}=\mathbf{s}\odot \mathbf{y}^i_{G} \\
                &\mathbf{s}=SP(\mathbf{y}^i_{G}) \\
			\end{aligned}
		\end{cases},
\end{equation}
where $\odot$ denotes the element-wise multiplication for two vectors, $SP(\cdot)$ is composed of four sparse convolutional layers with non-linear activation functions. The purpose of using the importance weights is to emphasize more important features and to reduce the weights of less informative features.

Then, the weighted geometry features $\dot{\mathbf{y}}^{i}_{G}$, the aligned attribute $\dot{\mathbf{y}}^{i}_{A2G}$ and the weights $\mathbf{s}$ are input to {a} feature exchange module for feature exchange and fusion. In this module, if the importance score $\mathbf{s}$ of features in $\dot{\mathbf{y}}^{i}_{G}$ is larger than the threshold $\alpha$, it is considered to be informative and kept. Otherwise, the features in $\dot{\mathbf{y}}^{i}_{G}$ that have low weights are exchanged with those of $\dot{\mathbf{y}}^{i}_{A2G}$ as they are less informative. Finally, the features after exchanging are added with the $\mathbf{y}^i_{G}$ to generate the final attribute assisted geometry feature $\mathbf{y}^i_{AG}$, which is presented as
\begin{equation}
	\mathbf{y}^i_{AG}=\dot{\mathbf{y}}^{i}_{G} \odot \mathbf{I}_{\mathbf{s} \geq {\alpha}}+ \dot{\mathbf{y}}^{i}_{A2G} \odot \mathbf{I}_{\mathbf{s} < {\alpha}}+\mathbf{y}^i_{G},
\end{equation}
where $\mathbf{I}_{\mathbf{s} < {\alpha}}$ is a binary matrix where the elementary value is 1 when $s_i<\alpha$, otherwise it is 0; $\mathbf{I}_{\mathbf{s} \geq \alpha}$ is a complementary set of $\mathbf{I}_{\mathbf{s} < \alpha}$ , i.e., $\overline{\mathbf{I}_{\mathbf{s} < \alpha}}$. This AIFM firstly incorporates the attribute features in presenting the geometry based on correlation between attribute and geometry. Secondly, the more informative geometry features are enhanced with the feature weighting. Overall, attribute assisted geometry feature $\mathbf{y}^i_{AG}$ are more effective for geometry coding.


\subsection{Re-colorization Module for Attribute Coding}
\label{sec:recolor}
\begin{figure}[t]
	\centering
	\subfigure[] {\label{fig6a}
		\includegraphics[width=1\columnwidth]{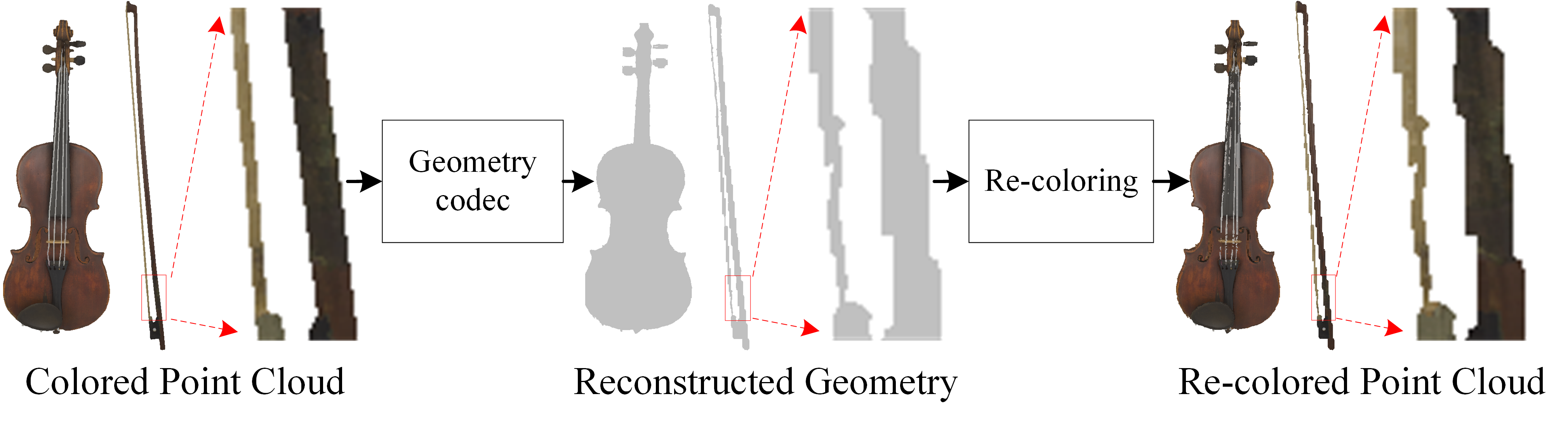}}
	\subfigure[] {\label{fig6b}
		\includegraphics[width=1\columnwidth]{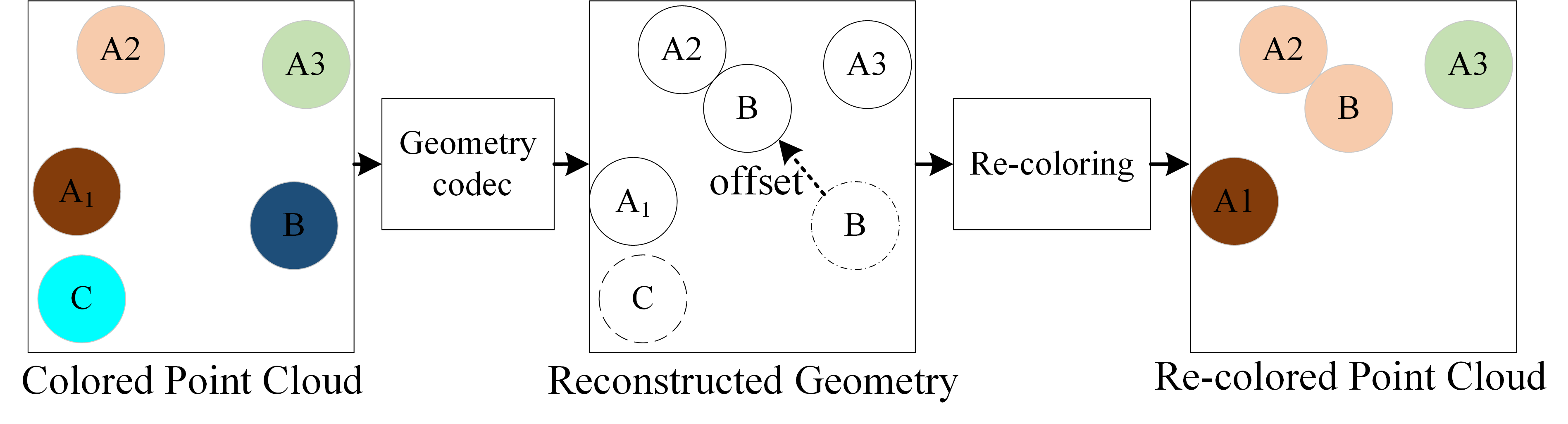}}
	\caption{Geometrical mismatch caused by geometry compression. (a) Visualization geometrical mismatch. (b) Example of geometrical mismatch and re-colorization, where point C is missing and point B is geometry distorted.}
	\label{fig6}
\end{figure}

The geometry coding is performed before the attribute coding, and the reconstructed geometry shall be used as coordinate support for the attribute coding. However, the compression distortion in geometry will cause the geometrical distortions, including the displacement of the points and point missing. As shown in Fig. \ref{fig6a}, we can observe that some points are displaced with geometrical offsets and some points are missing in violin strings. Fig. \ref{fig6b} shows an example of the geometry distortion. {Points (A1 to A3) have no or very limited geometrical distortion, but the bottom blue point C is missing and the dark blue point B moves to another place due to serious geometrical distortion. However, there is no attribute information for this point B at the current position in the source pristine point clouds. The attribute given to this point B} will no longer be the dark blue, but more probably be the orange that {is} close to its neighboring point A2. To assign the attribute value to geometrically mismatched points, we therefore propose a re-colorization module which facilitates the subsequent attribute coding in Deep-JGAC.

\begin{figure}[t]
	\centering
	\subfigure[] {\label{fig5a}
		\includegraphics[width=0.8\columnwidth]{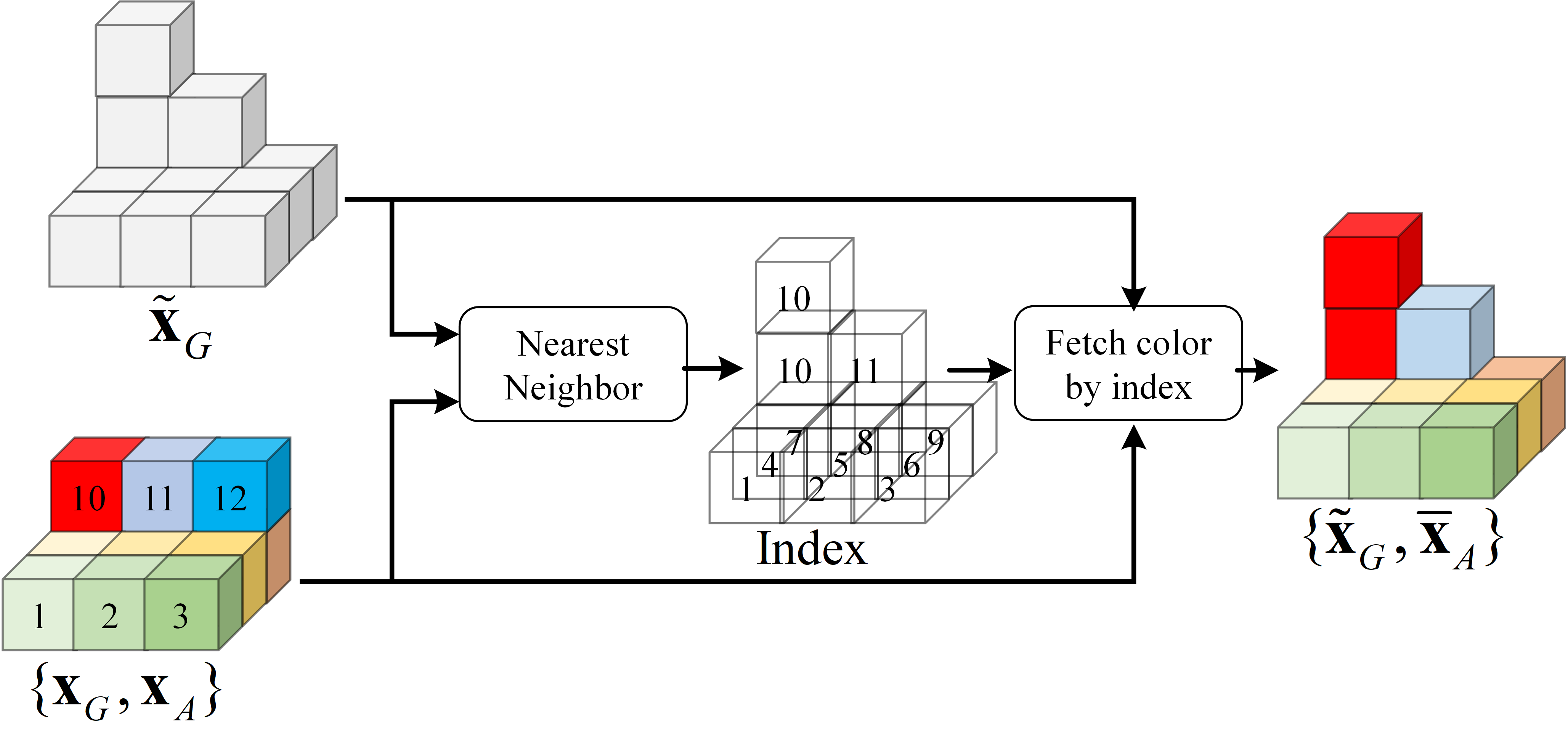}}
	\subfigure[] {\label{fig5b}
		\includegraphics[width=0.8\columnwidth]{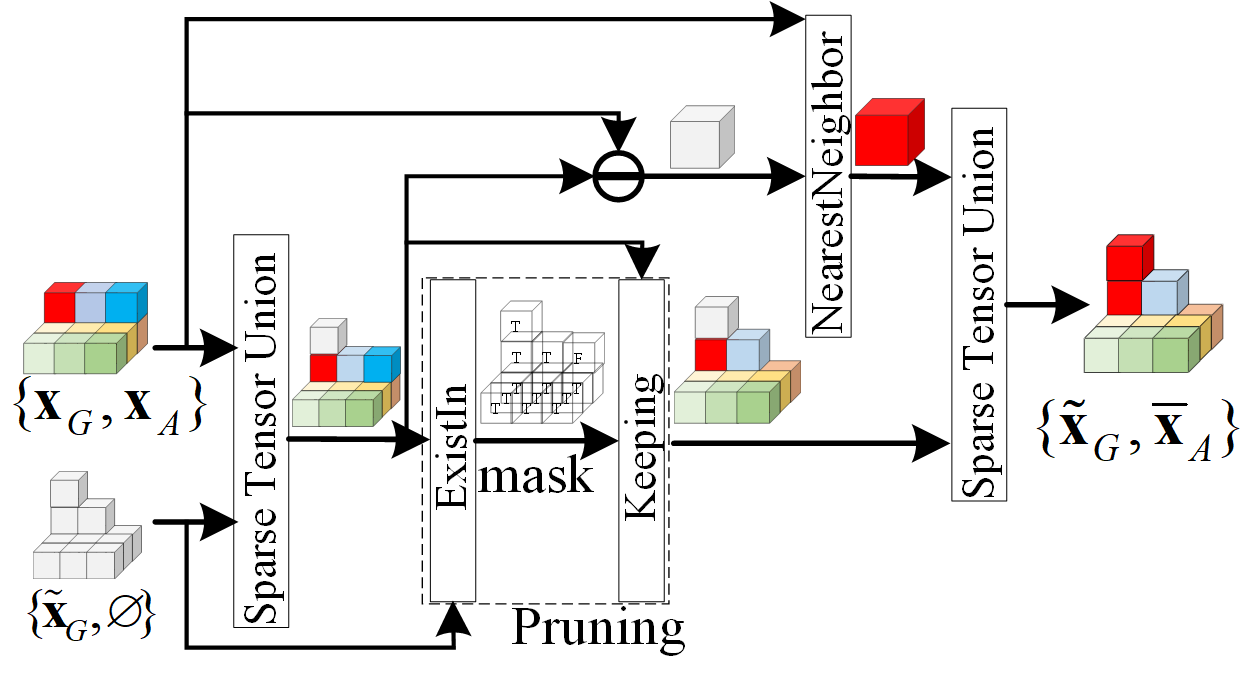}}
	\caption{ Flowcharts of the re-colorization. (a) Conventional re-colorization. (b) The optimized re-colorization in Deep-JGAC. Similar to \textit{Pruning}, the $\boldsymbol{\ominus}$ function requires an additional reverse operation to invert the mask. { The \textit{Sparse Tensor Union} operation adds features with the same geometry position while retaining features with different geometry positions.} }
	\label{fig5}
\end{figure}

The re-colorization module is to re-color the reconstructed point cloud geometry $\tilde{\mathbf{x}}_{G}$ based {on} the original colored point cloud $\{\mathbf{x}_A,\mathbf{x}_{G}\}$, resulting in a re-colored point cloud $\{\overline{\mathbf{x}}_{A},\tilde{\mathbf{x}}_{G}\}$. Then, the updated attribute $\overline{\mathbf{x}}_{A}$ will be encoded with an attribute encoder. Fig. \ref{fig5} shows the flowchart of the re-colorization module, including the conventional re-colorization \cite{mammou2019g} and the proposed re-colorization. In Fig. \ref{fig5a}, the Nearest Neighbor Algorithm (NNA) is performed for each point based on the geometrical coordinate $\tilde{\mathbf{x}}_{G}$ and calculated closeness indices of the neighboring points. Then, attribute $\mathbf{x}_A$ of nearest neighboring points, i.e., the smallest distance between $\mathbf{x}_{G}$ and $\tilde{\mathbf{x}}_{G}$, is assigned to the points at coordinate $\tilde{\mathbf{x}}_{G}$, which generates the re-colored point cloud $\{\overline{\mathbf{x}}_{A},\tilde{\mathbf{x}}_{G}\}$. However, calculating the NNA for each point is time-consuming, especially to process large-scale point cloud with millions or even billions of points.

To reduce the computational complexity of the re-colorization while maintaining quality, we proposed an optimized re-colorization module, as shown in Fig. \ref{fig5b}. The proposed Deep-JGAC uses a sparse convolution engine MinkowskiEngine \cite{11}, which maintains an internal geometrical coordinate manager. Thus, we leverage the coordinate manager to perform a faster re-colorization. After geometry coding, there usually {is} a difference between the source $\mathbf{x}_G$ and reconstructed geometry $\tilde{\mathbf{x}}_{G}$ due to the compression distortion. So, points from $\mathbf{x}_G$ and $\tilde{\mathbf{x}}_{G}$ are mostly overlapped, which can be exploited for low complexity optimization. For the overlapped points, the correspondence between $\mathbf{x}_A$ and $\mathbf{x}_G$ can be directly mapped to $\mathbf{x}_A$ and $\tilde{\mathbf{x}}_{G}$ based on MinkowskiEngine. For the non-overlapped points caused by geometrical displacement, the NNA is applied and the {nearest}  $\mathbf{x}_A$ is assigned to them. Finally, the attribute of the overlapped points and non-overlapped points are combined to generate the updated attribute $\overline{\mathbf{x}}_{A}$. Since there is a large ratio of overlapping points and only a small number of points need updates with NNA, significant computational complexity can be reduced. Meanwhile, the re-colorization quality of $\overline{\mathbf{x}}_{A}$ remains the same as that of the conventional re-colorization in Fig. \ref{fig5a}.

\begin{table}[!t]
	\caption{Processing Time Comparison between Conventional and Optimized Re-colorizations. [Uint:s]}
	\label{tab21}
	\centering
	\setlength{\tabcolsep}{0.2mm}
	\begin{tabular}{|c|c|c|c|c|}
		\hline
		\multirow{2}{*}{Sequence} & \multicolumn{2}{c|}{Conventional Re-colorization} &  \multicolumn{2}{c|}{Proposed Re-colorization} \\
\cline{2-5}
		&       Re-colorization                 &        Total Enc. Time                        &     Re-colorization      &    Total Enc. Time                                                                              \\ \hline
		Basket.                   & 70.22     &      76.64                                                                                & 6.16         &    11.84                                                                           \\ \hline
		Boxer                     & 10.64      &      12.73                                                                               & 0.90         &     3.50                                                                          \\ \hline
		Dancer                    & 61.79     &       67.62                                                                               & 5.37        &      9.14                                                                          \\ \hline
		Soldier                   & 10.25     &     11.92                                                                                 & 1.30        &     4.13                                                                          \\ \hline
		\textbf{Average}          & \textbf{38.23}   &\textbf{42.23}                                                                       & \textbf{3.43}   &   \textbf{7.15   }                                                                      \\ \hline
	\end{tabular}
\end{table}
	\subsection{Loss Function}

To validate the computational complexity of the optimized re-colorization, we performed a re-colorization for {reconstructed
point cloud geometry, whose geometry is compressed with attribute-assisted geometry encoder and reconstructed}. Four colored point clouds were tested, which are $Basket.$, $Boxer$, $Dancer$ and $Soldier$. Table \ref{tab21} presents the computing time comparison for the conventional re-colorization and optimized re-colorization. We can observe that the average processing time of the conventional ANN based re-colorization and total encoding time are 38.23s and 42.23s, respectively. The conventional re-colorization occupies 90.53\% of the total encoding time, which is highly complex. The proposed optimized re-colorization only requires 3.43s on average to process a point cloud, where 91.03\% computational complexity of the re-colorization is reduced.

The proposed Deep-JGAC aims to compress both the geometry and attribute of point clouds effectively and the optimization objective is {to} minimize the total loss of geometry and attribute, which is
\begin{equation}
	\label{all}
	\begin{gathered}
		L=L_G+\omega L_A,
	\end{gathered}
\end{equation}
where $\omega$ is an importance weight of attribute as comparing with the geometry. For the geometry coding, the geometry Rate-Distortion (RD) loss is calculated as
\begin{equation}
		\label{G}
	 	\begin{gathered}
		 L_G=R(\hat{\mathbf{y}}_G)+	{\lambda_G} D_G(\mathbf{x}_G,\tilde{\mathbf{x}}_G),
	 	\end{gathered}
\end{equation}
where $D_G(\mathbf{x}_G,\tilde{\mathbf{x}}_G)$ denotes the distortion between the original and reconstructed point cloud geometry, in which the Binary Cross-Entropy (BCE) loss is used. $\lambda_G$ is a multiplier to balance the bitrate and distortion of compressed geometry. $R(\hat{\mathbf{y}}_G)$ represents the geometry bitrate. For the attribute coding, the RD loss is calculated as
\begin{equation}
\label{A}
\begin{gathered}
	  L_A=R(\hat{\mathbf{y}}_A)+	{\lambda_A} D_A(\mathbf{x}_A,\tilde{\mathbf{x}}_A),
\end{gathered}
  \end{equation}
where $ D_A(\mathbf{x}_A,\tilde{\mathbf{x}}_A)$ denotes the distortion between the original and reconstructed point cloud attribute, in which the Mean Squared Error (MSE) is used.  $ R(\hat{\mathbf{y}}_A)$ represents the attribute bitrate, $\lambda_A$ is used to balance $D_A$ and $R(\hat{\mathbf{y}}_A)$.
	
In training the Deep-JGAC, we firstly trained the geometry encoder $E_G$ and decoder $D_G$  by using geometry loss in Eq.\ref{G}. Then, the attribute encoder $E_A$ and decoder $D_A$ were trained by using the attribute loss in Eq.\ref{A}. Finally, the overall networks in Deep-JGAC model were jointly trained by using the joint loss in Eq.\ref{all}. In this paper, $\omega$ was set as 1 to give equal importance to attribute and geometry.
\section{Experimental Results and Analyses}
\label{exp}
\subsection{Experimental Settings}
\subsubsection{Benchmark Coding Methods}
The Deep-JGAC can be used stand-alone for geometry coding and compared with other PCGC schemes. Four state-of-the-art PCGC schemes, PCGCv2 \cite{4}, GRASP \cite{pang2022grasp}, V-PCC v18\cite{mammou2017video}, and G-PCC v23 \cite{mammou2019g}, were used to compare with the geometry coding in the proposed Deep-JGAC. Then, Joint PCC schemes, including IT-DL-PCC \cite{guarda2023deep}, V-PCC, and G-PCC were compared with our Deep-JGAC. In IT-DL-PCC \cite{guarda2023deep}, it consists of the joint PCC and {an} enhancement network to reduce the compression artifact. For fair comparison, the post-enhancement network of the IT-DL-PCC was de-activated as the proposed Deep-JGAC, V-PCC, and G-PCC did not have.
In G-PCC, the geometry coding mode of G-PCC was set as octree mode, and the attribute coding mode was RAHT transform mode. The QP pairs $\{QP_G,QP_A\}$ for geometry and attribute in G-PCC were set as \{0.25, 36\}, \{0.5, 32\}, \{0.6, 30\}, \{0.75, 28\}, \{0.875, 26\}, respectively. The V-PCC was set intra coding, and HEVC was used as the base 2D encoder, where $\{QP_G,QP_A\}$ were set as \{27, 38\}, \{22, 32\}, \{17, 28\}, \{12, 24\}, \{7, 20\}, respectively. Meanwhile, attribute coding efficiency was evaluated by counting the quality and bit rate of attribute for joint PCC. Sparse-PCAC was additionally compared.
\subsubsection{Databases}
\begin{figure}[t]
		\centering
		\subfigure[] {\label{fig7a}
			\includegraphics[width=0.9\columnwidth]{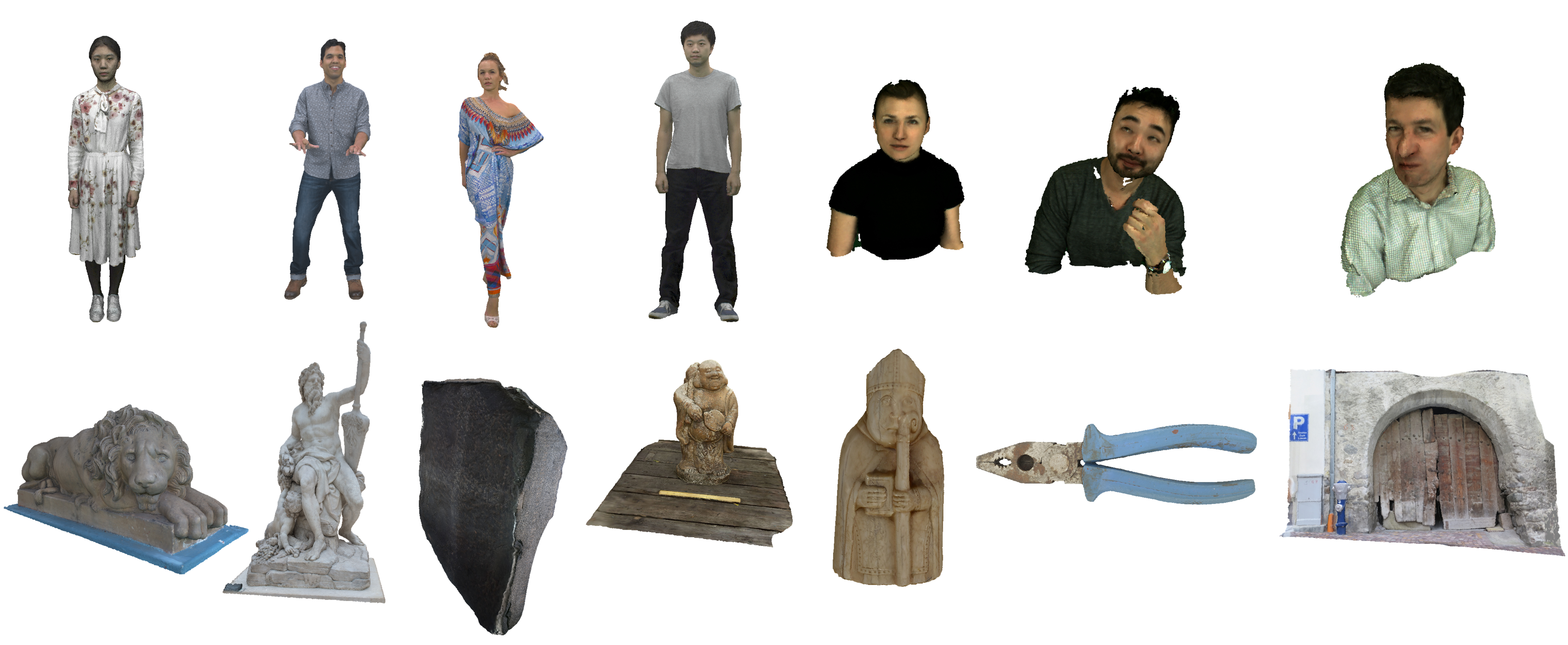}}
		\subfigure[] {\label{fig7b}
			\includegraphics[width=0.9\columnwidth]{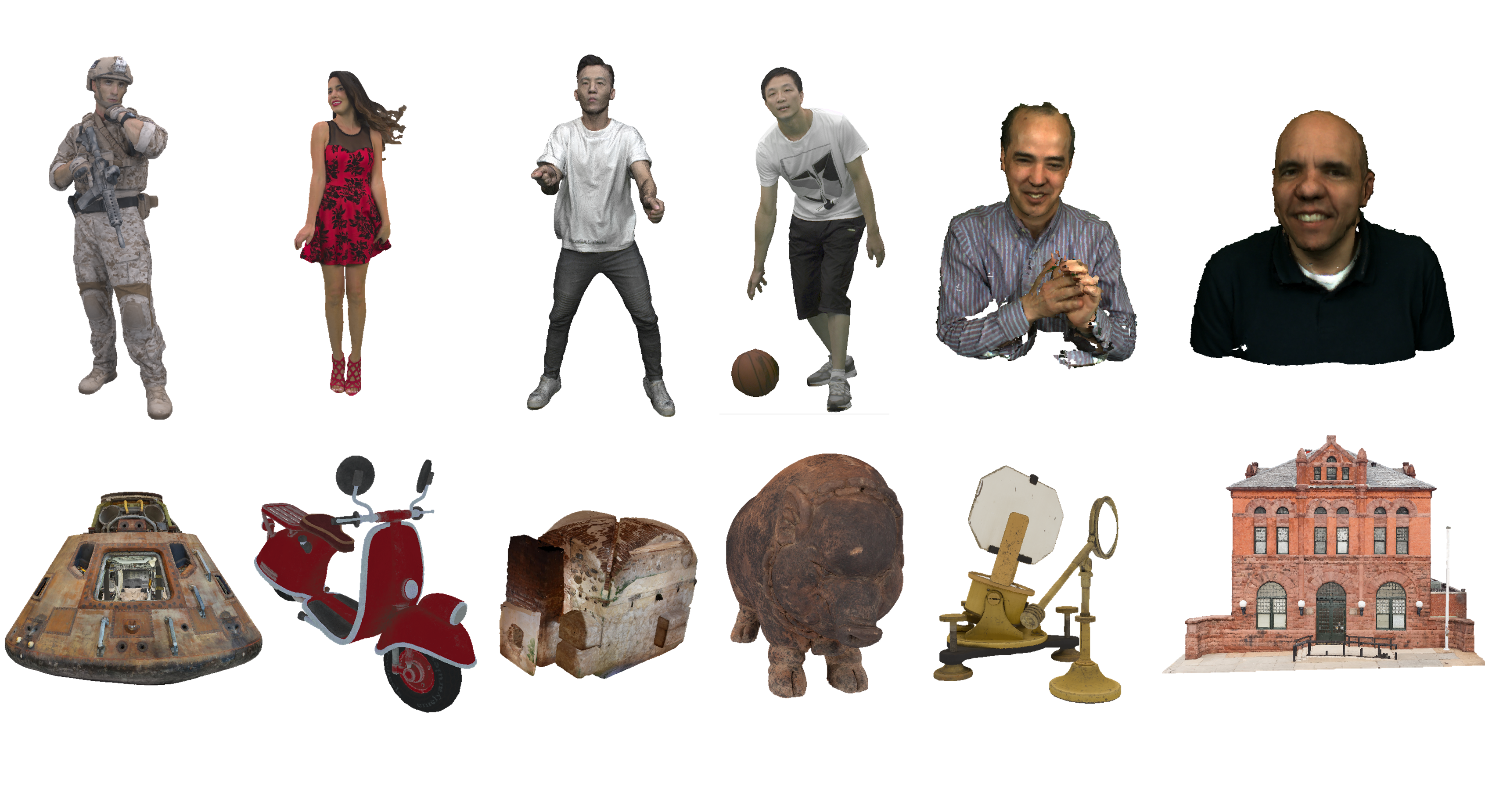}}
		\caption{ Snapshots of the point clouds. (a) Training set (b) Testing set.}
		\label{fig7}
	\end{figure}

We used the training and testing sets from \cite{10693649}, referred to training set1 and testing set1. Additionally, to increase the color diversity of the training data, we augmented the training set with the Real World Textured Things (RWTT) dataset \cite{maggiordomo2020real}, which consists of 568 textured meshes. We selected the first 561 meshes and sampled them into 10-bit colored point clouds. We used the RWTT dataset plus the training set1 as the total training set. Since testing set1 only contains human point clouds, to fully validate the proposed Deep-JGAC, we utilized the recently released textured mesh dataset from AVS \cite{AVS}. The dataset comprises meshes from different categories (e.g., inanimate objects and buildings). We selected some mesh sequences, sampled them into 10-bit colored point clouds, and added them to the testing set. We refer this dataset as `AVS-PC'. Overall, the datasets we used include 8iVFB \cite{30}, Owlii \cite{31}, 8iVSLF \cite{krivokuca20188i}, Volograms \cite{pages2021volograms}, MVUB \cite{32}, RWTT \cite{maggiordomo2020real}, and the mesh dataset launched by AVS \cite{AVS}. Some examples of the training and testing point clouds are shown in Figs. \ref{fig7a} and \ref{fig7b}.
	\label{section3}

\begin{table*}[!t]
	\caption{BDBR Comparison between the Deep-JGAC and benchmark schemes on different quality metrics. [Unit: \%]. }
	\label{tab11}
	\centering
	\setlength{\tabcolsep}{0.2mm}
	\scalebox{0.9}{
	\begin{tabular}{|cc|cccc|cccc|cccc|}
		\hline
		\multicolumn{1}{|c|}{\multirow{2}{*}{\textbf{Dataset}}}   & \multirow{2}{*}{\textbf{Sequence}} & \multicolumn{4}{c|}{\textbf{Deep-JGAC vs G-PCC}}                                                                                            & \multicolumn{4}{c|}{\textbf{Deep-JGAC   vs IT-DL-PCC}}                                                                                      & \multicolumn{4}{c|}{\textbf{Deep-JGAC vs V-PCC}}                                                                                            \\ \cline{3-14}
		\multicolumn{1}{|c|}{}                                    &                                    & \multicolumn{1}{c|}{\textbf{MPED}}   & \multicolumn{1}{c|}{\textbf{TCDM}}   & \multicolumn{1}{c|}{\textbf{GraphSIM}} & \textbf{MS-GraphSIM} & \multicolumn{1}{c|}{\textbf{MPED}}   & \multicolumn{1}{c|}{\textbf{TCDM}}   & \multicolumn{1}{c|}{\textbf{GraphSIM}} & \textbf{MS-GraphSIM} & \multicolumn{1}{c|}{\textbf{MPED}}   & \multicolumn{1}{c|}{\textbf{TCDM}}   & \multicolumn{1}{c|}{\textbf{GraphSIM}} & \textbf{MS-GraphSIM} \\ \hline
		\multicolumn{1}{|c|}{\multirow{3}{*}{\textbf{8iVFB}}}     & Red.                               & \multicolumn{1}{c|}{-69.04}          & \multicolumn{1}{c|}{-49.45}          & \multicolumn{1}{c|}{-39.18}            & -47.28               & \multicolumn{1}{c|}{-50.57}          & \multicolumn{1}{c|}{-60.95}          & \multicolumn{1}{c|}{-63.47}            & -66.56               & \multicolumn{1}{c|}{-5.23}           & \multicolumn{1}{c|}{-4.35}           & \multicolumn{1}{c|}{2.58}              & -13.75               \\ \cline{2-14}
		\multicolumn{1}{|c|}{}                                    & Soldier                            & \multicolumn{1}{c|}{-72.25}          & \multicolumn{1}{c|}{-58.32}          & \multicolumn{1}{c|}{-56.36}            & -59.66               & \multicolumn{1}{c|}{-53.43}          & \multicolumn{1}{c|}{-61.33}          & \multicolumn{1}{c|}{-63.23}            & -62.64               & \multicolumn{1}{c|}{-44.94}          & \multicolumn{1}{c|}{-29.54}          & \multicolumn{1}{c|}{-31.69}            & -34.10               \\ \cline{2-14}
		\multicolumn{1}{|c|}{}                                    & \textbf{Average}                   & \multicolumn{1}{c|}{\textbf{-70.25}} & \multicolumn{1}{c|}{\textbf{-55.21}} & \multicolumn{1}{c|}{\textbf{-49.03}}   & \textbf{-53.97}      & \multicolumn{1}{c|}{\textbf{-51.21}} & \multicolumn{1}{c|}{\textbf{-61.10}} & \multicolumn{1}{c|}{\textbf{-62.92}}   & \textbf{-64.18}      & \multicolumn{1}{c|}{\textbf{-27.68}} & \multicolumn{1}{c|}{\textbf{-18.50}} & \multicolumn{1}{c|}{\textbf{-17.29}}   & \textbf{-25.30}      \\ \hline
		\multicolumn{1}{|c|}{\multirow{3}{*}{\textbf{8iVSLF}}}    & Thai.                              & \multicolumn{1}{c|}{-72.92}          & \multicolumn{1}{c|}{-57.85}          & \multicolumn{1}{c|}{-48.95}            & -57.10               & \multicolumn{1}{c|}{-40.85}          & \multicolumn{1}{c|}{-58.34}          & \multicolumn{1}{c|}{-67.21}            & -64.93               & \multicolumn{1}{c|}{-40.23}          & \multicolumn{1}{c|}{-7.99}           & \multicolumn{1}{c|}{-27.62}            & -35.40               \\ \cline{2-14}
		\multicolumn{1}{|c|}{}                                    & Boxer                              & \multicolumn{1}{c|}{-80.48}          & \multicolumn{1}{c|}{-61.23}          & \multicolumn{1}{c|}{-43.50}            & -46.81               & \multicolumn{1}{c|}{-66.39}          & \multicolumn{1}{c|}{-65.49}          & \multicolumn{1}{c|}{-63.16}            & -61.31               & \multicolumn{1}{c|}{-38.02}          & \multicolumn{1}{c|}{-0.28}           & \multicolumn{1}{c|}{6.29}              & 3.42                 \\ \cline{2-14}
		\multicolumn{1}{|c|}{}                                    & \textbf{Average}                   & \multicolumn{1}{c|}{\textbf{-75.42}} & \multicolumn{1}{c|}{\textbf{-60.51}} & \multicolumn{1}{c|}{\textbf{-44.98}}   & \textbf{-52.50}      & \multicolumn{1}{c|}{\textbf{-51.06}} & \multicolumn{1}{c|}{\textbf{-61.98}} & \multicolumn{1}{c|}{\textbf{-63.29}}   & \textbf{-62.22}      & \multicolumn{1}{c|}{\textbf{-38.29}} & \multicolumn{1}{c|}{\textbf{-14.18}} & \multicolumn{1}{c|}{\textbf{-16.45}}   & \textbf{-25.49}      \\ \hline
		\multicolumn{1}{|c|}{\multirow{3}{*}{\textbf{MVUB}}}      & Phil                               & \multicolumn{1}{c|}{-64.67}          & \multicolumn{1}{c|}{-51.97}          & \multicolumn{1}{c|}{-54.32}            & -49.54               & \multicolumn{1}{c|}{-30.64}          & \multicolumn{1}{c|}{-60.76}          & \multicolumn{1}{c|}{-67.13}            & -61.44               & \multicolumn{1}{c|}{-26.97}          & \multicolumn{1}{c|}{-9.86}           & \multicolumn{1}{c|}{-24.55}            & -20.31               \\ \cline{2-14}
		\multicolumn{1}{|c|}{}                                    & Ricardo                            & \multicolumn{1}{c|}{-63.18}          & \multicolumn{1}{c|}{-62.59}          & \multicolumn{1}{c|}{-49.74}            & -48.80               & \multicolumn{1}{c|}{-50.26}          & \multicolumn{1}{c|}{-74.02}          & \multicolumn{1}{c|}{-67.92}            & -66.26               & \multicolumn{1}{c|}{8.49}            & \multicolumn{1}{c|}{-25.36}          & \multicolumn{1}{c|}{-1.40}             & -14.54               \\ \cline{2-14}
		\multicolumn{1}{|c|}{}                                    & \textbf{Average}                   & \multicolumn{1}{c|}{\textbf{-63.80}} & \multicolumn{1}{c|}{\textbf{-55.92}} & \multicolumn{1}{c|}{\textbf{-53.33}}   & \textbf{-49.62}      & \multicolumn{1}{c|}{\textbf{-40.59}} & \multicolumn{1}{c|}{\textbf{-67.12}} & \multicolumn{1}{c|}{\textbf{-68.39}}   & \textbf{-64.55}      & \multicolumn{1}{c|}{\textbf{-13.67}} & \multicolumn{1}{c|}{\textbf{-18.96}} & \multicolumn{1}{c|}{\textbf{-15.73}}   & \textbf{-19.81}      \\ \hline
		\multicolumn{1}{|c|}{\multirow{3}{*}{\textbf{Owlii}}}     & Basket.                            & \multicolumn{1}{c|}{-78.44}          & \multicolumn{1}{c|}{-60.23}          & \multicolumn{1}{c|}{-50.91}            & -54.59               & \multicolumn{1}{c|}{-68.29}          & \multicolumn{1}{c|}{-71.54}          & \multicolumn{1}{c|}{-68.77}            & -69.14               & \multicolumn{1}{c|}{-12.58}          & \multicolumn{1}{c|}{3.22}            & \multicolumn{1}{c|}{24.08}             & 9.22                 \\ \cline{2-14}
		\multicolumn{1}{|c|}{}                                    & Dancer                             & \multicolumn{1}{c|}{-76.65}          & \multicolumn{1}{c|}{-60.52}          & \multicolumn{1}{c|}{-53.77}            & -56.29               & \multicolumn{1}{c|}{-66.42}          & \multicolumn{1}{c|}{-70.68}          & \multicolumn{1}{c|}{-69.56}            & -68.72               & \multicolumn{1}{c|}{-17.78}          & \multicolumn{1}{c|}{-3.91}           & \multicolumn{1}{c|}{25.74}             & 11.36                \\ \cline{2-14}
		\multicolumn{1}{|c|}{}                                    & \textbf{Average}                   & \multicolumn{1}{c|}{\textbf{-77.42}} & \multicolumn{1}{c|}{\textbf{-60.40}} & \multicolumn{1}{c|}{\textbf{-52.47}}   & \textbf{-55.56}      & \multicolumn{1}{c|}{\textbf{-67.22}} & \multicolumn{1}{c|}{\textbf{-71.11}} & \multicolumn{1}{c|}{\textbf{-69.22}}   & \textbf{-68.95}      & \multicolumn{1}{c|}{\textbf{-15.22}} & \multicolumn{1}{c|}{\textbf{-0.59}}  & \multicolumn{1}{c|}{\textbf{24.92}}    & \textbf{10.36}       \\ \hline
		\multicolumn{1}{|c|}{\multirow{3}{*}{\textbf{Volograms}}} & Sir Fre.                           & \multicolumn{1}{c|}{-69.55}          & \multicolumn{1}{c|}{-55.72}          & \multicolumn{1}{c|}{-48.51}            & -50.76               & \multicolumn{1}{c|}{-64.73}          & \multicolumn{1}{c|}{-61.28}          & \multicolumn{1}{c|}{-66.46}            & -62.14               & \multicolumn{1}{c|}{-18.17}          & \multicolumn{1}{c|}{3.61}            & \multicolumn{1}{c|}{7.18}              & -2.52                \\ \cline{2-14}
		\multicolumn{1}{|c|}{}                                    & Rafa.                              & \multicolumn{1}{c|}{-75.21}          & \multicolumn{1}{c|}{-59.68}          & \multicolumn{1}{c|}{-44.04}            & -49.11               & \multicolumn{1}{c|}{-59.77}          & \multicolumn{1}{c|}{-64.13}          & \multicolumn{1}{c|}{-62.33}            & -58.02               & \multicolumn{1}{c|}{-31.19}          & \multicolumn{1}{c|}{-10.04}          & \multicolumn{1}{c|}{2.52}              & -1.48                \\ \cline{2-14}
		\multicolumn{1}{|c|}{}                                    & \textbf{Average}                   & \multicolumn{1}{c|}{\textbf{-73.30}} & \multicolumn{1}{c|}{\textbf{-57.66}} & \multicolumn{1}{c|}{\textbf{-46.49}}   & \textbf{-49.68}      & \multicolumn{1}{c|}{\textbf{-62.33}} & \multicolumn{1}{c|}{\textbf{-62.69}} & \multicolumn{1}{c|}{\textbf{-64.89}}   & \textbf{-60.73}      & \multicolumn{1}{c|}{\textbf{-26.08}} & \multicolumn{1}{c|}{\textbf{-3.25}}  & \multicolumn{1}{c|}{\textbf{5.21}}     & \textbf{-1.20}       \\ \hline
		\multicolumn{1}{|c|}{\multirow{23}{*}{\textbf{AVS-PC}}}   & Apollo.                            & \multicolumn{1}{c|}{-52.85}          & \multicolumn{1}{c|}{-27.51}          & \multicolumn{1}{c|}{-48.23}            & -52.33               & \multicolumn{1}{c|}{62.21}           & \multicolumn{1}{c|}{-26.45}          & \multicolumn{1}{c|}{-31.57}            & -23.36               & \multicolumn{1}{c|}{-35.38}          & \multicolumn{1}{c|}{-8.42}           & \multicolumn{1}{c|}{-23.49}            & -34.27               \\ \cline{2-14}
		\multicolumn{1}{|c|}{}                                    & Armillary.                         & \multicolumn{1}{c|}{-60.99}          & \multicolumn{1}{c|}{-41.51}          & \multicolumn{1}{c|}{-45.41}            & -50.89               & \multicolumn{1}{c|}{-41.22}          & \multicolumn{1}{c|}{-60.52}          & \multicolumn{1}{c|}{-52.77}            & -54.43               & \multicolumn{1}{c|}{-45.93}          & \multicolumn{1}{c|}{13.57}           & \multicolumn{1}{c|}{-24.60}            & -35.57               \\ \cline{2-14}
		\multicolumn{1}{|c|}{}                                    & Buste.                             & \multicolumn{1}{c|}{-47.04}          & \multicolumn{1}{c|}{-25.76}          & \multicolumn{1}{c|}{-10.38}            & -26.96               & \multicolumn{1}{c|}{-33.24}          & \multicolumn{1}{c|}{-60.17}          & \multicolumn{1}{c|}{-49.49}            & -49.94               & \multicolumn{1}{c|}{27.91}           & \multicolumn{1}{c|}{0.79}            & \multicolumn{1}{c|}{16.44}             & 1.11                 \\ \cline{2-14}
		\multicolumn{1}{|c|}{}                                    & Butter.                            & \multicolumn{1}{c|}{-27.00}          & \multicolumn{1}{c|}{2.03}            & \multicolumn{1}{c|}{-17.37}            & -29.71               & \multicolumn{1}{c|}{2.63}            & \multicolumn{1}{c|}{-31.95}          & \multicolumn{1}{c|}{-41.55}            & -33.51               & \multicolumn{1}{c|}{6.54}            & \multicolumn{1}{c|}{31.69}           & \multicolumn{1}{c|}{-22.32}            & -29.76               \\ \cline{2-14}
		\multicolumn{1}{|c|}{}                                    & Candle.                            & \multicolumn{1}{c|}{-69.56}          & \multicolumn{1}{c|}{-49.43}          & \multicolumn{1}{c|}{-54.67}            & -63.55               & \multicolumn{1}{c|}{-32.90}          & \multicolumn{1}{c|}{-57.12}          & \multicolumn{1}{c|}{-56.57}            & -56.94               & \multicolumn{1}{c|}{-20.46}          & \multicolumn{1}{c|}{0.16}            & \multicolumn{1}{c|}{-11.65}            & -22.37               \\ \cline{2-14}
		\multicolumn{1}{|c|}{}                                    & Ruin.                              & \multicolumn{1}{c|}{-62.34}          & \multicolumn{1}{c|}{-45.24}          & \multicolumn{1}{c|}{-38.38}            & -45.69               & \multicolumn{1}{c|}{-42.05}          & \multicolumn{1}{c|}{-58.99}          & \multicolumn{1}{c|}{-62.72}            & -59.60               & \multicolumn{1}{c|}{-16.28}          & \multicolumn{1}{c|}{15.93}           & \multicolumn{1}{c|}{11.82}             & 7.03                 \\ \cline{2-14}
		\multicolumn{1}{|c|}{}                                    & Boot.                              & \multicolumn{1}{c|}{-65.80}          & \multicolumn{1}{c|}{-53.39}          & \multicolumn{1}{c|}{-51.11}            & -56.09               & \multicolumn{1}{c|}{-17.21}          & \multicolumn{1}{c|}{-60.60}          & \multicolumn{1}{c|}{-57.76}            & -57.06               & \multicolumn{1}{c|}{41.20}           & \multicolumn{1}{c|}{25.14}           & \multicolumn{1}{c|}{35.50}             & 18.67                \\ \cline{2-14}
		\multicolumn{1}{|c|}{}                                    & Dead Rose                          & \multicolumn{1}{c|}{-64.11}          & \multicolumn{1}{c|}{-52.06}          & \multicolumn{1}{c|}{-46.77}            & -55.35               & \multicolumn{1}{c|}{-46.49}          & \multicolumn{1}{c|}{-62.29}          & \multicolumn{1}{c|}{-69.42}            & -66.54               & \multicolumn{1}{c|}{-32.89}          & \multicolumn{1}{c|}{-13.30}          & \multicolumn{1}{c|}{-17.85}            & -30.85               \\ \cline{2-14}
		\multicolumn{1}{|c|}{}                                    & Electro.                           & \multicolumn{1}{c|}{-92.02}          & \multicolumn{1}{c|}{-24.29}          & \multicolumn{1}{c|}{-3.60}             & -15.17               & \multicolumn{1}{c|}{56.46}           & \multicolumn{1}{c|}{-57.23}          & \multicolumn{1}{c|}{-43.72}            & -38.52               & \multicolumn{1}{c|}{/}               & \multicolumn{1}{c|}{18.05}           & \multicolumn{1}{c|}{-11.01}            & 8.57                 \\ \cline{2-14}
		\multicolumn{1}{|c|}{}                                    & Gramo.                             & \multicolumn{1}{c|}{-54.19}          & \multicolumn{1}{c|}{-38.15}          & \multicolumn{1}{c|}{-44.91}            & -51.38               & \multicolumn{1}{c|}{-26.74}          & \multicolumn{1}{c|}{-48.06}          & \multicolumn{1}{c|}{-54.23}            & -47.72               & \multicolumn{1}{c|}{40.24}           & \multicolumn{1}{c|}{104.37}          & \multicolumn{1}{c|}{53.82}             & 44.50                \\ \cline{2-14}
		\multicolumn{1}{|c|}{}                                    & Heilig.                            & \multicolumn{1}{c|}{-47.69}          & \multicolumn{1}{c|}{-37.37}          & \multicolumn{1}{c|}{-33.05}            & -45.24               & \multicolumn{1}{c|}{/}               & \multicolumn{1}{c|}{-44.26}          & \multicolumn{1}{c|}{-36.81}            & -28.97               & \multicolumn{1}{c|}{/}               & \multicolumn{1}{c|}{/}          & \multicolumn{1}{c|}{7.25}              & 35.52                \\ \cline{2-14}
		\multicolumn{1}{|c|}{}                                    & Heliostat                          & \multicolumn{1}{c|}{-51.72}          & \multicolumn{1}{c|}{-42.16}          & \multicolumn{1}{c|}{-43.35}            & -50.60               & \multicolumn{1}{c|}{-46.05}          & \multicolumn{1}{c|}{-52.33}          & \multicolumn{1}{c|}{-57.31}            & -55.78               & \multicolumn{1}{c|}{33.81}           & \multicolumn{1}{c|}{55.13}           & \multicolumn{1}{c|}{50.15}             & 15.21                \\ \cline{2-14}
		\multicolumn{1}{|c|}{}                                    & Hussar                             & \multicolumn{1}{c|}{-71.58}          & \multicolumn{1}{c|}{-52.15}          & \multicolumn{1}{c|}{-52.74}            & -56.85               & \multicolumn{1}{c|}{18.51}           & \multicolumn{1}{c|}{-49.85}          & \multicolumn{1}{c|}{-49.22}            & -43.16               & \multicolumn{1}{c|}{-25.86}          & \multicolumn{1}{c|}{-34.98}          & \multicolumn{1}{c|}{-46.73}            & -47.81               \\ \cline{2-14}
		\multicolumn{1}{|c|}{}                                    & Marble.                            & \multicolumn{1}{c|}{-78.26}          & \multicolumn{1}{c|}{-52.94}          & \multicolumn{1}{c|}{-48.89}            & -54.83               & \multicolumn{1}{c|}{-64.59}          & \multicolumn{1}{c|}{-65.01}          & \multicolumn{1}{c|}{-65.54}            & -63.75               & \multicolumn{1}{c|}{-22.59}          & \multicolumn{1}{c|}{29.11}           & \multicolumn{1}{c|}{16.10}             & 6.77                 \\ \cline{2-14}
		\multicolumn{1}{|c|}{}                                    & Bowl.                              & \multicolumn{1}{c|}{-77.39}          & \multicolumn{1}{c|}{-62.47}          & \multicolumn{1}{c|}{-50.97}            & -56.78               & \multicolumn{1}{c|}{-68.39}          & \multicolumn{1}{c|}{-71.94}          & \multicolumn{1}{c|}{-66.71}            & -66.06               & \multicolumn{1}{c|}{17.83}           & \multicolumn{1}{c|}{31.36}           & \multicolumn{1}{c|}{51.27}             & 32.47                \\ \cline{2-14}
		\multicolumn{1}{|c|}{}                                    & Motor.                             & \multicolumn{1}{c|}{-52.60}          & \multicolumn{1}{c|}{-25.71}          & \multicolumn{1}{c|}{-40.73}            & -43.73               & \multicolumn{1}{c|}{-27.20}          & \multicolumn{1}{c|}{-52.48}          & \multicolumn{1}{c|}{-68.57}            & -65.94               & \multicolumn{1}{c|}{-9.70}           & \multicolumn{1}{c|}{22.08}           & \multicolumn{1}{c|}{2.44}              & -4.74                \\ \cline{2-14}
		\multicolumn{1}{|c|}{}                                    & Pig.                               & \multicolumn{1}{c|}{-82.46}          & \multicolumn{1}{c|}{-53.04}          & \multicolumn{1}{c|}{-48.14}            & -53.40               & \multicolumn{1}{c|}{-62.56}          & \multicolumn{1}{c|}{-62.98}          & \multicolumn{1}{c|}{-61.61}            & -61.45               & \multicolumn{1}{c|}{-39.83}          & \multicolumn{1}{c|}{12.84}           & \multicolumn{1}{c|}{6.48}              & -4.27                \\ \cline{2-14}
		\multicolumn{1}{|c|}{}                                    & Ashtray.                           & \multicolumn{1}{c|}{-66.83}          & \multicolumn{1}{c|}{-45.48}          & \multicolumn{1}{c|}{-21.93}            & -24.63               & \multicolumn{1}{c|}{-74.25}          & \multicolumn{1}{c|}{-70.97}          & \multicolumn{1}{c|}{-70.56}            & -66.96               & \multicolumn{1}{c|}{59.20}           & \multicolumn{1}{c|}{61.13}           & \multicolumn{1}{c|}{107.58}            & 88.08                \\ \cline{2-14}
		\multicolumn{1}{|c|}{}                                    & Police.                            & \multicolumn{1}{c|}{-63.58}          & \multicolumn{1}{c|}{-49.00}          & \multicolumn{1}{c|}{-40.77}            & -55.01               & \multicolumn{1}{c|}{-2.51}           & \multicolumn{1}{c|}{-31.46}          & \multicolumn{1}{c|}{-43.82}            & -27.58               & \multicolumn{1}{c|}{-30.24}          & \multicolumn{1}{c|}{-14.34}          & \multicolumn{1}{c|}{-7.10}             & -37.25               \\ \cline{2-14}
		\multicolumn{1}{|c|}{}                                    & Stereo.                            & \multicolumn{1}{c|}{-8.61}           & \multicolumn{1}{c|}{-25.65}          & \multicolumn{1}{c|}{-27.81}            & -45.00               & \multicolumn{1}{c|}{-36.70}          & \multicolumn{1}{c|}{-58.35}          & \multicolumn{1}{c|}{-61.01}            & -59.52               & \multicolumn{1}{c|}{/}               & \multicolumn{1}{c|}{65.40}           & \multicolumn{1}{c|}{35.90}             & 7.28                 \\ \cline{2-14}
		\multicolumn{1}{|c|}{}                                    & Violin                             & \multicolumn{1}{c|}{-44.47}          & \multicolumn{1}{c|}{-31.79}          & \multicolumn{1}{c|}{-33.00}            & -39.36               & \multicolumn{1}{c|}{-27.34}          & \multicolumn{1}{c|}{-55.51}          & \multicolumn{1}{c|}{-61.63}            & -56.91               & \multicolumn{1}{c|}{5.15}            & \multicolumn{1}{c|}{5.32}            & \multicolumn{1}{c|}{-26.79}            & -23.77               \\ \cline{2-14}
		\multicolumn{1}{|c|}{}                                    & Chair                              & \multicolumn{1}{c|}{-72.58}          & \multicolumn{1}{c|}{-44.13}          & \multicolumn{1}{c|}{-31.85}            & -40.18               & \multicolumn{1}{c|}{-53.28}          & \multicolumn{1}{c|}{-60.79}          & \multicolumn{1}{c|}{-58.55}            & -58.45               & \multicolumn{1}{c|}{-60.84}          & \multicolumn{1}{c|}{33.24}           & \multicolumn{1}{c|}{-5.68}             & -23.27               \\ \cline{2-14}
		\multicolumn{1}{|c|}{}                                    & \textbf{Average}                   & \multicolumn{1}{c|}{\textbf{-52.92}} & \multicolumn{1}{c|}{\textbf{-41.61}} & \multicolumn{1}{c|}{\textbf{-39.92}}   & \textbf{-47.52}      & \multicolumn{1}{c|}{\textbf{-34.12}} & \multicolumn{1}{c|}{\textbf{-56.01}} & \multicolumn{1}{c|}{\textbf{-56.47}}   & \textbf{-54.12}      & \multicolumn{1}{c|}{\textbf{-14.11}} & \multicolumn{1}{c|}{\textbf{11.90}}  & \multicolumn{1}{c|}{\textbf{-3.57}}    & \textbf{-13.64}      \\ \hline
		\multicolumn{2}{|c|}{\textbf{Total Average}}                                                   & \multicolumn{1}{c|}{\textbf{-58.01}} & \multicolumn{1}{c|}{\textbf{-47.16}} & \multicolumn{1}{c|}{\textbf{-42.42}}   & \textbf{-48.72}      & \multicolumn{1}{c|}{\textbf{-35.61}} & \multicolumn{1}{c|}{\textbf{-58.74}} & \multicolumn{1}{c|}{\textbf{-59.18}}   & \textbf{-57.14}      & \multicolumn{1}{c|}{\textbf{-11.86}} & \multicolumn{1}{c|}{\textbf{3.29}}   & \multicolumn{1}{c|}{\textbf{-5.08}}    & \textbf{-14.67}      \\ \hline
		\end{tabular}}
\end{table*}
\begin{figure*}[!t]
	\centering
	\subfigure[] {
		\label{fig88b}
		\includegraphics[width=0.22\linewidth]{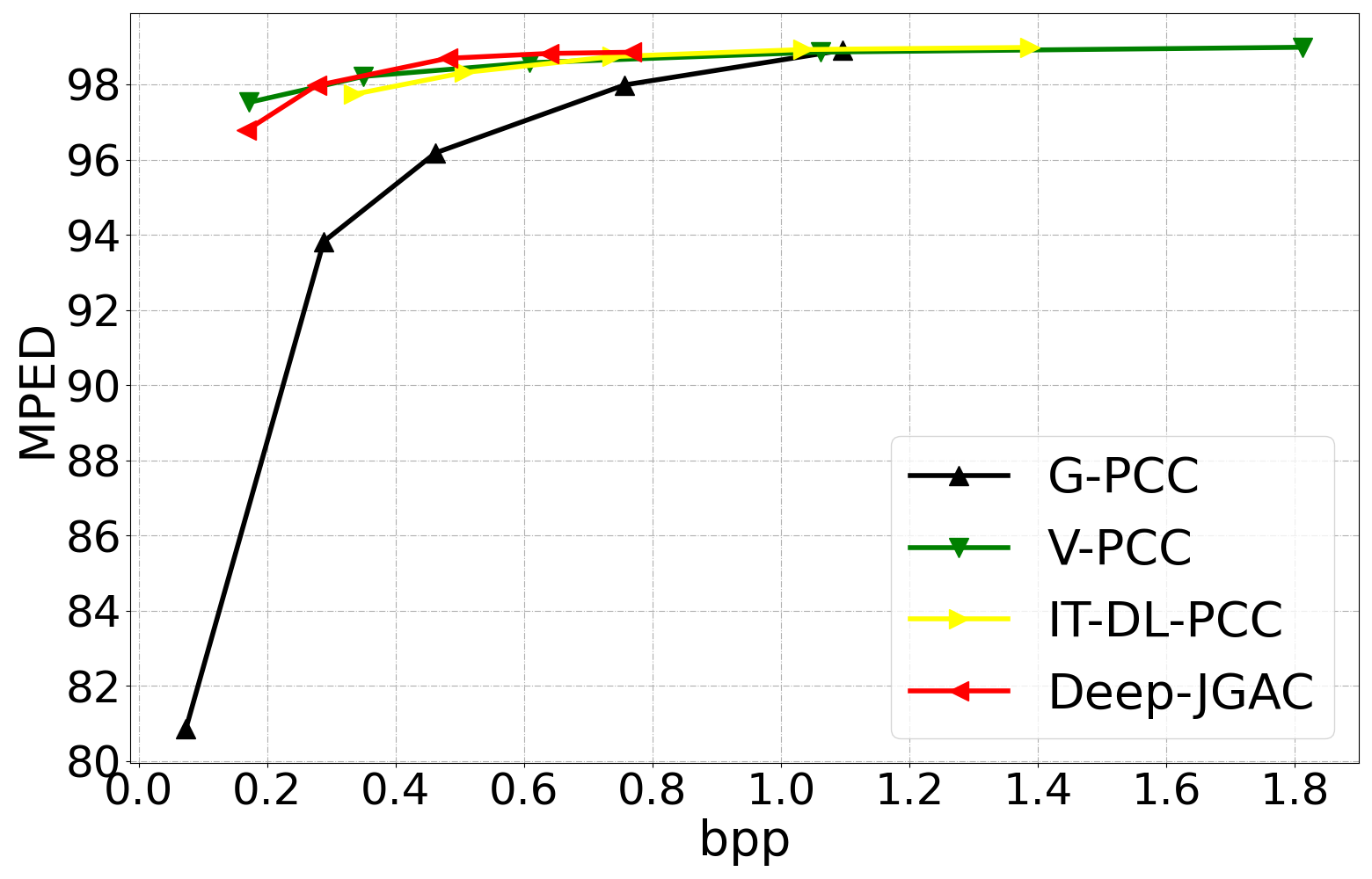}
		}
	\subfigure[] {
		\label{fig888d}
		\includegraphics[width=0.22\linewidth]{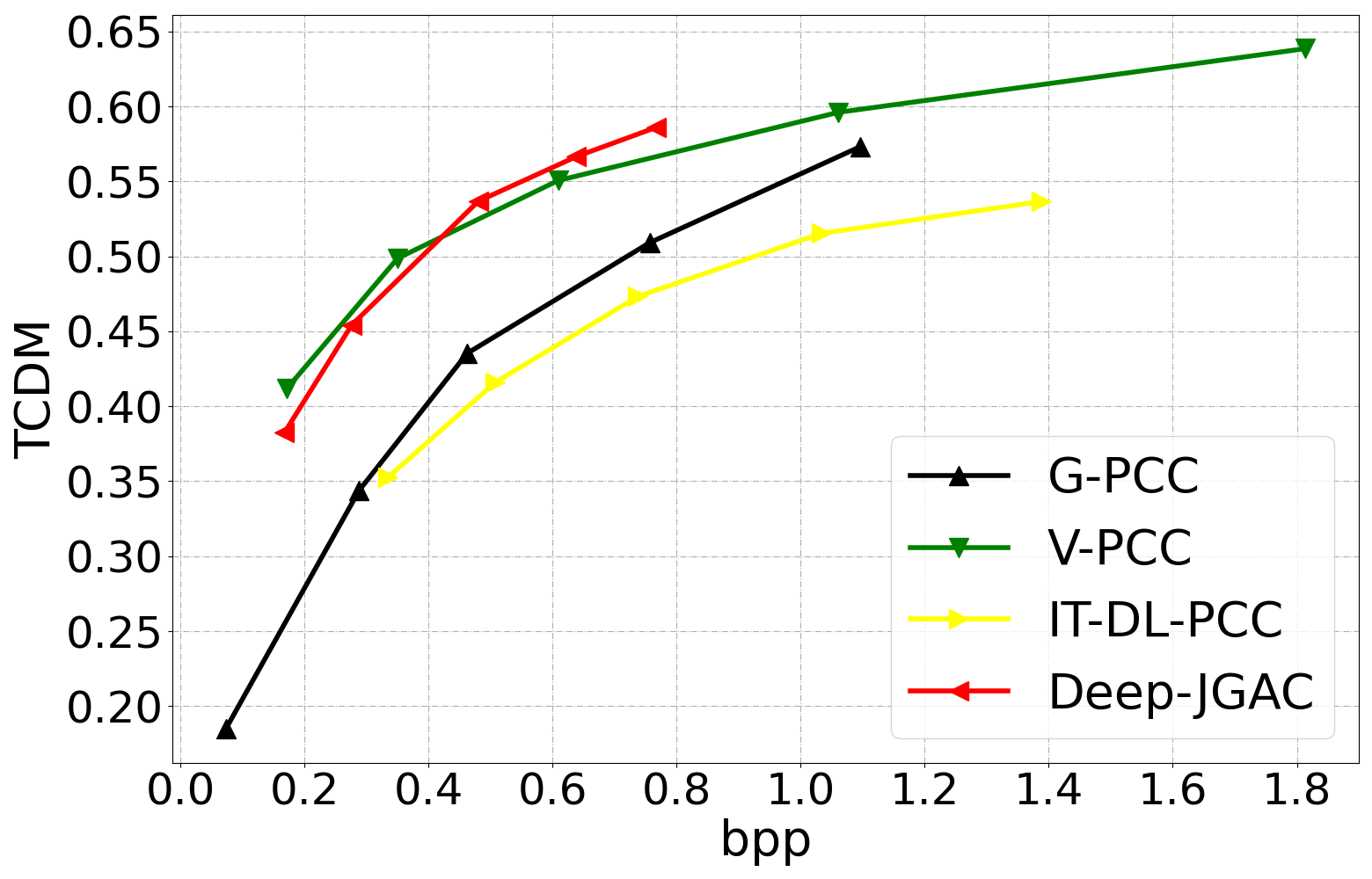}
	}	
	\subfigure[] {
		\label{fig88c}
		\includegraphics[width=0.22\linewidth]{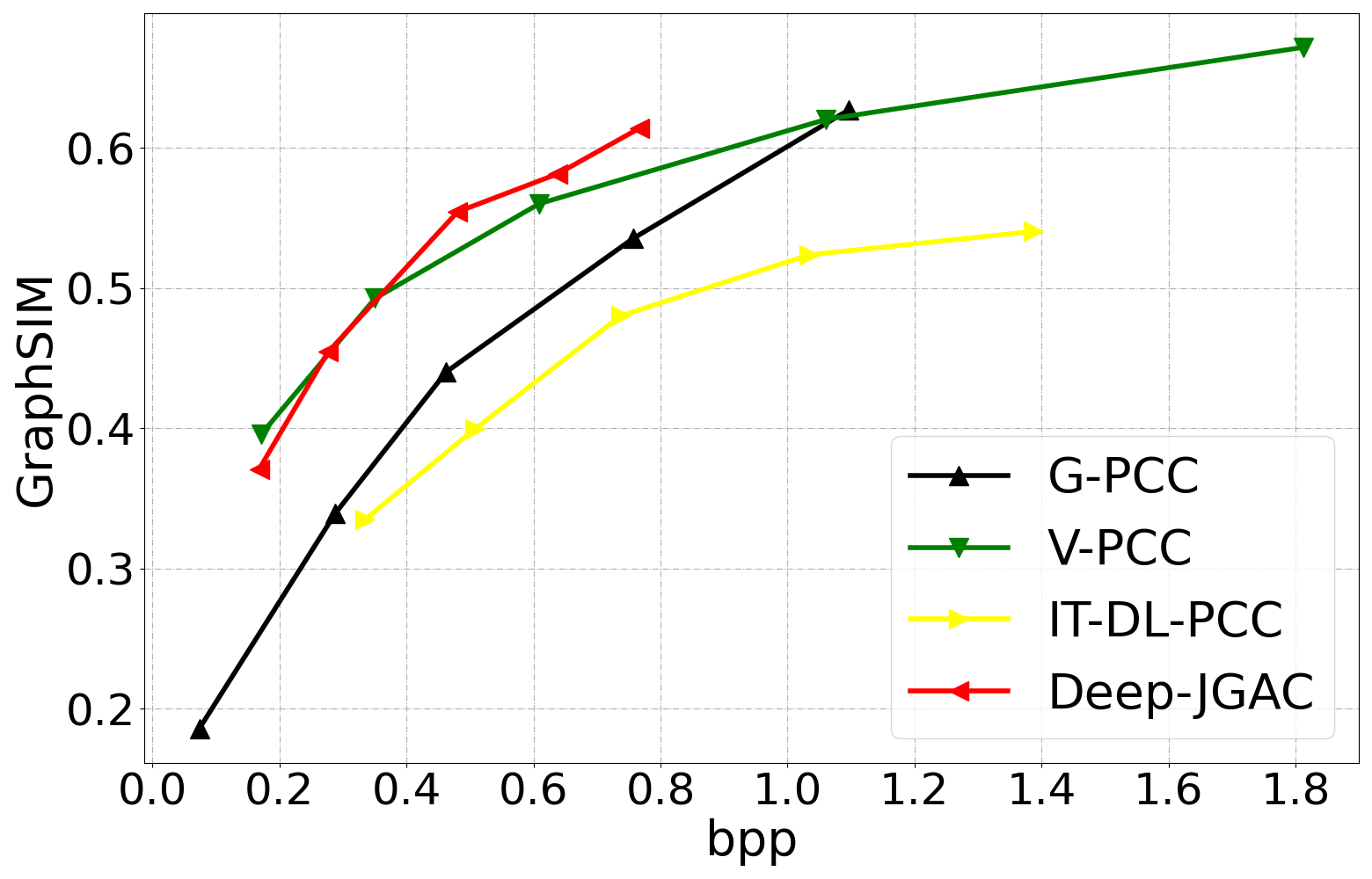}
	}
	\subfigure[] {
		\label{fig88d}
		\includegraphics[width=0.22\linewidth]{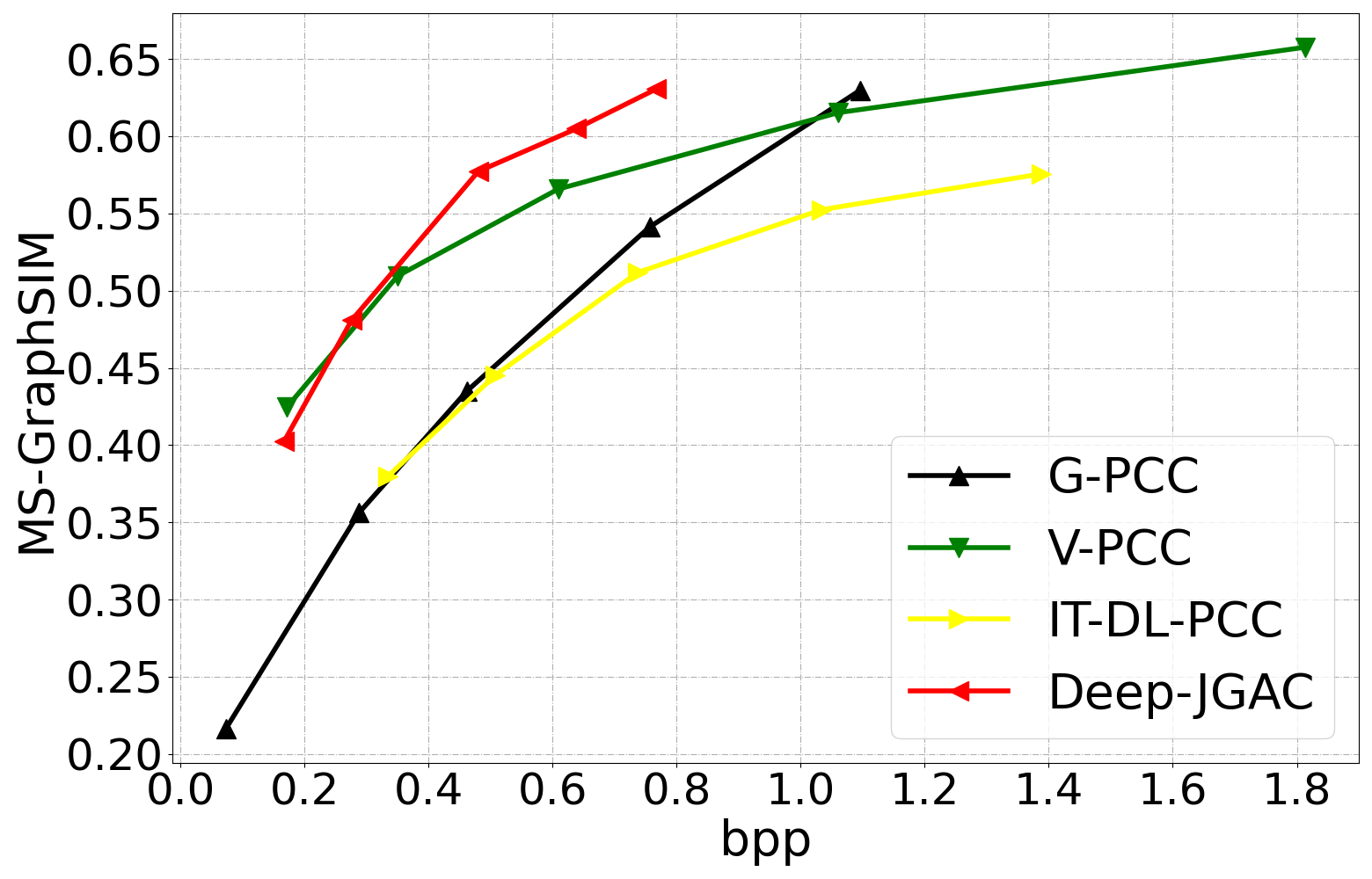}
	}
	
	\caption{Average RD curves comparisons among joint PCC schemes using different quality metrics. (a)  MPED, (b) TCDM, (c) GraphSIM, (d) MS-GraphSIM.}
	\label{fig88}
\end{figure*}
\subsubsection{Quality Metrics}
The reconstructed point cloud exhibits both geometry and color distortions. Geometry distortion is measured using point-to-point Peak Signal-to-Noise Ratio (D1-PSNR) and point-to-plane PSNR (D2-PSNR) \cite{33}. Additionally, since it requires the point cloud normal in computing D2 PSNR, we used the open3D library to generate the point cloud normal with a neighborhood search radius of 30.
Attribute color distortion was measured using luminance component, called Y-PSNR, by using MPEG software \cite{33}.
To evaluate the visual quality of colored point clouds, four perceptual quality metrics, including Multiscale Potential Energy Discrepancy (MPED) \cite{yang2022mped}, Transformational Complexity based Distortion Metric (TCDM) \cite{10337742}, Graph Similarity (GraphSIM) \cite{9306905}, and Multiscale
Graph Similarity (MS-GraphSIM) \cite{Zhang2021MSGraphSIMIP}, were used. For better readability, MPED value was calibrated as 100-MPED, such that a higher MPED value indicates a better quality. The coding bit rate was measured with bits per point ($bpp$). Given a quality metric, we used Bjøntegaard Delta Bit Rate (BDBR) to evaluate the average bit reduction, where a negative BDBR value indicates a bit rate saving.
\subsubsection{Training Settings}
The authors of GRASP, IT-DL-PCC, and PCGCv2 have provided their pre-trained models, which were used for coding directly.
Since the proposed geometry codec in Deep-JGAC was developed based on the PCGCv2, we retrained PCGCv2 with the same training settings as the Deep-JGAC for further comparison, which {is} referred as PCGCv2*.
In Deep-JGAC, $\{\lambda_G,\lambda_A\}$ were set to \{3, 16000\}, \{2, 8000\}, \{1, 4000\}, \{0.5, 1000\}, \{0.25, 400\} to achieve different rate points. We set $\alpha$ in AIFM as 0.02 \cite{multimodal}. The attribute encoder $E_A(\cdot)$ and decoder $D_A(\cdot)$ were adopted from TSC-PCAC \cite{10693649}. We first trained a highest rate point, and then trained the remaining rate points for 50 epochs. In training the highest rate point, we had three training phases including the geometry training, the attribute training and the joint training. In training geometry codec, we used the RWTT dataset \cite{maggiordomo2020real} to train for 70 epochs with an initial learning rate of $8\times10^{-4}$, then continued the training for 30 epochs by using the total training dataset with a learning rate of $10^{-5}$. In training the attribute codec, we loaded the model weights provided by \cite{10693649} and trained for 200 epochs by using the total training dataset with a learning rate of $10^{-5}$. Finally, the Deep-JGAC was trained jointly for 30 epochs with a learning rate of $10^{-5}$.

	\begin{table*}[]
 	\caption{BDBR Comparison between geometry coding of Deep-JGAC and benchmarks, where the geometry quality was measured with D1-PSNR and D2-PSNR. [Unit: \%].}
 	\label{tab1}
 	\centering
 	\setlength{\tabcolsep}{0.6mm}
 	\scalebox{0.9}{
 		\begin{tabular}{|cc|cc|cc|cc|cc|cc|}
 			\hline
 			\multicolumn{1}{|c|}{\multirow{2}{*}{\textbf{Dataset}}} & \multirow{2}{*}{\textbf{Sequence}} & \multicolumn{2}{c|}{\textbf{Deep-JGAC vs G-PCC}}       & \multicolumn{2}{c|}{\textbf{Deep-JGAC vs V-PCC}}       & \multicolumn{2}{c|}{\textbf{Deep-JGAC vs GRASP}}       & \multicolumn{2}{c|}{\textbf{Deep-JGAC vs PCGCv2}}      & \multicolumn{2}{c|}{\textbf{Deep-JGAC vs PCGCv2*}}     \\ \cline{3-12}
 			\multicolumn{1}{|c|}{}                                  &                                    & \multicolumn{1}{c|}{\textbf{D1}}     & \textbf{D2}     & \multicolumn{1}{c|}{\textbf{D1}}     & \textbf{D2}     & \multicolumn{1}{c|}{\textbf{D1}}     & \textbf{D2}     & \multicolumn{1}{c|}{\textbf{D1}}     & \textbf{D2}     & \multicolumn{1}{c|}{\textbf{D1}}     & \textbf{D2}     \\ \hline
 			\multicolumn{1}{|c|}{\multirow{3}{*}{\textbf{8iVFB}}}   & Red.                               & \multicolumn{1}{c|}{-86.96}          & -82.62          & \multicolumn{1}{c|}{-62.00}          & -71.71          & \multicolumn{1}{c|}{-52.10}          & -59.87          & \multicolumn{1}{c|}{-26.94}          & -31.32          & \multicolumn{1}{c|}{-3.18}           & -4.43           \\ \cline{2-12}
 			\multicolumn{1}{|c|}{}                                  & Soldier                            & \multicolumn{1}{c|}{-88.86}          & -83.26          & \multicolumn{1}{c|}{-64.83}          & -68.51          & \multicolumn{1}{c|}{-58.62}          & -60.65          & \multicolumn{1}{c|}{-34.20}          & -34.60          & \multicolumn{1}{c|}{-3.65}           & -3.26           \\ \cline{2-12}
 			\multicolumn{1}{|c|}{}                                  & \textbf{Average}                   & \multicolumn{1}{c|}{\textbf{-88.00}} & \textbf{-82.98} & \multicolumn{1}{c|}{\textbf{-63.53}} & \textbf{-70.19} & \multicolumn{1}{c|}{\textbf{-55.57}} & \textbf{-60.30} & \multicolumn{1}{c|}{\textbf{-30.85}} & \textbf{-33.11} & \multicolumn{1}{c|}{\textbf{-3.50}}  & \textbf{-3.84}  \\ \hline
 			\multicolumn{1}{|c|}{\multirow{3}{*}{\textbf{8iVSLF}}}  & Boxer                              & \multicolumn{1}{c|}{-91.21}          & -86.18          & \multicolumn{1}{c|}{-59.55}          & -63.36          & \multicolumn{1}{c|}{-62.53}          & -64.37          & \multicolumn{1}{c|}{-41.35}          & -39.73          & \multicolumn{1}{c|}{-4.01}           & -5.34           \\ \cline{2-12}
 			\multicolumn{1}{|c|}{}                                  & Thai.                              & \multicolumn{1}{c|}{-86.71}          & -80.22          & \multicolumn{1}{c|}{-57.52}          & -65.66          & \multicolumn{1}{c|}{-50.66}          & -56.30          & \multicolumn{1}{c|}{-23.64}          & -28.95          & \multicolumn{1}{c|}{-2.21}           & -2.52           \\ \cline{2-12}
 			\multicolumn{1}{|c|}{}                                  & \textbf{Average}                   & \multicolumn{1}{c|}{\textbf{-89.18}} & \textbf{-83.50} & \multicolumn{1}{c|}{\textbf{-59.06}} & \textbf{-65.02} & \multicolumn{1}{c|}{\textbf{-57.02}} & \textbf{-60.53} & \multicolumn{1}{c|}{\textbf{-33.17}} & \textbf{-34.56} & \multicolumn{1}{c|}{\textbf{-3.13}}  & \textbf{-3.93}  \\ \hline
 			\multicolumn{1}{|c|}{\multirow{3}{*}{\textbf{MVUB}}}    & Phil                               & \multicolumn{1}{c|}{-84.16}          & -77.31          & \multicolumn{1}{c|}{-72.16}          & -70.33          & \multicolumn{1}{c|}{-29.82}          & -44.71          & \multicolumn{1}{c|}{7.30}            & -18.92          & \multicolumn{1}{c|}{-0.23}           & -4.23           \\ \cline{2-12}
 			\multicolumn{1}{|c|}{}                                  & Ricardo                            & \multicolumn{1}{c|}{-87.07}          & -80.85          & \multicolumn{1}{c|}{-74.88}          & -60.71          & \multicolumn{1}{c|}{-38.76}          & -45.49          & \multicolumn{1}{c|}{0.69}            & -24.51          & \multicolumn{1}{c|}{-4.54}           & -6.15           \\ \cline{2-12}
 			\multicolumn{1}{|c|}{}                                  & \textbf{Average}                   & \multicolumn{1}{c|}{\textbf{-85.61}} & \textbf{-79.15} & \multicolumn{1}{c|}{\textbf{-73.61}} & \textbf{-66.13} & \multicolumn{1}{c|}{\textbf{-34.46}} & \textbf{-45.11} & \multicolumn{1}{c|}{\textbf{4.11}}   & \textbf{-21.68} & \multicolumn{1}{c|}{\textbf{-2.37}}  & \textbf{-5.19}  \\ \hline
 			\multicolumn{1}{|c|}{\multirow{3}{*}{\textbf{OW}}}      & Basket.                            & \multicolumn{1}{c|}{-95.28}          & -91.57          & \multicolumn{1}{c|}{-64.61}          & -65.83          & \multicolumn{1}{c|}{-67.12}          & -65.38          & \multicolumn{1}{c|}{-40.82}          & -41.67          & \multicolumn{1}{c|}{-5.24}           & -6.00           \\ \cline{2-12}
 			\multicolumn{1}{|c|}{}                                  & Dancer                             & \multicolumn{1}{c|}{-95.00}          & -90.83          & \multicolumn{1}{c|}{-66.00}          & -67.49          & \multicolumn{1}{c|}{-66.47}          & -65.22          & \multicolumn{1}{c|}{-41.24}          & -41.78          & \multicolumn{1}{c|}{-4.98}           & -5.66           \\ \cline{2-12}
 			\multicolumn{1}{|c|}{}                                  & \textbf{Average}                   & \multicolumn{1}{c|}{\textbf{-95.14}} & \textbf{-91.21} & \multicolumn{1}{c|}{\textbf{-65.33}} & \textbf{-66.69} & \multicolumn{1}{c|}{\textbf{-66.80}} & \textbf{-65.30} & \multicolumn{1}{c|}{\textbf{-41.04}} & \textbf{-41.73} & \multicolumn{1}{c|}{\textbf{-5.11}}  & \textbf{-5.83}  \\ \hline
 			\multicolumn{1}{|c|}{\multirow{3}{*}{\textbf{VOLO}}}    & Sir Fre.                           & \multicolumn{1}{c|}{-95.42}          & -90.74          & \multicolumn{1}{c|}{-73.51}          & -72.38          & \multicolumn{1}{c|}{-71.12}          & -66.06          & \multicolumn{1}{c|}{-37.98}          & -38.90          & \multicolumn{1}{c|}{-4.68}           & -5.27           \\ \cline{2-12}
 			\multicolumn{1}{|c|}{}                                  & Rafa.                              & \multicolumn{1}{c|}{-94.42}          & -89.26          & \multicolumn{1}{c|}{-68.90}          & -68.94          & \multicolumn{1}{c|}{-67.80}          & -63.51          & \multicolumn{1}{c|}{-34.61}          & -35.98          & \multicolumn{1}{c|}{-4.63}           & -5.10           \\ \cline{2-12}
 			\multicolumn{1}{|c|}{}                                  & \textbf{Average}                   & \multicolumn{1}{c|}{\textbf{-94.95}} & \textbf{-90.02} & \multicolumn{1}{c|}{\textbf{-71.27}} & \textbf{-70.70} & \multicolumn{1}{c|}{\textbf{-69.45}} & \textbf{-64.79} & \multicolumn{1}{c|}{\textbf{-36.29}} & \textbf{-37.45} & \multicolumn{1}{c|}{\textbf{-4.66}}  & \textbf{-5.18}  \\ \hline
 			\multicolumn{1}{|c|}{\multirow{23}{*}{\textbf{AVS-PC}}} & Apollo.                            & \multicolumn{1}{c|}{-69.30}          & -62.84          & \multicolumn{1}{c|}{-24.09}          & -57.38          & \multicolumn{1}{c|}{30.67}           & -32.48          & \multicolumn{1}{c|}{3.21}            & 2.80            & \multicolumn{1}{c|}{-7.99}           & -3.01           \\ \cline{2-12}
 			\multicolumn{1}{|c|}{}                                  & Armillary.                         & \multicolumn{1}{c|}{-78.79}          & -74.16          & \multicolumn{1}{c|}{-50.40}          & -62.86          & \multicolumn{1}{c|}{-46.16}          & -61.08          & \multicolumn{1}{c|}{-34.95}          & -23.65          & \multicolumn{1}{c|}{-21.12}          & -21.62          \\ \cline{2-12}
 			\multicolumn{1}{|c|}{}                                  & Buste.                             & \multicolumn{1}{c|}{-73.81}          & -66.41          & \multicolumn{1}{c|}{-39.18}          & -50.73          & \multicolumn{1}{c|}{-45.71}          & -54.05          & \multicolumn{1}{c|}{-39.00}          & -30.38          & \multicolumn{1}{c|}{-16.79}          & /               \\ \cline{2-12}
 			\multicolumn{1}{|c|}{}                                  & Butter.                            & \multicolumn{1}{c|}{-53.27}          & -55.06          & \multicolumn{1}{c|}{1.74}            & -27.13          & \multicolumn{1}{c|}{6.91}            & -43.50          & \multicolumn{1}{c|}{-18.01}          & -33.66          & \multicolumn{1}{c|}{-21.71}          & -10.76          \\ \cline{2-12}
 			\multicolumn{1}{|c|}{}                                  & Candle.                            & \multicolumn{1}{c|}{-80.55}          & -75.45          & \multicolumn{1}{c|}{-24.17}          & -42.73          & \multicolumn{1}{c|}{-35.10}          & -49.21          & \multicolumn{1}{c|}{1.17}            & -3.45           & \multicolumn{1}{c|}{-10.88}          & -9.58           \\ \cline{2-12}
 			\multicolumn{1}{|c|}{}                                  & Ruin.                              & \multicolumn{1}{c|}{-85.08}          & -79.52          & \multicolumn{1}{c|}{-30.36}          & -46.34          & \multicolumn{1}{c|}{-54.21}          & -65.14          & \multicolumn{1}{c|}{-55.45}          & -45.44          & \multicolumn{1}{c|}{-1.79}           & -2.04           \\ \cline{2-12}
 			\multicolumn{1}{|c|}{}                                  & Boot.                              & \multicolumn{1}{c|}{-87.76}          & -82.40          & \multicolumn{1}{c|}{-15.36}          & -32.00          & \multicolumn{1}{c|}{-45.21}          & -55.68          & \multicolumn{1}{c|}{-14.74}          & -23.65          & \multicolumn{1}{c|}{-1.01}           & -1.99           \\ \cline{2-12}
 			\multicolumn{1}{|c|}{}                                  & Dead Rose                          & \multicolumn{1}{c|}{-88.31}          & -83.01          & \multicolumn{1}{c|}{-61.74}          & -69.29          & \multicolumn{1}{c|}{-50.82}          & -55.21          & \multicolumn{1}{c|}{-19.04}          & -23.04          & \multicolumn{1}{c|}{-2.57}           & -2.85           \\ \cline{2-12}
 			\multicolumn{1}{|c|}{}                                  & Electro.                           & \multicolumn{1}{c|}{-57.17}          & -48.07          & \multicolumn{1}{c|}{8.65}            & -27.81          & \multicolumn{1}{c|}{-5.83}           & -46.57          & \multicolumn{1}{c|}{-42.74}          & -28.78          & \multicolumn{1}{c|}{/}               & -6.71           \\ \cline{2-12}
 			\multicolumn{1}{|c|}{}                                  & Gramo.                             & \multicolumn{1}{c|}{-69.33}          & -61.12          & \multicolumn{1}{c|}{-10.54}          & -21.63          & \multicolumn{1}{c|}{0.97}            & -31.28          & \multicolumn{1}{c|}{33.88}           & 19.12           & \multicolumn{1}{c|}{-17.32}          & -1.40           \\ \cline{2-12}
 			\multicolumn{1}{|c|}{}                                  & Heilig.                            & \multicolumn{1}{c|}{-18.00}          & 7.28            & \multicolumn{1}{c|}{/}               & 150.56          & \multicolumn{1}{c|}{125.57}          & 23.62           & \multicolumn{1}{c|}{50.42}           & 29.27           & \multicolumn{1}{c|}{/}               & /               \\ \cline{2-12}
 			\multicolumn{1}{|c|}{}                                  & Heliostat                          & \multicolumn{1}{c|}{-72.90}          & -62.97          & \multicolumn{1}{c|}{-27.23}          & -29.51          & \multicolumn{1}{c|}{-19.31}          & -36.82          & \multicolumn{1}{c|}{-8.01}           & 6.83            & \multicolumn{1}{c|}{-26.31}          & -10.21          \\ \cline{2-12}
 			\multicolumn{1}{|c|}{}                                  & Hussar                             & \multicolumn{1}{c|}{-77.67}          & -69.90          & \multicolumn{1}{c|}{-63.85}          & -72.23          & \multicolumn{1}{c|}{-34.20}          & -44.72          & \multicolumn{1}{c|}{6.74}            & -7.48           & \multicolumn{1}{c|}{-3.92}           & -3.33           \\ \cline{2-12}
 			\multicolumn{1}{|c|}{}                                  & Marble.                            & \multicolumn{1}{c|}{-97.72}          & -95.36          & \multicolumn{1}{c|}{-51.38}          & -54.40          & \multicolumn{1}{c|}{-70.23}          & -73.54          & \multicolumn{1}{c|}{-54.20}          & -43.72          & \multicolumn{1}{c|}{/}               & /               \\ \cline{2-12}
 			\multicolumn{1}{|c|}{}                                  & Bowl.                              & \multicolumn{1}{c|}{-96.16}          & -93.06          & \multicolumn{1}{c|}{-28.28}          & -31.46          & \multicolumn{1}{c|}{-62.30}          & -62.54          & \multicolumn{1}{c|}{-22.87}          & -22.26          & \multicolumn{1}{c|}{-7.22}           & -5.00           \\ \cline{2-12}
 			\multicolumn{1}{|c|}{}                                  & Motor.                             & \multicolumn{1}{c|}{-69.11}          & -63.49          & \multicolumn{1}{c|}{-28.79}          & -47.88          & \multicolumn{1}{c|}{3.05}            & -31.69          & \multicolumn{1}{c|}{55.81}           & 28.13           & \multicolumn{1}{c|}{-2.77}           & -4.82           \\ \cline{2-12}
 			\multicolumn{1}{|c|}{}                                  & Pig.                               & \multicolumn{1}{c|}{-95.18}          & -90.86          & \multicolumn{1}{c|}{-55.52}          & -55.61          & \multicolumn{1}{c|}{-64.26}          & -64.86          & \multicolumn{1}{c|}{-36.82}          & -33.37          & \multicolumn{1}{c|}{-7.00}           & -7.25           \\ \cline{2-12}
 			\multicolumn{1}{|c|}{}                                  & Ashtray.                           & \multicolumn{1}{c|}{-96.93}          & -92.00          & \multicolumn{1}{c|}{-49.54}          & -47.52          & \multicolumn{1}{c|}{-72.63}          & -74.09          & \multicolumn{1}{c|}{-22.60}          & -42.75          & \multicolumn{1}{c|}{-26.07}          & -28.03          \\ \cline{2-12}
 			\multicolumn{1}{|c|}{}                                  & Police.                            & \multicolumn{1}{c|}{-65.81}          & -54.10          & \multicolumn{1}{c|}{-9.56}           & -30.08          & \multicolumn{1}{c|}{-9.96}           & -41.13          & \multicolumn{1}{c|}{-15.38}          & -21.68          & \multicolumn{1}{c|}{-11.44}          & -4.81           \\ \cline{2-12}
 			\multicolumn{1}{|c|}{}                                  & Stereo.                            & \multicolumn{1}{c|}{-54.87}          & -84.51          & \multicolumn{1}{c|}{30.95}           & 9.58            & \multicolumn{1}{c|}{-35.98}          & -72.97          & \multicolumn{1}{c|}{-60.36}          & -52.34          & \multicolumn{1}{c|}{5.78}            & -9.91           \\ \cline{2-12}
 			\multicolumn{1}{|c|}{}                                  & Violin                             & \multicolumn{1}{c|}{-58.72}          & -51.86          & \multicolumn{1}{c|}{-25.94}          & -49.51          & \multicolumn{1}{c|}{-4.75}           & -38.30          & \multicolumn{1}{c|}{-2.18}           & -11.17          & \multicolumn{1}{c|}{-19.09}          & -12.54          \\ \cline{2-12}
 			\multicolumn{1}{|c|}{}                                  & Chair                              & \multicolumn{1}{c|}{-78.28}          & -68.94          & \multicolumn{1}{c|}{-45.82}          & -65.34          & \multicolumn{1}{c|}{-50.05}          & -62.09          & \multicolumn{1}{c|}{-49.89}          & -38.77          & \multicolumn{1}{c|}{-21.17}          & -13.75          \\ \cline{2-12}
 			\multicolumn{1}{|c|}{}                                  & \textbf{Average}                   & \multicolumn{1}{c|}{\textbf{-77.81}} & \textbf{-73.19} & \multicolumn{1}{c|}{\textbf{-24.28}} & \textbf{-44.62} & \multicolumn{1}{c|}{\textbf{-32.62}} & \textbf{-52.16} & \multicolumn{1}{c|}{\textbf{-30.58}} & \textbf{-23.53} & \multicolumn{1}{c|}{\textbf{-26.38}} & \textbf{-17.01} \\ \hline
 			\multicolumn{2}{|c|}{\textbf{Total Average}}                                                 & \multicolumn{1}{c|}{\textbf{-82.96}} & \textbf{-78.16} & \multicolumn{1}{c|}{\textbf{-36.46}} & \textbf{-53.50} & \multicolumn{1}{c|}{\textbf{-41.72}} & \textbf{-54.90} & \multicolumn{1}{c|}{\textbf{-31.16}} & \textbf{-26.62} & \multicolumn{1}{c|}{\textbf{-22.06}} & \textbf{-14.06} \\ \hline
 			\end{tabular}}
 \end{table*}

\subsection{Coding Efficiency Evaluation}	
\label{efficiency}
To validate the coding efficiency of the proposed Deep-JGAC, we firstly evaluated the joint geometry and attribute coding performance. Four perceptual quality metrics were used to measure the quality of compressed point clouds. Total geometry and attribute bit rates were counted. Secondly, the geometry coding performance of the proposed Deep-JGAC was also evaluated and compared. D1-PSNR and D2-PSNR were used to measure the geometry coded point clouds. Geometry bit rate was counted. Thirdly, the attribute coding of the proposed Deep-JGAC was also evaluated, where Y-PNSR was used as quality metric of the attribute coding.

\subsubsection{Joint Geometry and Attribute Coding}	
Table \ref{tab11} shows the BDBR between the Deep-JGAC and benchmark schemes on four quality metrics, where `/' indicates the BDBR is not available because the gap is too large. Note that the average BDBR values were calculated based on the average bit rate and average quality of multiple point clouds. We can have the following three key observations. Firstly, compared to the G-PCC, Deep-JGAC achieves BDBR gain from -8.61\% to -92.02\%, and -58.01\% on average, for all test sequences when quality is measured with the MPED. Similarly, while the compressed point clouds are measured with the TCDM, GraphSIM and MS-GraphSIM, the Deep-JGAC achieves an average of -42.61\%, -42.42\%, and -48.72\% BDBR bit reductions, respectively, which are significant. Secondly, compared with the IT-DL-PCC, the Deep-JGAC achieves an average of -35.61\%, -58.74\%, -59.18\% and -57.14\% BDBR bit reductions, respectively, in terms of MPED, TCDM, GraphSIM and MS-GraphSIM. It indicates the proposed Deep-JGAC significantly {outperforms} the IT-DL-PCC and G-PCC in the joint geometry and attribute coding.

\begin{figure*}[!t]
	\centering
	
	\subfigure[] {
		\label{fig98b}
		\includegraphics[width=0.2\linewidth]{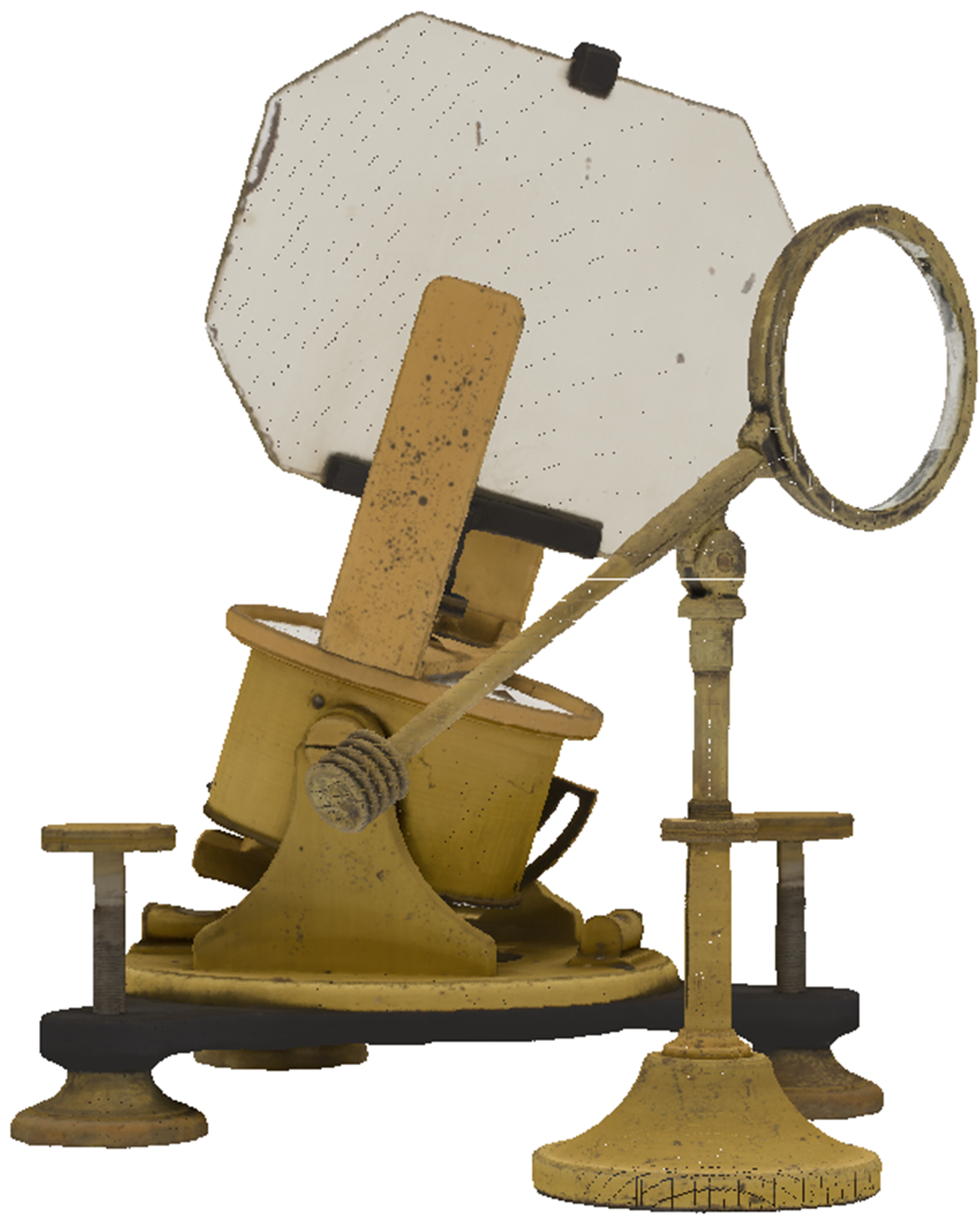}
	}
	\subfigure[] {
		\label{fig988d}
		\includegraphics[width=0.2\linewidth]{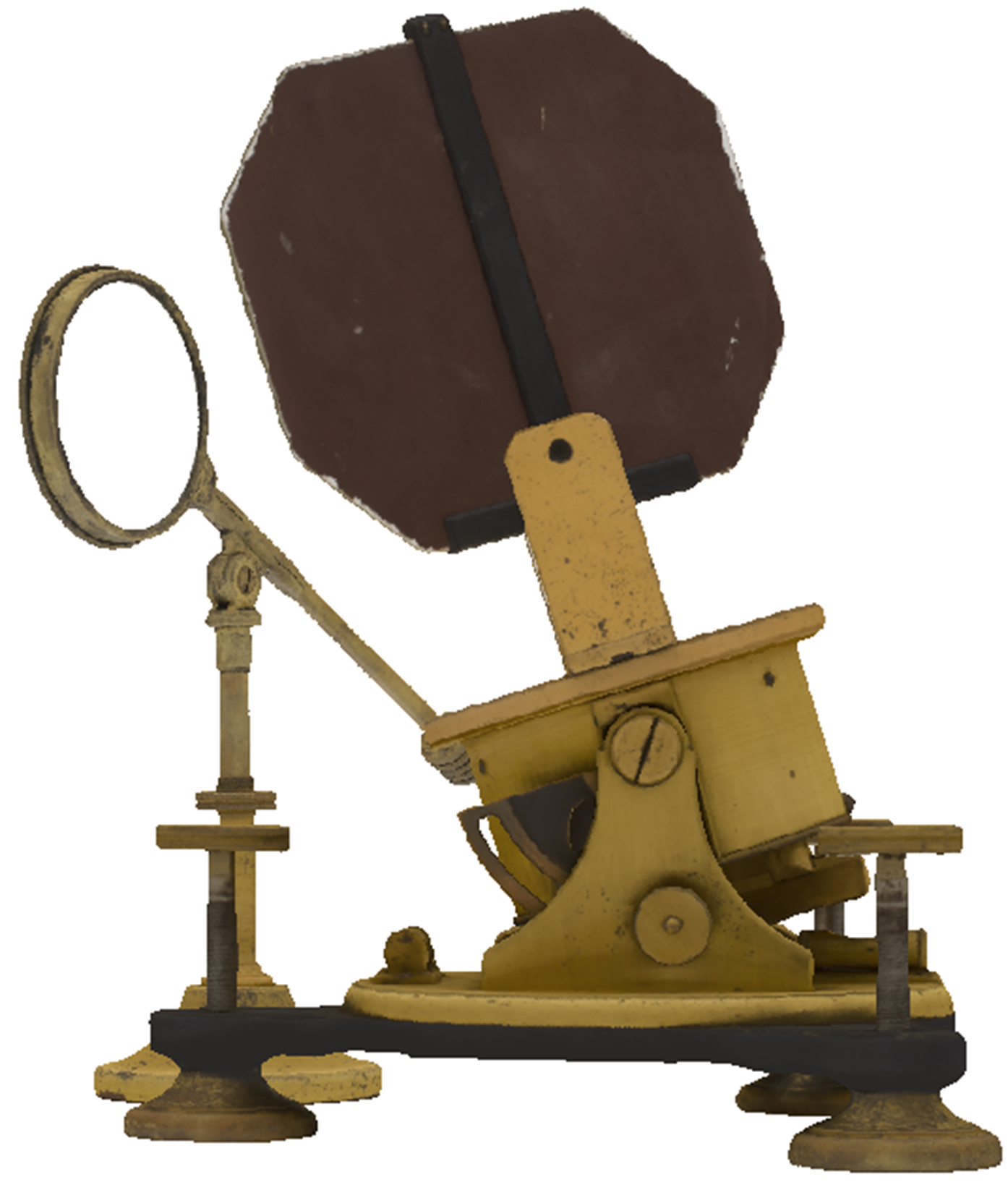}
	}
	\subfigure[] {
		\label{fig98c}
		\includegraphics[width=0.2\linewidth]{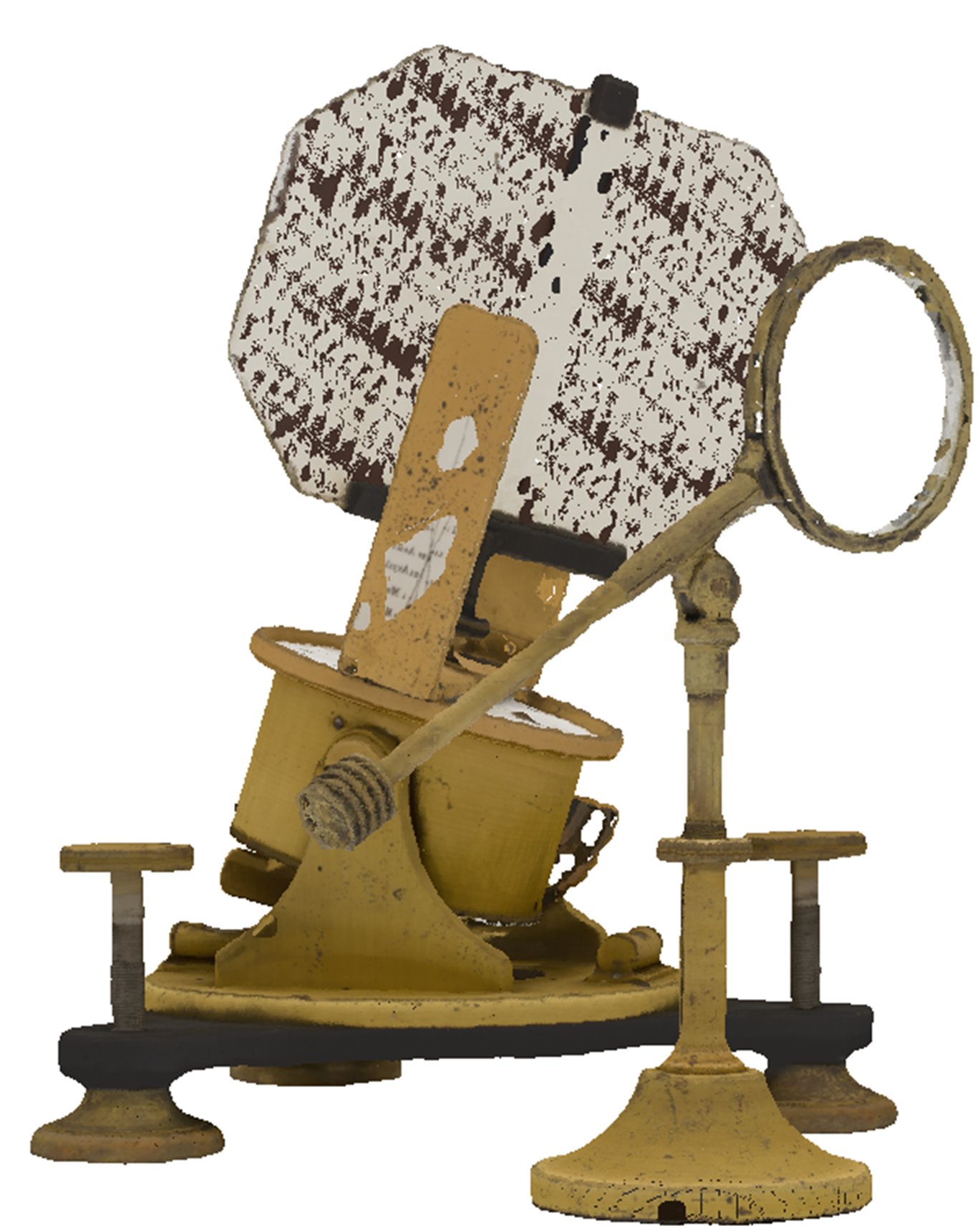}
	}
	\subfigure[] {
		\label{fig98d}
		\includegraphics[width=0.2\linewidth]{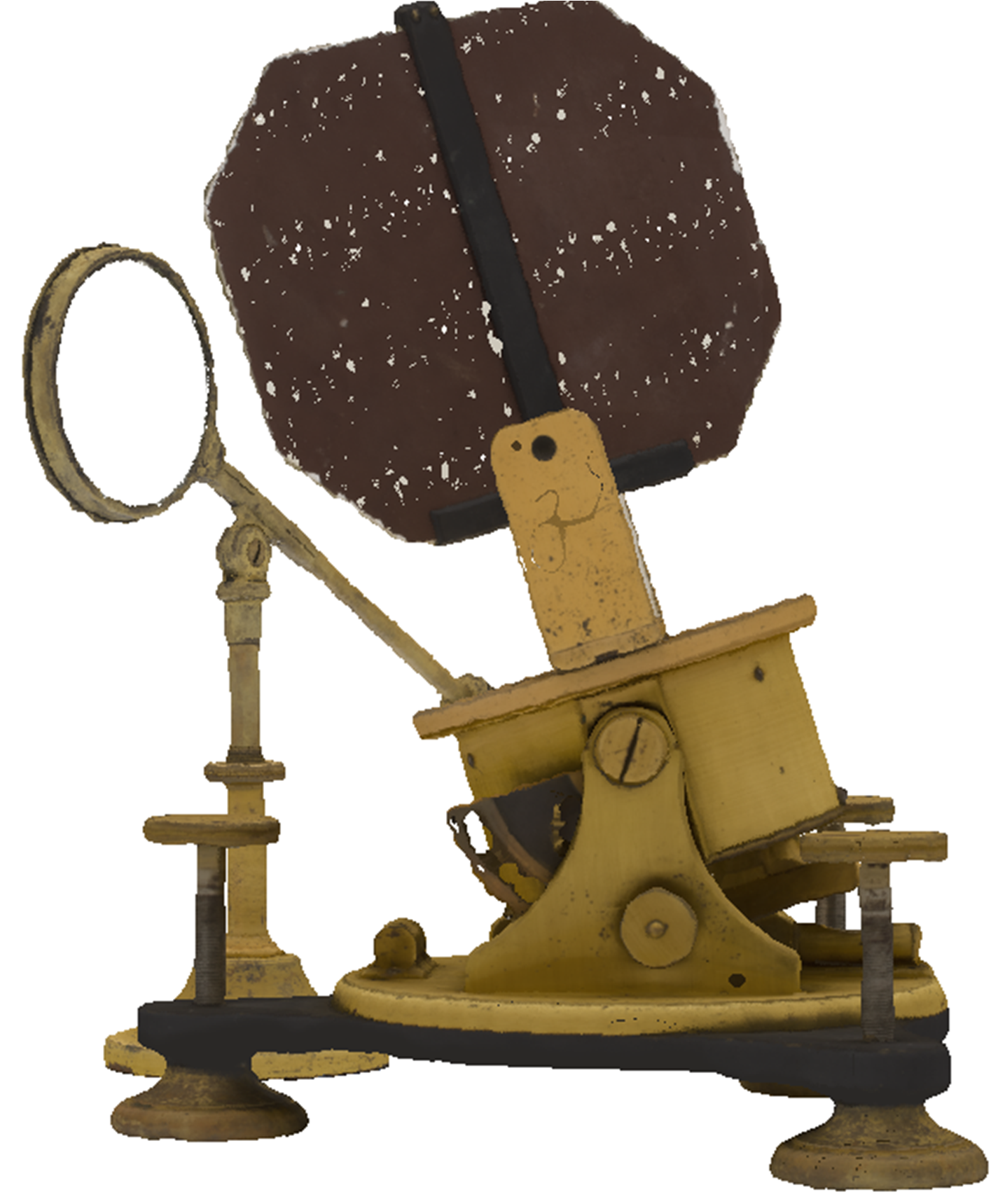}
	}
	
	\caption{Visualization {of} the source and re-colored point cloud \textit{Heliostat.} (a)-(b) Front and back views of the source point cloud. (c)-(d) Front and back view of the re-colored point cloud, where geometry is distorted.}
	\label{fig98}
\end{figure*}

Thirdly, compared with the V-PCC, Deep-JGAC achieves an average of -11.86\%, -5.08\% and -14.67\% BDBR bit reductions, respectively, in terms of MPED, GraphSIM and MS-GraphSIM. While measured with the TCDM, the BDBR of the Deep-JGAC is 3.29\%, which is inferior to that of V-PCC.
The proposed Deep-JGAC performs worse than V-PCC for some sequences, such as \textit{Heliostat.} and \textit{Gramo.}. This is mainly because the severe geometry distortion at low bitrates may significantly affect the performance of attribute compression. Fig. {\ref{fig98}} shows the visualization results of the original and geometry distorted point cloud \textit{Heliostat.}. We can observe that there are noticeable attribute differences on both sides of thin areas, such as mirrors, where front pixels are moved to back views due to the geometry distortion. It means that some geometry distortion may cause the considerable attribute distortion, although the attribute is not yet compressed. The Deep-JGAC performs relatively worse on the AVS-PC dataset compared to other datasets because most of training point clouds are from 8iVFB and Owlii, whose distributions are not well aligned with those in the AVS-PC dataset, reducing coding efficiency comparing to the V-PCC on the TCDM.

Fig. \ref{fig88} shows the average RD curves of the joint PCC. We can observe that the IT-DL-PCC performs the worst on the TCDM, GraphSIM, and MS-GraphSIM metrics, while G-PCC performs slightly better than IT-DL-PCC. On the MPED metric, IT-DL-PCC performs better than G-PCC. The possible reason is that MPED is sensitive to the number of points, and the G-PCC significantly reduces the number of points. V-PCC shows significantly better compression performance across all metrics compared to the G-PCC and IT-DL-PCC. Our Deep-JGAC has compression performance similar to V-PCC and even outperforms V-PCC in certain metrics, such as MS-GraphSIM. Overall, the proposed Deep-JGAC performs better on average than G-PCC and IT-DL-PCC across all joint metrics and outperforms the V-PCC on most quality metrics.

 \begin{figure}[!t]
 	\centering
 	\subfigure[] {
 		\label{fig8a}
 		\includegraphics[width=0.6\linewidth]{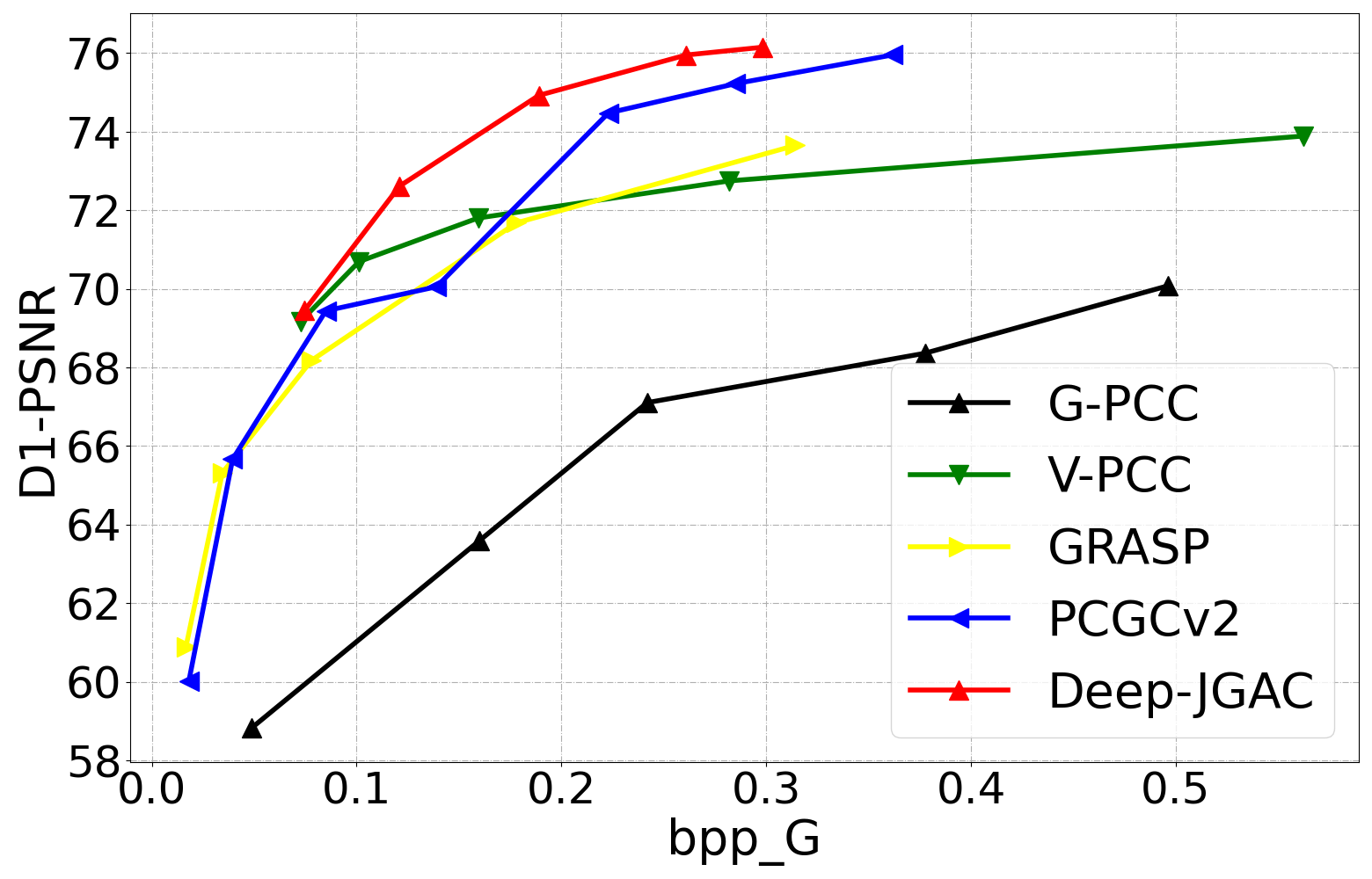}
 	}
 	\subfigure[] {
 		\label{fig8b}
 		\includegraphics[width=0.6\linewidth]{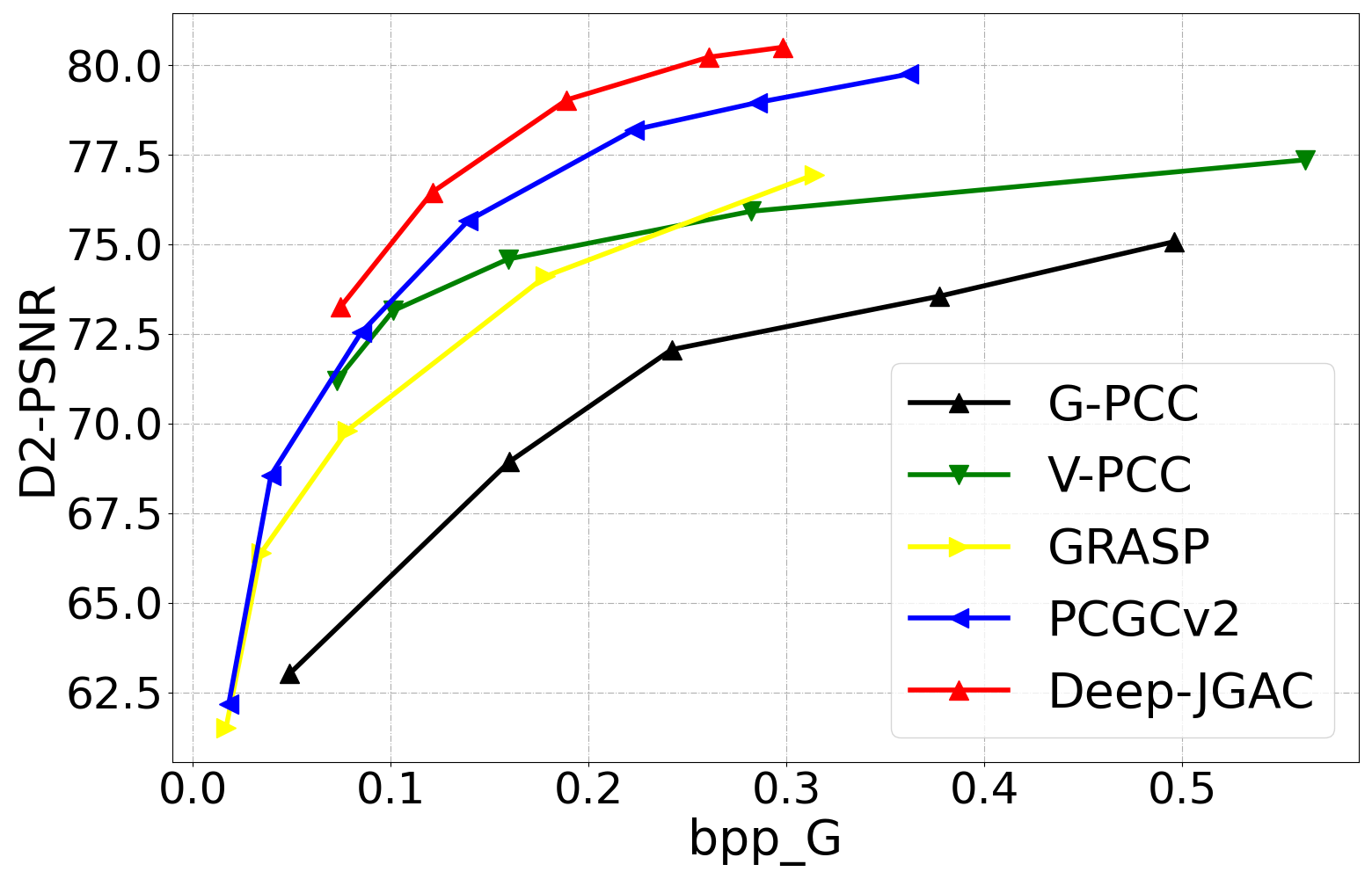}
 	}
 	\caption{The average RD curves of geometry codecs, where quality of reconstructed point clouds are measured with D1-PSNR and D2-PSNR. (a) D1-PSNR. (b) D2-PSNR.}
 	\label{fig8}
 \end{figure}

\begin{figure}[!t]
	\centering
	
	\includegraphics[width=0.6\linewidth]{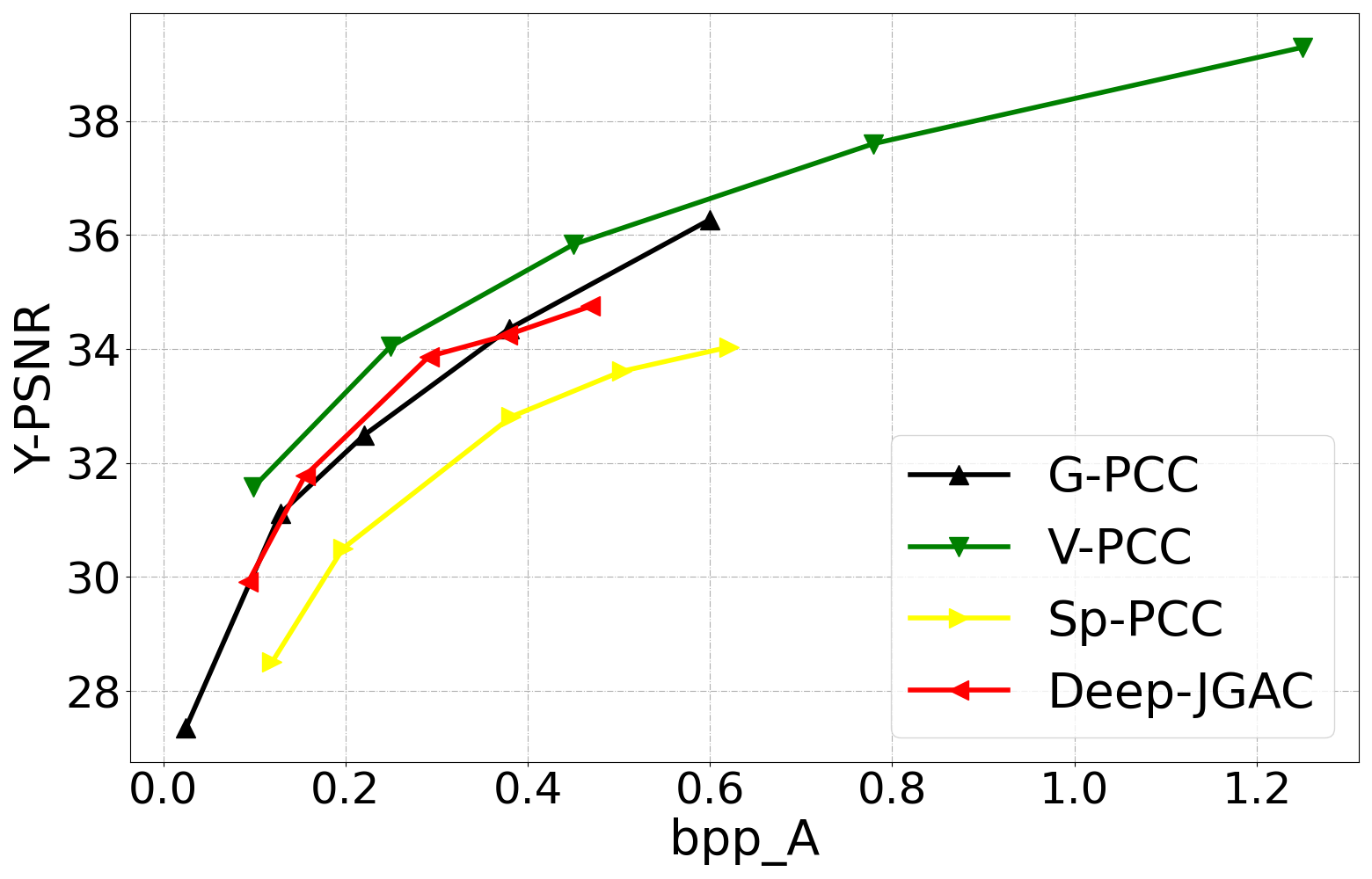}
	\caption{The average RD curves of attribute codecs, where quality of reconstructed point clouds are measured with Y-PSNR.}
	\label{figaa}
\end{figure}

\subsubsection{Geometry Coding}	
In addition to the joint coding, the geometry coding performance was evaluated and compared.
Table \ref{tab1} presents the geometry coding BDBR comparison between the Deep-JGAC and benchmarks, where the geometry quality was measured with D1-PSNR and D2-PSNR.
Compared with the G-PCC, the Deep-JGAC reduces bit rate from {-18.00\% to -97.72\% and -82.96\%} on average, when the quality is measured with D1-PSNR. In terms of the D2-PSNR, the Deep-JGAC achieves an average of -78.16\% BDBR reductions. Similarly, as compared with V-PCC, the Deep-JGAC achieves an average of -36.46\% and -53.50\% BDBR reductions while measured with D1-PSNR and D2-PSNR, respectively. The Deep-JGAC significantly {outperforms} the traditional non-learning based PCC, G-PCC and V-PCC. In addition, as comparing with GRASP, PCGCv2 and PCGCv2*, the Deep-JGAC achieves an average of -41.72\%, {-31.16\%} and -22.06\% respectively when measured with D1-PSNR. Moreover, it achieves an average of -54.90\%, {-26.26\%} and -14.06\% respectively when the geometry quality is measured with D2-PSNR. Overall, the Deep-JGAC significantly outperforms the learning based PCGC for most sequences. The Deep-JGAC performs poorly for some point clouds, such as { \textit{Gramo.} and \textit{Motor.}}, as comparing with the PCGCv2. The main reason is {that the differences in training datasets lead to inconsistent optimization directions, which can be further validated by the results of PCGCv2*}.

\begin{figure*}[!t]
	\centering
	\subfigure[] {
		\label{fig6d}
		\includegraphics[width=0.18\linewidth]{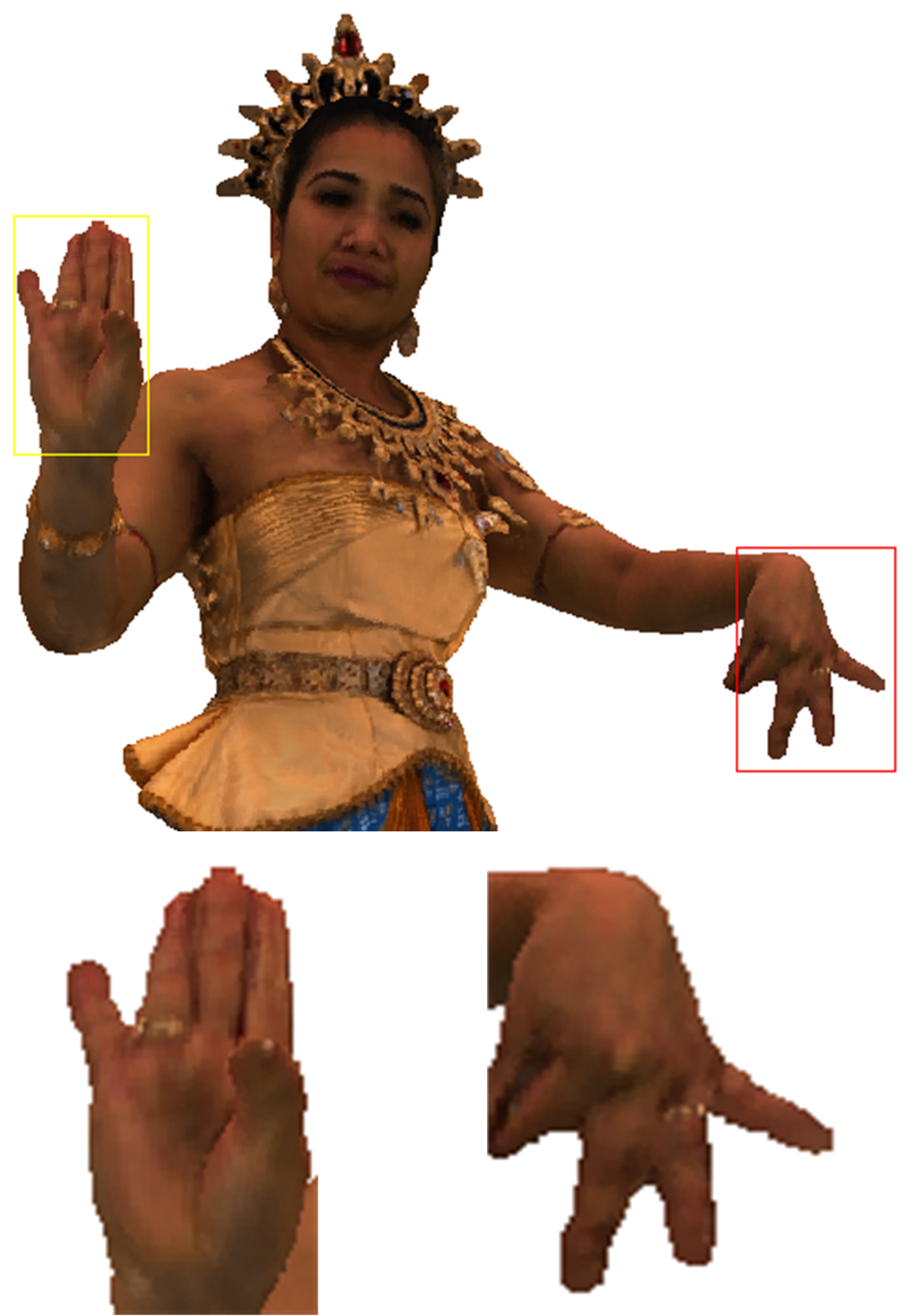}
	}
	\subfigure[] {
		\label{fig6e}
		\includegraphics[width=0.18\linewidth]{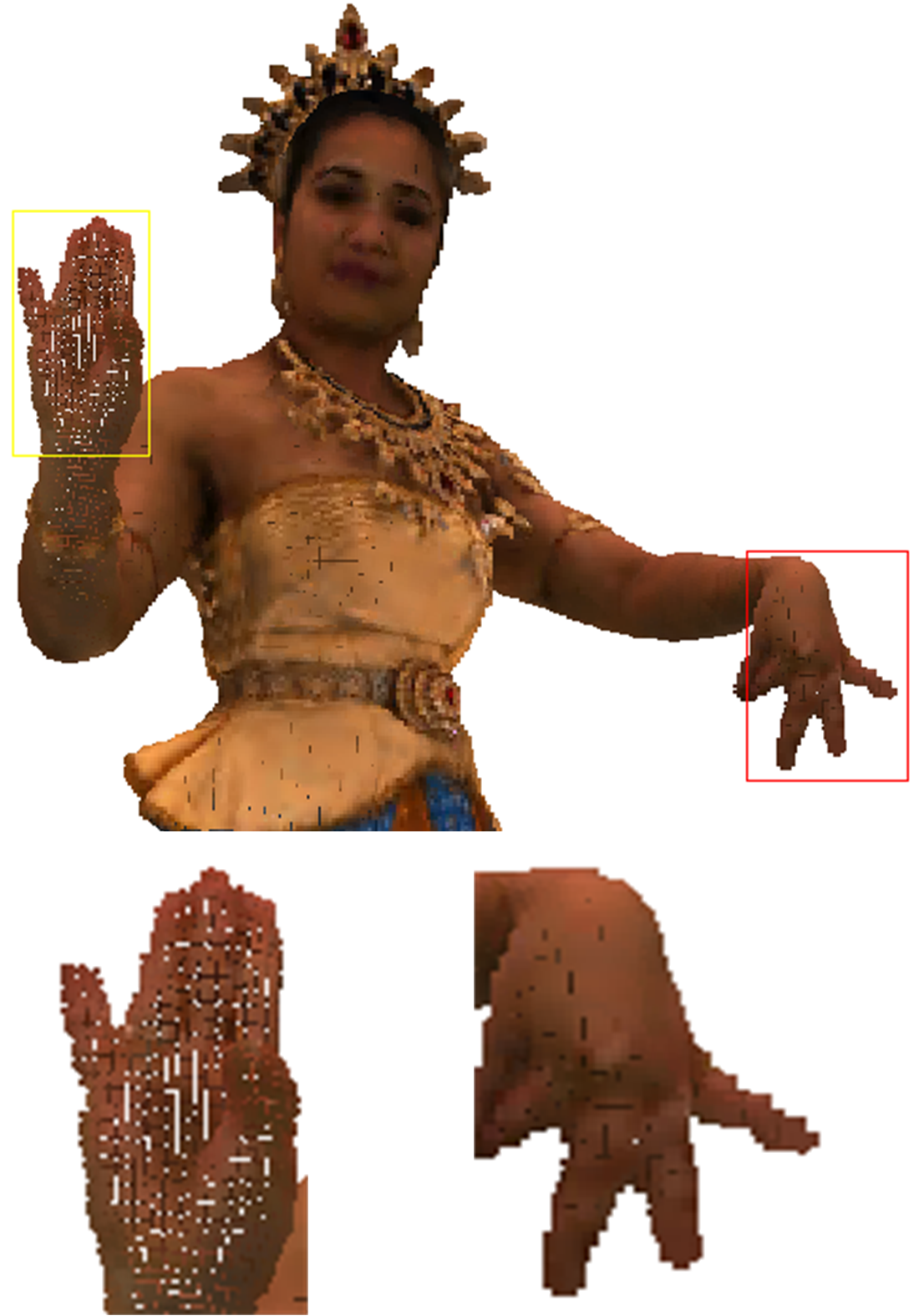}
	}
	\subfigure[] {
		\label{fig6f}
		\includegraphics[width=0.18\linewidth]{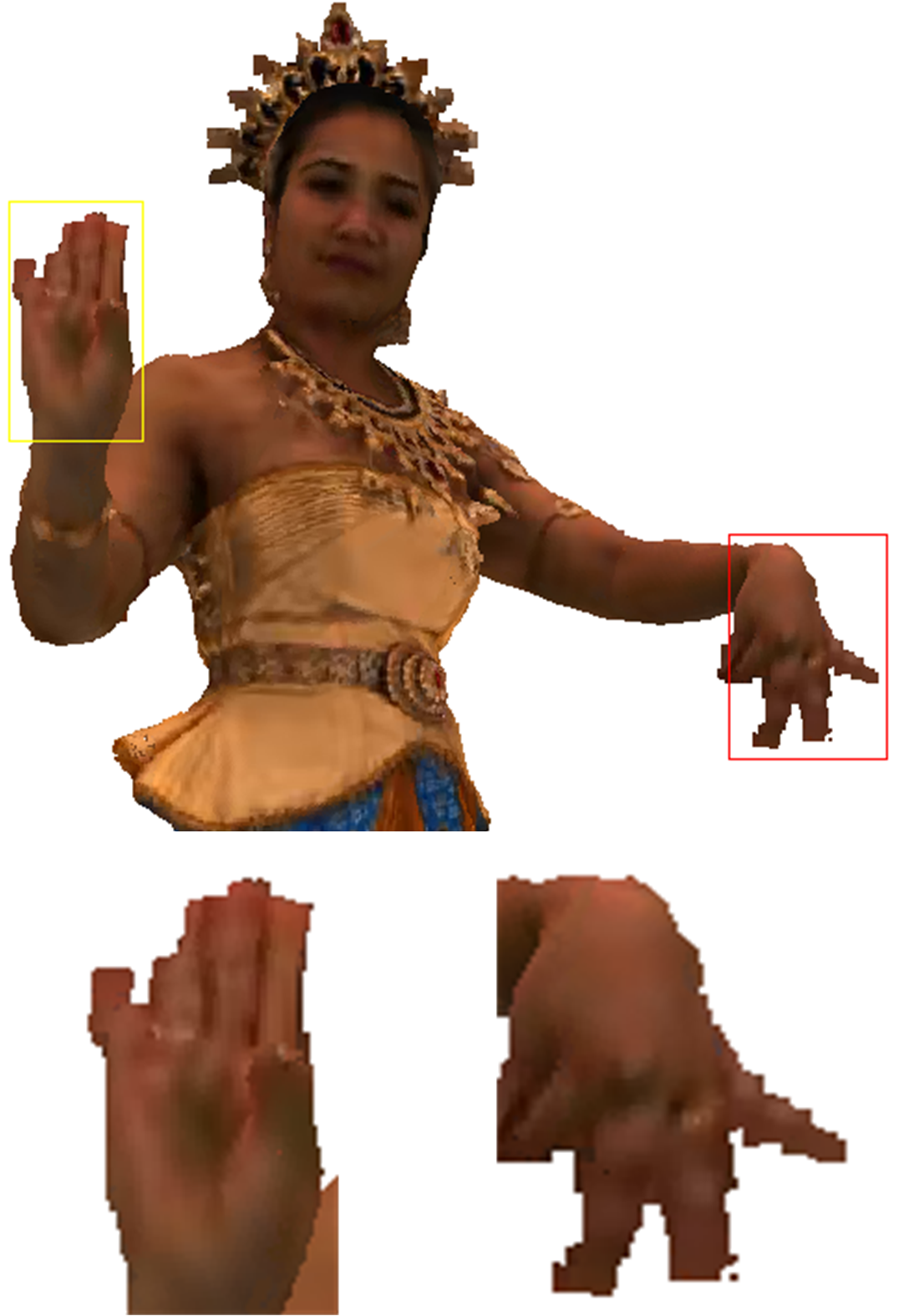}
	}
	\subfigure[] {
		\label{fig6g}
		\includegraphics[width=0.18\linewidth]{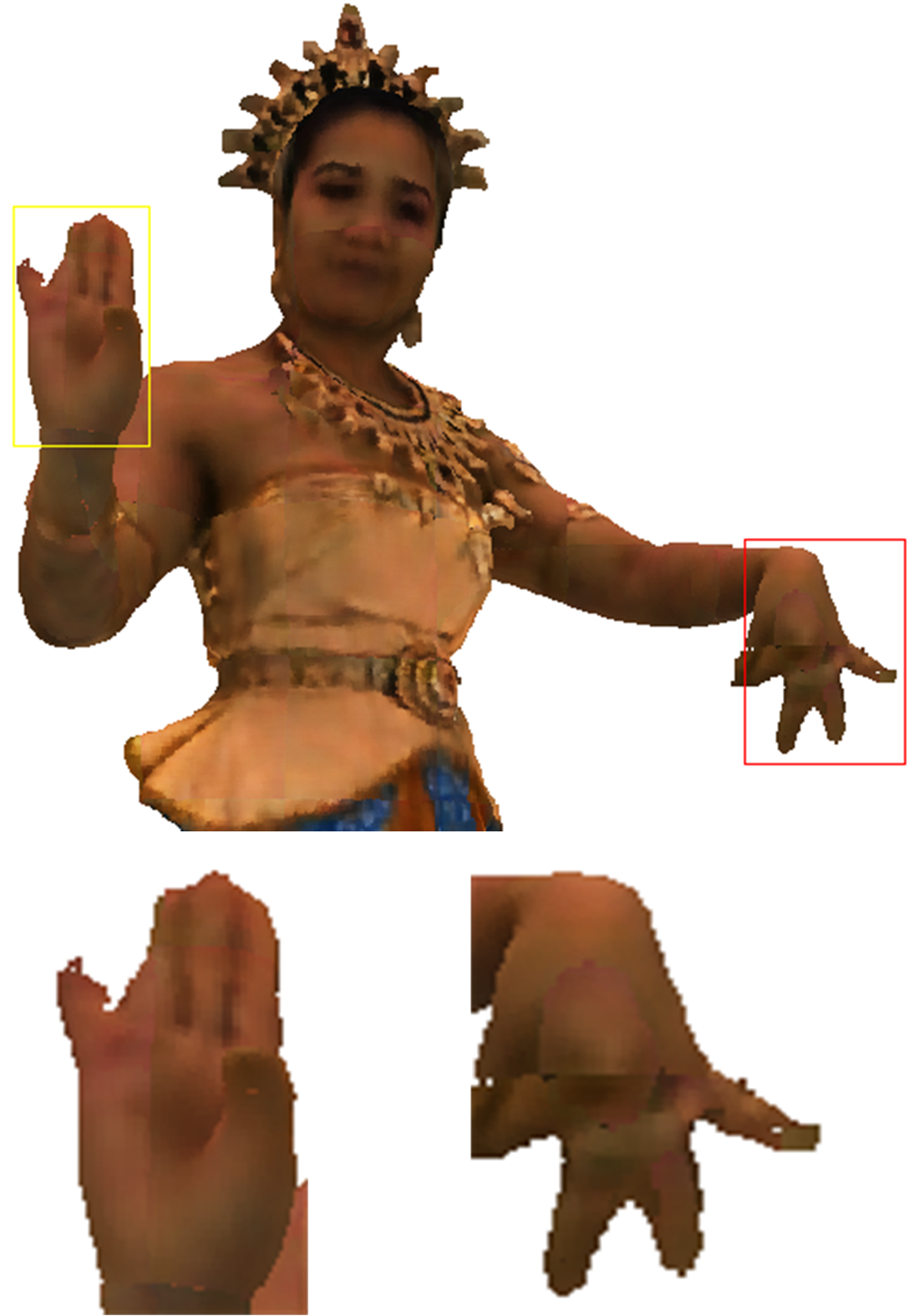}
	}
	\subfigure[] {
		\label{fig6so}
		\includegraphics[width=0.18\linewidth]{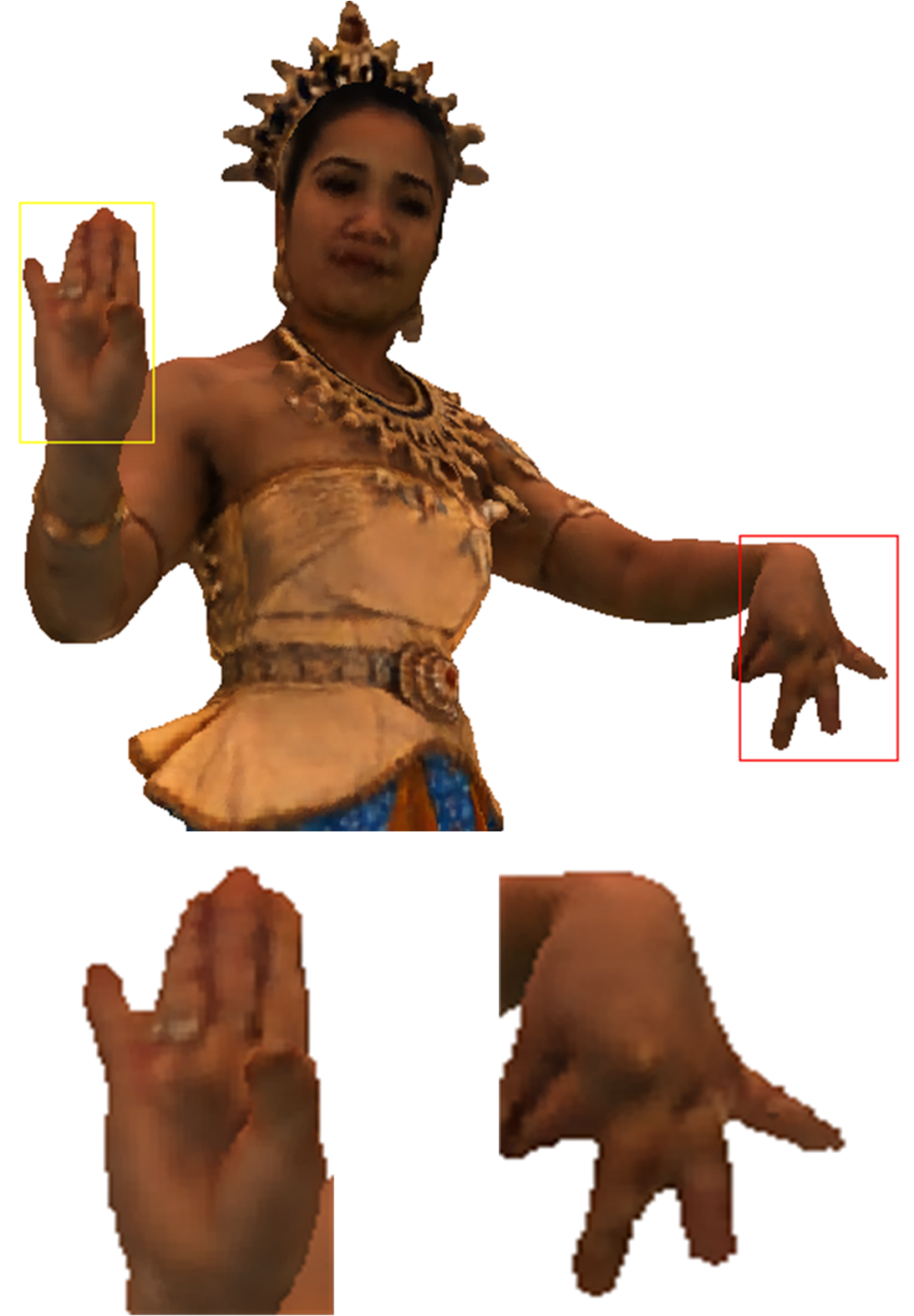}}
	
	\subfigure[] {
		\label{fig6dd}
		\includegraphics[width=0.18\linewidth]{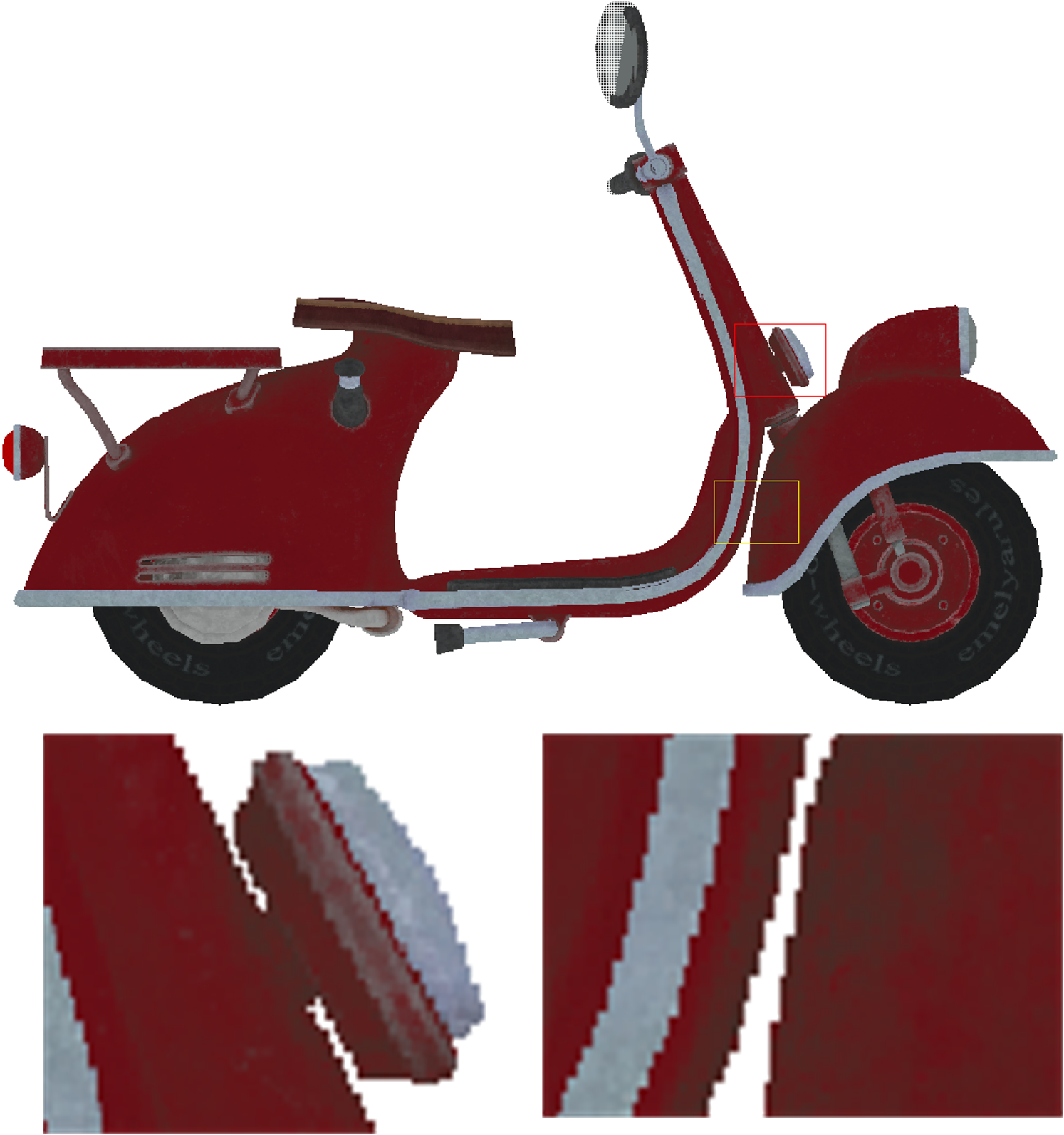}
	}
	\subfigure[] {
		\label{fig6ed}
		\includegraphics[width=0.18\linewidth]{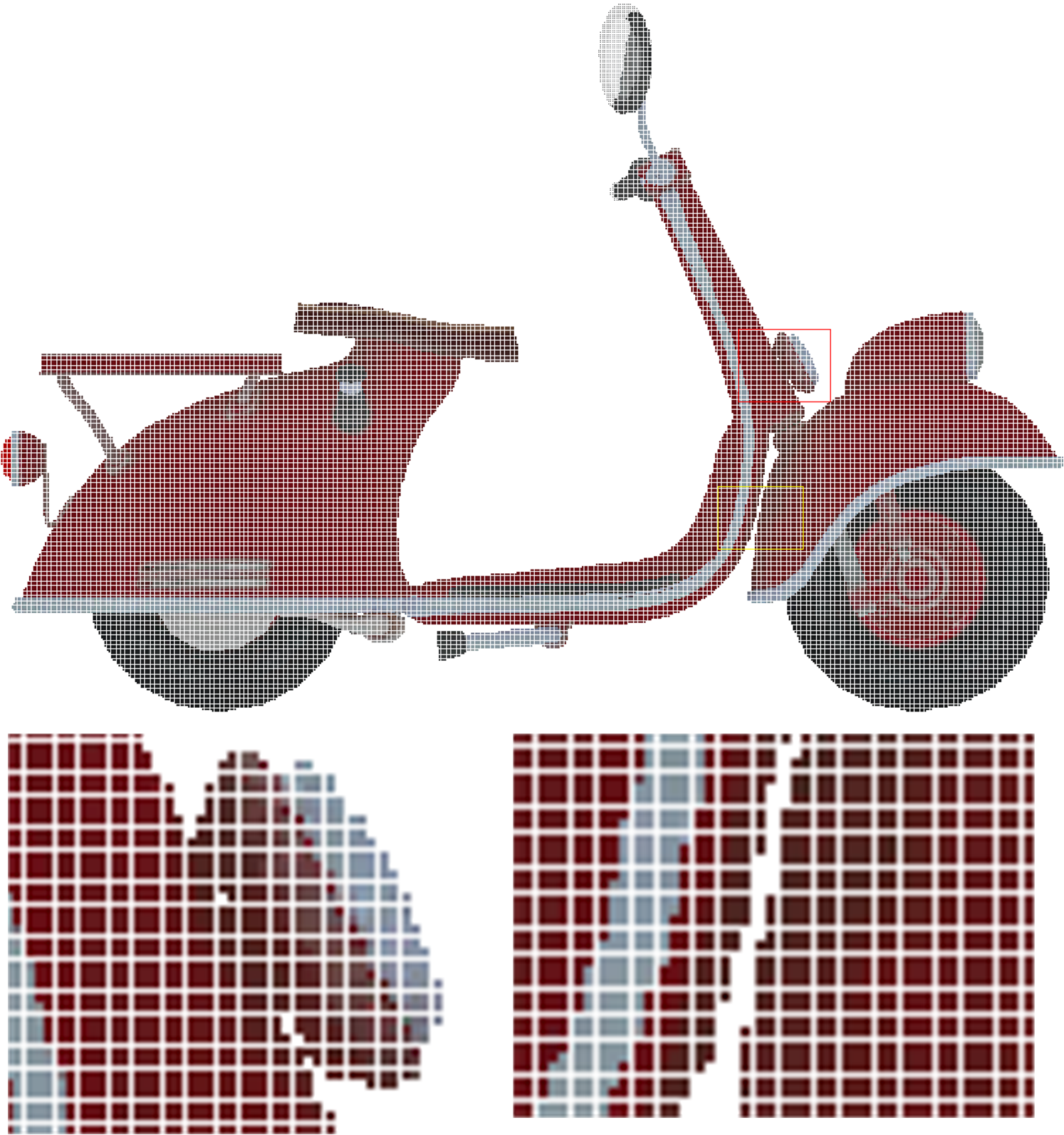}
	}
	\subfigure[] {
		\label{fig6df}
		\includegraphics[width=0.18\linewidth]{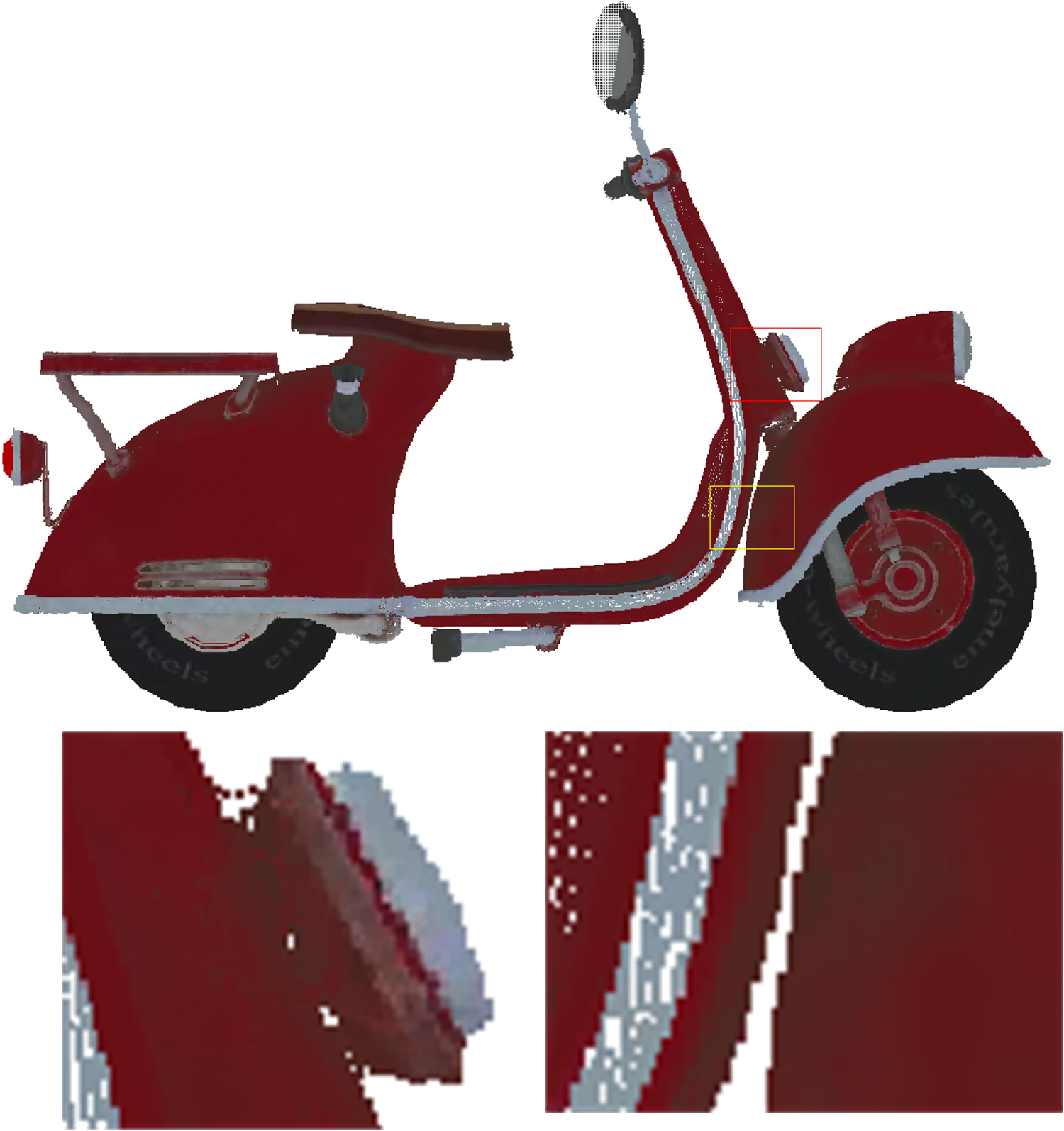}
	}
	\subfigure[] {
		\label{fig6dg}
		\includegraphics[width=0.18\linewidth]{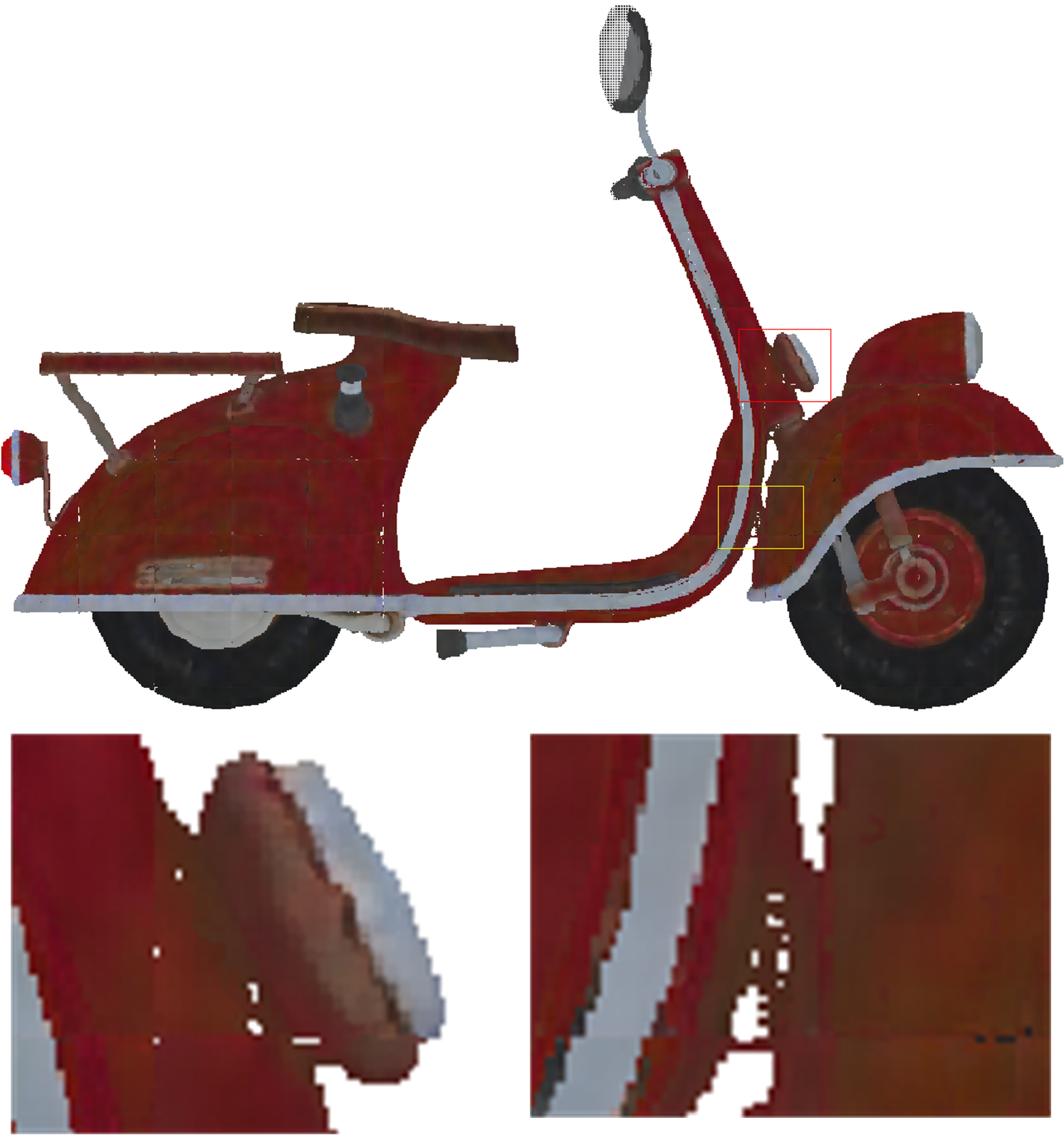}
	}
	\subfigure[] {
		\label{fig6sdo}
		\includegraphics[width=0.18\linewidth]{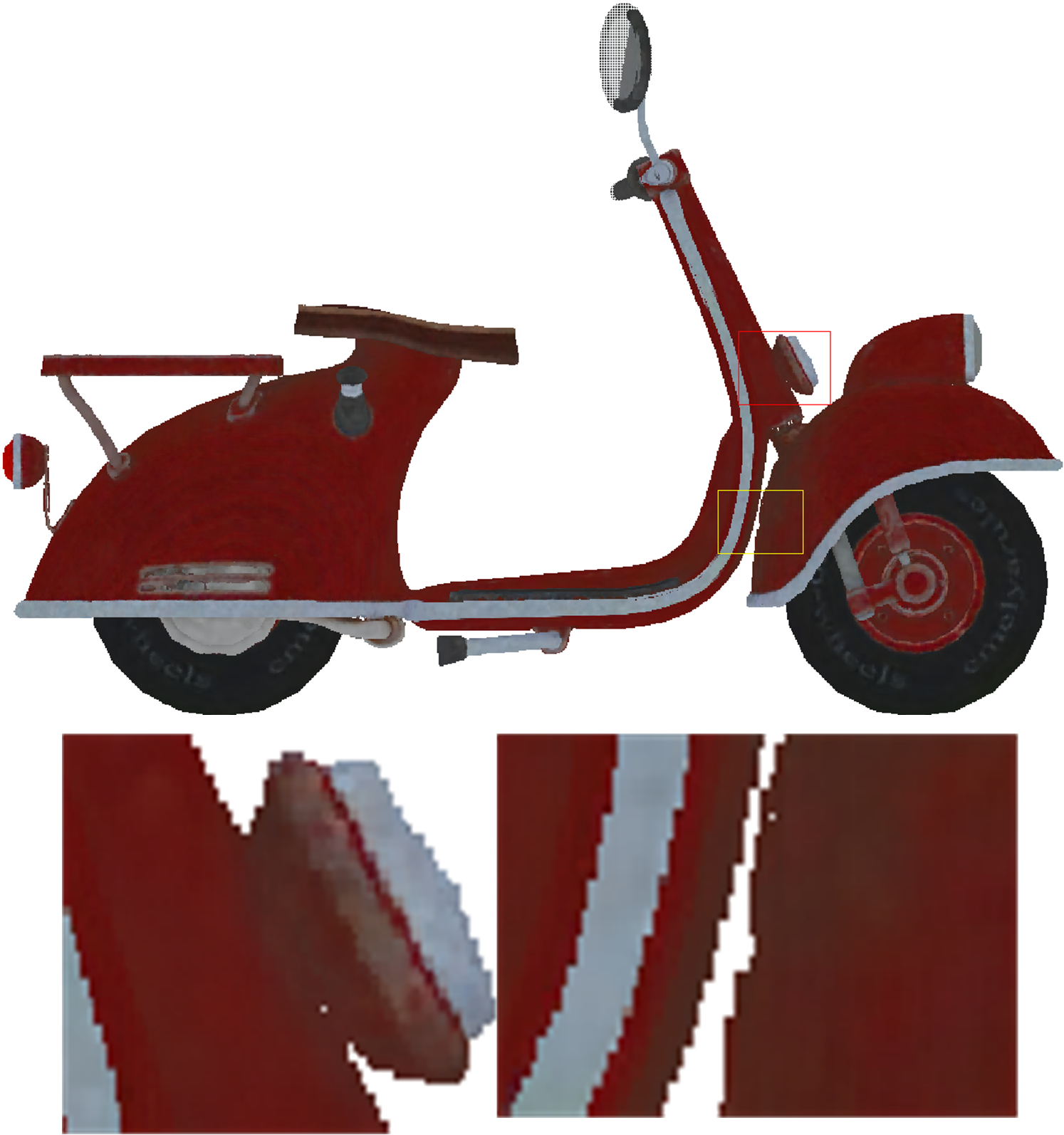}}
	
	\caption{{Visual quality comparison of the compressed point clouds, where rectangle regions are enlarged. {Digitals} denote the bit rate and visual quality, i.e. (bpp, MS-GraphSIM value).  (a) Original \textit{Thaidancer}, (b)  G-PCC (0.711, 0.444), (c)  V-PCC (0.564, 0.513), (d)  IT-DL-PCC (0.656, 0.429), (e)  Deep-JGAC (0.626, 0.560),   (f) Original \textit{Motor}, (g) G-PCC (0.472, 0.407), (h) V-PCC (0.502, 0.524), (i) IT-DL-PCC (0.470, 0.359),  (j) Deep-JGAC (0.444, 0.519).}
	}
	\label{fig6666}
\end{figure*}
\begin{figure}[!t]
	\centering
	
	\includegraphics[width=0.8\columnwidth]{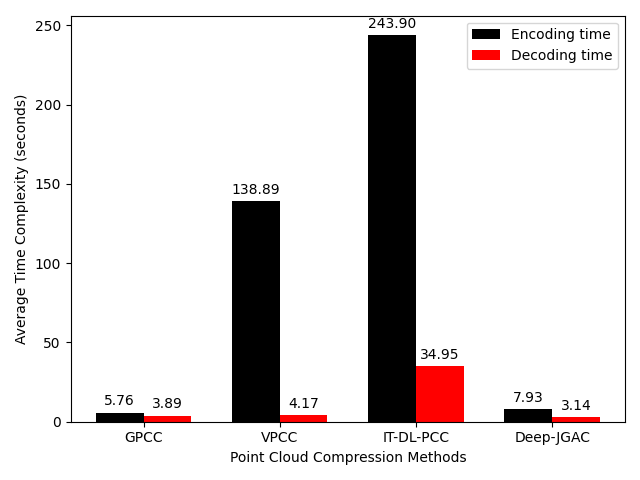}
	\caption{Encoding and decoding time of joint PCCs.}
	\label{fig9}
\end{figure}

Figs. \ref{fig8a} and \ref{fig8b} show the average RD curves of different geometry compression methods using D1-PSNR and D2-PSNR as geometry metrics. It can be seen that G-PCC has poor compression performance, followed by GRASP. PCGCv2 and V-PCC perform better than G-PCC and GRASP, with PCGCv2 outperforming V-PCC.  Our Deep-JGAC achieves the better results. Overall, our Deep-JGAC achieves better performance compared to other methods.
\subsubsection{Attribute Coding}
Attribute is attached to geometry in point clouds, which can not be compressed independently. Therefore, to evaluate the performance of attribute coding, we performed joint geometry and attribute coding, {and} then only bit rate and quality of attribute were measured and counted.
Fig. \ref{figaa} shows the average RD curves of attribute coding schemes, where $y$-axis is the attribute quality measured with Y-PSNR and $x$-axis is the attribute bit rate. In addition to G-PCC and V-PCC,  we replaced the attribute codec in Deep-JGAC with the attribute codec proposed by Sparse-PCAC \cite{8}, which is a variant combining Deep-JGAC and Sparse-PCAC, denoted as Sp-PCC.
We can observe that in terms of attribute coding, the Deep-JGAC is superior to Sp-PCC and slightly superior to the G-PCC. In addition, the Deep-JGAC is inferior to V-PCC in attribute coding. The main reason is the V-PCC is mature and effective in compressing attribute. There are still some rooms to be improved for the learning based attribute coding. The experimental results of Sp-PCC, which combines Deep-JGAC and Sparse-PCAC, indicate that the proposed Deep-JGAC framework is flexible and plug-and-play for geometry or attribute coding. This coding experiment demonstrates the {flexibility} of the proposed Deep-JGAC framework and superiority of attribute coding.

\subsection{Visual Quality Evaluation}
We also compare the visual quality of reconstructed point clouds from different PCC schemes. Similar bitrates were selected for fair comparisons. Fig. \ref{fig6666} shows the reconstructed point clouds \textit{Thaidancer} and \textit{Motor} from different codecs. As seen in Fig. \ref{fig6e}, the $Thaidancer$ compressed by G-PCC lost many points, resulting in many holes and poor visual quality. The bitrate and MS-GraphSIM are 0.711 and 0.444. In addition, the point clouds coded by V-PCC and IT-DL-PCC are shown in Figs. \ref{fig6f} and \ref{fig6g}{,} where the bitrate and MS-GraphSIM values are 0.564 and 0.513, and 0.656 and 0.429, respectively. There are many edge artifacts, such as around the pinky finger, and color artifacts, such as in the gaps between fingers. Fig. \ref{fig6so} shows the reconstructed point clouds from Deep-JGAC. The bitrate and MS-GraphSIM values of Deep-JGAC are 0.626 and 0.560, having lower bitrate and higher quality as comparing with benchmarks. We can also observe that edge structures are smoother, and the internal textures are clear. Among these compression schemes, the point clouds compressed by Deep-JGAC are the closest to the original point clouds \ref{fig6d} at similar bit rate. Similar results can be found from Figs. \ref{fig6dd} to \ref{fig6sdo}. These results validate the superior reconstruction quality of the proposed Deep-JGAC.

\subsection{Computational Complexity Evaluation}
We also evaluate the computational complexity of the different joint PCCs in {tested point clouds.} The G-PCC and V-PCC were executed on the CPU, while the learning-based schemes were executed on the CPU plus GPU. The computing platform is workstation using Intel Core i9-10900 CPU and NVIDIA GeForce RTX 3090 GPU. Operating system is Ubuntu 22.04. Fig. \ref{fig9} shows the average encoding/decoding time of the joint PCCs, including G-PCC, V-PCC, IT-DL-PCC and Deep-JGAC. The average encoding/decoding time of G-PCC, V-PCC, and IT-DL-PCC are 5.76s/3.89s, 138.89s/4.17s, and 243.90s/34.85s, respectively. The IT-DL-PCC has the highest time complexity. The average encoding/decoding time of Deep-JGAC are 7.93s/3.14s, which are similar to those of G-PCC and significantly lower than those of V-PCC and IT-DL-PCC. The Deep-JGAC {reduces} an average of 96.75\%/91.02\% encoding/decoding time compared to the learning based IT-DL-PCC. There are two key reasons. One is that Deep-JGAC {uses} sparse convolution {which reduces} the time complexity and memory consumption of PCC. Secondly, the optimized re-colorization module in Subsection \ref{sec:recolor} significantly reduces the number of NNA operations for each point, which reduces an average of 83.07\% encoding time, as shown in Table \ref{tab21}.

\subsection{Ablation Study}
We conducted ablation experiments to verify the effectiveness of the proposed AIFM in Deep-JGAC. Specifically, we changed the interaction between geometry and attribute features with concatenation or addition, which are denoted as `Deep-JGAC (Concat+Attr)' and `Deep-JGAC (Add+Attr)'. Additionally, to validate the geometry coding gain achieved by using attribute information, we replace the attribute input in the attribute-assisted geometry coding with the point cloud geometry, denoted as `Deep-JGAC (AIFM+Geo)', where no attribute was used. All the variant Deep-JGAC models were retrained with the RWTT dataset with the same settings. {The test dataset used is consistent with the previous experiments.}

	\begin{table}[!t]
		\caption{Average BDBR achieved by Deep-JGAC geometry coding comparing with the PCGCv2. [Unit: \%]. }
		\label{tab22}
		\centering
		\begin{tabular}{|c|cc|}
			\hline
			\multirow{2}{*}{Methods} & \multicolumn{2}{c|}{Deep-JGAC vs PCGCv2}           \\ \cline{2-3}
			& \multicolumn{1}{c|}{D1-PSNR} & D2-PSNR \\ \hline
			Deep-JGAC (Add+Attr)                      & \multicolumn{1}{c|}{-8.43}        & -6.37        \\ \hline
			Deep-JGAC (Concat+Attr)                   & \multicolumn{1}{c|}{-19.45}       & -15.38       \\ \hline
			Deep-JGAC (AIFM+Geo)               & \multicolumn{1}{c|}{-10.90}        & -14.29       \\ \hline
			Deep-JGAC              & \multicolumn{1}{c|}{-20.31}       & -26.55       \\ \hline
		\end{tabular}
	\end{table}
	
Table \ref{tab22} shows geometry coding efficiency comparison between Deep-JGAC, its variants and the PCGCv2. We have two key findings. Firstly, we can observe that the Deep-JGAC(Add+Attr) saves an average of -8.43\% and -6.37\% bit rate, respectively, in terms of D1-PSNR and D2-PSNR, which are the lowest ones. The Deep-JGAC (Concat+Attr) achieves an average of -19.45\% and -15.38\% BDBR gain, respectively, which indicates the concatenation is more effective than the addition in feature fusion. Moreover, The Deep-JGAC using the AIFM in the last row achieves an average of -20.31\% and -26.55\% BDBR gains, which are higher than those of using concatenation and addition. It indicates the proposed AIFM is more effective. Secondly, while measured with D1-PSNR and D2-PSNR, the `Deep-JGAC(AIFM+Geo)' achieves an average of -10.90\% and -14.29\% BDBR gains, which are almost half of the Deep-JGAC. It indicates that the geometry coding is significantly improved by introducing the attribute information in the proposed Deep-JGAC.

\section{Conclusions}
\label{section4}
	We propose an end-to-end Deep Joint Geometry and Attribute point cloud Compression (Deep-JGAC) framework for dense colored point clouds. Firstly, we propose a flexible Deep-JGAC framework, where the
geometry and attribute sub-encoders are compatible to either learning or non-learning based geometry and attribute encoders. We propose an attribute-assisted deep geometry encoder that enhances the geometry latent representation with the help of attribute. We present an optimized re-colorization module to attach the attribute to the geometrically distorted point cloud for attribute coding, which maintains high quality and {reduces} the computational complexity. The proposed Deep-JGAC outperforms the state-of-the-art non-learning (G-PCC, V-PCC) and learning based point cloud compression schemes (IT-DL-PCC, GRASP, PCGCv2) in terms of joint coding and geometry coding.

	\bibliographystyle{IEEEtran}
	\bibliography{IEEEabrv,aaaaa}

\begin{IEEEbiography}[{\includegraphics[width=1in,height=1.25in,clip,keepaspectratio]{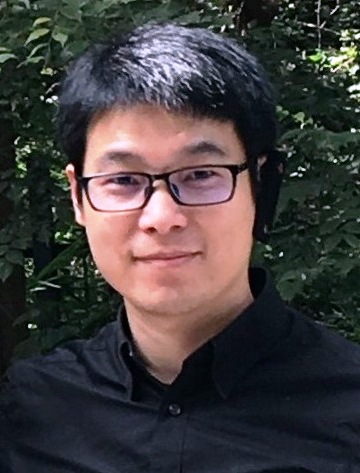}}]{Yun Zhang (Senior Member, IEEE)}
 received the B.S. and M.S. degrees in electrical engineering from Ningbo University, Ningbo, China, in 2004 and 2007, respectively, and the Ph.D. degree in computer science from the Institute of Computing Technology (ICT), Chinese Academy of Sciences (CAS), Beijing, China, in 2010. From 2009 to 2014, he was a Visiting Scholar with the Department of Computer Science, City University of Hong Kong, Kowloon, Hong Kong. From 2010 to 2022, he was a Professor/Associate Professor with the Shenzhen Institutes of Advanced Technology (SIAT), CAS, Shenzhen, China. Since 2022, he is a Professor with the School of Electronics and Communication Engineering, Sun Yat-sen University, Shenzhen Campus, Guangdong, China. His research interests mainly include 3D visual signal processing, image/video compression, visual perception, and machine learning.
\end{IEEEbiography}

\begin{IEEEbiography}[{\includegraphics[width=1in,height=1.25in,clip,keepaspectratio]{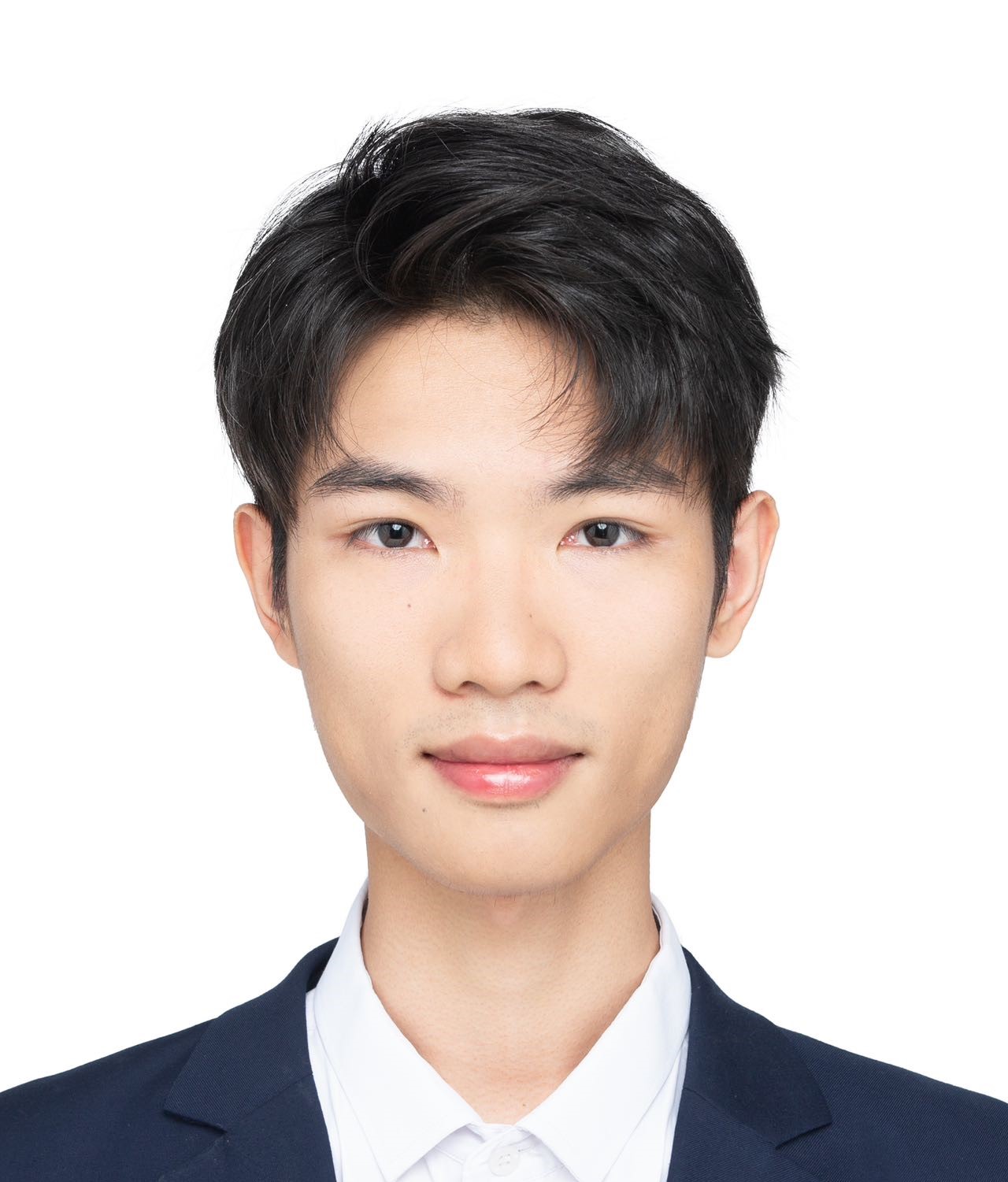}}]{Zixi Guo}
received the B.E. degree in information engineering from Guangdong University of Technology, China, in 2022. Since 2022, he has been pursuing the master’s degree in electronic information with the School of Electronics and Communication Engineering, Sun Yat-sen University, Shenzhen, China. His current research interests include video compression, point cloud compression, and deep learning.
\end{IEEEbiography}

\begin{IEEEbiography}[{\includegraphics[width=1in,height=1.25in,clip,keepaspectratio]{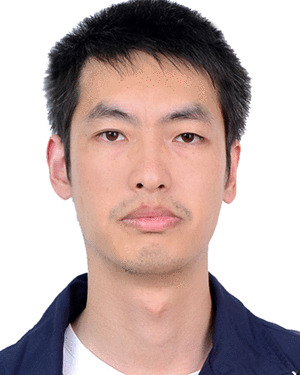}}]{Linwei Zhu}
received the B.S. degree in applied physics from the Tianjin University of Technology, Tianjin, China, in 2010, the M.S. degree in signal and information processing from Ningbo University, Ningbo, China, in 2013, and the Ph.D. degree from the Department of Computer Science, City University of Hong Kong, Hong Kong SAR, China, in 2019. He is currently an Associate Professor with Shenzhen Institute of Advanced Technology, Chinese Academy of Sciences, Shenzhen, China. His research interests mainly include depth image-based rendering, depth estimation, and machine learning based video coding/transcoding.
\end{IEEEbiography}

\begin{IEEEbiography}[{\includegraphics[width=1in,height=1.25in,clip,keepaspectratio]{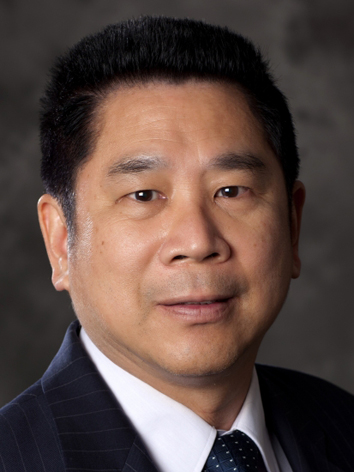}}]{C.-C. Jay Kuo (Life Fellow, IEEE)} received the B.S. degree in electrical engineering from the National Taiwan University, Taipei, Taiwan, in 1980, and the M.S. and Ph.D. degrees in electrical engineering from the Massachusetts Institute of Technology, Cambridge, in 1985 and 1987, respectively. He is currently the holder of Ming Hsieh Chair Professorship and is a Distinguished Professor of Electrical and Computer Engineering and Computer Science and the Director of the USC Multimedia Communication Laboratory (MCL), University of Southern California, Los Angeles, CA, USA. Dr. Kuo is a Fellow of the American Association for the Advancement of Science (AAAS), the Association for Computing Machinery (ACM), the Institute of Electrical and Electronics Engineers (IEEE), the National Academy of Inventors (NAI), the International Society for Optical Engineers (SPIE), and an Academician of Academia Sinica in Taiwan.
\end{IEEEbiography}

\end{document}